\documentclass[preprint,aps]{revtex4}
\usepackage{longtable}
\usepackage{epsfig}
\usepackage{graphicx}
\usepackage{subfigure}
\usepackage{amssymb}
\usepackage{amsmath}

\def\be{\begin{equation}}
\def\ee{\end{equation}}
\def\bea{\begin{eqnarray}}
\def\eea{\end{eqnarray}}
\def\ba{\begin{array}}
\def\ea{\end{array}}
\newcommand{\beq}{\begin{equation}}
\newcommand{\eeq}{\end{equation}}

\begin{document}
\title{Dynamics of Entanglement in one and two-dimensional spin systems}
\author{Gehad Sadiek\footnote{E-mail: gehad@ksu.edu.sa}}
\affiliation{Department of Physics, King Saud University, Riyadh, Saudi Arabia \\ Department of Physics, Ain Shams University, Cairo 11566, Egypt}
\author{Qing Xu}
\affiliation{Department of Chemistry, Purdue University, West Lafayette, IN 47907}
\author{Sabre Kais}
\affiliation{Department of Chemistry and Birck Nanotechnology center, Purdue University, West Lafayette, IN 47907}
\maketitle
\tableofcontents

\section{Introduction}
The state of a classical composite system is described in the phase space as a product of its individual constituents separate states. On the other hand, the state of a composite quantum system is expressed in the Hilbert space as a superposition of tensor products of its individual subsystems states. This means that the state of the quantum composite system is not necessarily expressible as a product of the individual quantum subsystems states. This peculiar property of quantum systems is called Entanglement, which has no classical analog \cite{Peres1993}. The phenomenon of entanglement was first introduced by Schrodinger \cite{Schrodinger1935} who called it ''Verschrankung'' and stated: ''For an entangled state the best possible knowledge of the whole does not include the best possible knowlege of its parts''. Quantum entanglement is a nonlocal correlation between two (or more) quantum systems such that the description of their states has to be done with reference to each other even if they are spatially well separated. In the early days of the quantum theory the notion of entanglement was first noted and introduced by Einstein et al. \cite{EPR} as a paradox in the formalism of the quantum theory. Einestien, Podolsky and Rosen in their famous EPR paper proposed a thought experiment to demonstrate that the quantum theory is not a complete physical theory as it lacks the elements of reality needed for such a theory. It needed about three decades before performing an experiment that invalidated the EPR argument and guaranteed victory to the quantum theory. The experiment was based on a set of inequalities derived by John Bell \cite{Bell1964}, which relate correlated measurements of two physical quantities that should be obeyed by any local theory. He demonstrated that the outcomes in the case of quantum entangled states violate the Bell inequality. This result emphasizes that entanglement is a quantum mechanical property that can not be simulated using a classical formalism. It took about three decades for the theoretical results confirming the Bell inequality violation to be obtained \cite{Gisin1991}.

Recently the interest in studying quantum entanglement was sparked by the development in the field of quantum computing which was initiated in the eighties by the pioneering work of Benioff, Bennett, Deutsch, Feynman and Landauer \cite{Benioff1981,Benioff1982, Bennett1985, Deutsch1985,Deutsch1989, Feynman1982, Landauer1961}. This interest gained a huge boost in 1994 after the distinguishable work of Peter shore where he developed a quantum computer algorithm for efficiently prime factorizing the composite integers \cite{Shor1994}. Other fields where entanglement plays a major role are quantum teleportation \cite{Bouwmeester1997,Bouwmeester1998}, dense coding \cite{Bennett1992-2, Mattle1996}, quantum communication \cite{Schumacher1995} and quantum
cryptography \cite{Bennett1993}.

Different physical systems have been proposed as reliable candidates for the underlying technology of quantum
computing and quantum information processing \cite{Barenco1995, Vandersypen2001, Chuang1998, Jones1998, Cirac1995, Monroe1995, Turchette1995, Averin1998, Shnirman1997}. The basic idea in each one of these systems is to define certain quantum degree of freedom to serve as a qubit, such as the charge, orbital, or spin angular momentum. This is usually followed by finding a controllable mechanism to form an entanglement between a two-qubit system in such a way to produce a fundamental quantum computing gate such as an exclusive Boolean XOR. In addition, we have to be able to coherently manipulate such an entangled state to provide an efficient computational process. Such coherent
manipulation of entangled states has been observed in different systems such as isolated trapped ions \cite{Chiaverini2005} and superconducting junctions \cite{Vion2002}. The coherent control of a two-electron spin state in a coupled quantum dot was achieved experimentally, in which the coupling mechanism is the Heisenberg exchange interaction between the electron spins \cite{Johnson2005, Koppens2005, Petta2005}.

Particularly, the solid state systems have been in the focus of interest as they facilitate the fabrication of large integrated networks that would be able to implement realistic quantum computing algorithms on a large scale. On the other hand, the strong coupling between a solid state system and its complex environment makes it a significantly challenging mission to achieve the high coherence control required to manipulate the system. Decoherence is considered as one of the main obstacles toward realizing an effective quantum computing system \cite{Zurek1991, Bacon2000, Shevni2005, deSousa2003}. The main effect of decoherence is to randomize the relative phases of the possible states of the isolated system as a result of coupling to the environment. By randomizing the relative phases, the system loses all quantum interference effects and its entanglement character and may end up behaving classically.

The interacting Heisenberg spin systems, in one, two and three-dimensions represent very reliable model for constructing quantum computing schemes in different solid-state systems and a very rich model for studying the novel physics of localized spin systems \cite{Loss1998, Burkard1999, Kane1998, Sorensen2001}. These spin systems can be experimentally realized, for instance, as a one-dimensional chain and lattices of coupled nano quantum dots.

Multi-particle systems are of central interest in the field of quantum information, as a quantum computer is considered as many body system by itself. Understanding, quantifying and exploring entanglement dynamics may provide an answer for many questions regarding the behaviour of complex quantum systems \cite{Kais2007}, particularly, quantum phase transitions and critical behaviour as entanglement is considered to be the physical property responsible for the long-range quantum correlations accompanying these phenomena \cite{Sondhi1997, Osborne2002, ZhangJ2009, Osenda2003, HuangZ2004}.

\subsection{Entanglement measures}
One of the central challenges in the theory of quantum computing and quantum information and their applications is the  preparation of entangled states and quantifying them. There has been an enormous number of approaches to tackle this problem both experimentally and theoretically. The term entanglement measure is used to describe any function that can be used to quantify entanglement. Unfortunately we are still far from a complete theory which can quantify entanglement of a general multipartite system in pure or mixed state \cite{Nielsen2000-book, HorodeckiM2001, HorodeckiP2001, Wootters2001-2}. There are very limited number of cases where we have successful entanglement measures. Two of these cases are (i) A bipartite system in a pure state and (ii) A bipartite system of two spin $1/2$ in a mixed state. Since these two particular cases are of special interest for our studies in spin systems we discuss them in more details in the coming two subsections.

To quantify entanglement, i.e. to find out how much entanglement is contained in a quantum state, Vedral et al. introduced the axiomatic approach to quantify entanglement \cite{Vedral1997}. They introduced the basic axioms that are necessary for an entanglement measure to satisfy. Before introducing these axioms one should first discuss the most common operations that can be performed on quantum systems and affect their entanglement \cite{Mintert2005}. (i) Local operation: is an operation that when applied to a quantum system consisting of two subsystems, each subsystem evolves independently. Therefore, possibly pre-existing correlations, whether classical or quantum, will not be affected. Hence, their entanglement also will not be affected. (ii) Global operation: is an operation that when applied to a quantum system consisting of two subsystems, the subsystems will evolve while interacting with each other. Therefore, correlations, both classical and quantum, may change under the effect of such operations. Hence, their entanglement will be affected by this operation. (iii) Local Operations with Classical Communications (LOCC): it is a special kind of global operations, at which the subsystems evolve independently with classical communications allowed between them. Information about local operations can be shared using the classical communications and then further local operations may be done according to the shared information. Therefore, classical, but not quantum, correlations may be changed by LOCC. It is reasonable then to require that an entanglement measure should not increase under LOCC.

Now let us briefly introduce a list of the most commonly accepted axioms that a function $E$ must obey to be considered an entanglement measure: (1) $E$ is a mapping from the density matrix of a system to a positive real number $\rho \rightarrow E(\rho) \in \textbf{R}$; (2) $E$ does not increase under LOCC only \cite{Mintert2005}; (3) $E$ is invariant under local unitary transformations; (4) For a pure state $\rho=\left|\psi\right\rangle \left\langle \psi\right|$, $E$ reduces to the entropy of entanglement \cite{Plenio2007} (will be discussed in more details latter); (5) $E$ = 0 iff the states are separable\cite{Horodecki2007-2}; (6) $E$ takes its maximum for maximally entangled states (Normalization); (7) $E$ is continous \cite{Keyl2002}; (8) $E$ should be a convex function, i.e. it cannot be obtained by mixing states [$\rho_{i}$ \cite{Keyl2002},$E(w_{i} \rho_{i}) \leq w_{i} E(\rho_{i})$]; (9) $E$ is Additive, i.e.  given two pairs of entangled particles in the total state $\sigma=\sigma_{1} \otimes \sigma_{2}$ then we have \cite{Plenio1998} $ E(\sigma)=E(\sigma_{1}) + E(\sigma_{2})$.

\subsubsection{Pure bipartite state}
A very effective approach to handle pure bipartite state is the Schmidt decomposition. For a pure state $|\psi\rangle$ of a composite system consists of two subsystems $A$ and $B$, with two orthonormal basis $\{| \phi_{a,i}\rangle \}$ and $\{|\phi_{b,i}\rangle \}$ respectively, the Schmidt decomposition of the state $|\psi\rangle$ is defined by
\beq
|\psi\rangle = \sum_{i} \lambda_i \; |\phi_{a,i}\rangle \; |\phi_{b,i}\rangle \;,
\eeq
where $\lambda_i$ are positive coefficients satisfying $\sum \lambda_{i}^{2}=1$ and are called Schmidt coefficients. Evaluating the reduced density operators
\beq
\rho_{A/B} = tr_{B/A}(|\psi\rangle) = \sum_{i} \lambda_{i}^{2} |\phi_{(a/b),i}\rangle \langle\phi_{(a/b),i}| \; ,
\eeq
shows that the two operators have the same spectrum ${\lambda_i}$, which means that the two subsystems will have many  properties in common. If the state $|\psi\rangle$ is a cross product of pure states, say $| \phi_{a,k}\rangle $ and $|\phi_{b,k}\rangle $, of the subsystems $A$ and $B$ respectively then $|\psi\rangle$ is a disentangled state and all Schmidt coefficients vanish except $\lambda_k=1$ in that case. In general the coefficients $\lambda_i$ can be used to quantify the entanglement in the composite system.

Entropy plays a major role in the classical and quantum information theory. Entropy is a measure of our uncertainty (lack of information) of the state of the system. The Shannon entropy quantifies the uncertainty associated with a classical distribution $\{P_i \}$ and is defined as $H(x)=-\sum_x P_x \log P_x$. The quantum analog of the Shannon entropy is the Von Neumann entropy where the classical probability distribution is replaced with the density operators. Considering the density operator $\rho$ representing the state of a quantum system, the Von Neumann entropy is defined as $S(\rho) \equiv -tr(\rho \log \rho)= -\sum_i \alpha_i \log \alpha_i$ where $\{\alpha_i\}$ are the eigenvalues of the matrix $\rho$ . For a bipartite system, the Von Neumann entropy of the reduced density matrix $\rho_{A/B}$ namely $S(\rho_{A/B})= -\sum_i \lambda_{i}^2 \log \lambda_{i}^2$ is a measure of entanglement which indeed satisfies all the above axioms. Although the entanglement content in a multipartite system is difficult to quantify, the bipartite entanglement of the different constituents of the system can provide a good insight about the entanglement of the whole system.

\subsubsection{Mixed bipartite state}
When the composite physical system is in a mixed state, which is more common, different entanglement measures are needed. In contrary to the pure state, which has only quantum correlations, the mixed state contains both classical and quantum correlations which the entanglement measure should discriminate between them. Developing and entanglement measure for mixed multipartite systems is a difficult task as it is very hard to discriminate the quantum and classical correlations in that case \cite{Plenio2007,Horodecki2007-2}. Nevertheless for bipartite systems different entanglement measures were introduced which can overcome the mathematical difficulty, particularly for subsystems with only two degrees of freedom. Among the most common entanglement measures in this case are the relative entropy $E_R$ \cite{Plenio1998, Schumacher2000}; Entanglement of distillation $E_D$ \cite{Schumacher2000}; Negativity and Logarithmic negativity \cite{Vidal2002}.

One of the most widely used measures is entanglement of formation, which was the first measure to appear in 1996 by C. Bennett et al. \cite{Bennett1996}. For a mixed state the entanglement of formation is defined as the minimum amount of entanglement needed to create the state. Any mixed state $\left|\psi\right\rangle$ can be decomposed into a mixture of pure states $\left|\psi_{i}\right\rangle$ with different probabilities $p_{i}$, which is called "ensemble". The entanglement of formation of a mixed state can be obtained by summing the entanglements of each pure state after multiplying each one by its probability $p_{i} E(\left|\psi_{i}\right\rangle)$. The entanglement of each pure state is expressed as the entropy of entanglement of that state. Hence, for an ensemble $\Gamma$ of pure states $\{p_{i},\left|\psi_{i}\right\rangle\}$ we have \cite{Myhr2004}:
\begin{equation}
E_{F}(\epsilon)=\sum_{i}p_{i}E_{vN}(\left|\psi_{i}\right\rangle)
\end{equation}

Since a mixed state can be decomposed into many different ensembles of pure states with different entanglements, the entanglement of formation is evaluated using what is called the most "economical" ensemble \cite{Bennett1996}, i.e.:
\begin{equation}
E_{F}(\rho)=\inf_{\Gamma}\sum_{i}p_{i}E_{vN}(\left|\psi_{i}\right\rangle) \; ,
\end{equation}
where the infimum is taken over all possible ensembles. $E_{F}$ is called a convex roof and the decomposition leading to this \textit{convex roof} is called the \textit{optimal decomposition} \cite{Amico2008}.

The minimum is selected because if there is a decomposition where the average entanglement is zero then this state can be created locally without the need for entangled pure states and therefore $E_{F}=0$ \cite{Amico2008}.

Performing minimization over all decompositions is a very difficult task because of the large number of terms involved \cite{Wootters2001-2}. Nevertheless, It was shown that only limited number of terms is sufficient to preform the minimization. However, finding explicit formula that does not need preforming the minimization would simplify the evaluation of $E_{F}$ significantly. Bennett et al. \cite{Bennett1996} evaluated $E_{F}$ for a mixture of Bell's states, which are completely entangled qubits. Hill and Wootters \cite{Hill1997} provided a closed form of $E_{F}$ as a function of the density matrix for two-level bipartite systems having only two non-zero eigenvalues in terms of the \textit{concurrence}, which was extended latter to the case of all two-level bipartite systems, i.e. two qubits \cite{Wootters2001}. The entanglement of formation satisfies the previously discussed axioms \cite{Plenio2007, Keyl2002}.
\subsection{Entanglement and quantum phase transitions}

Quantum phase transition (QPT) in many body systems, in contrary to classical phase transition, takes place at zero temperature. QPT are driven by quantum fluctuations as a consequence of Heisenberg uncertainty principle \cite{Kutzelnigg1968, Guevara2003}. Examples of quantum phase transition are Quantum Hall transitions, magnetic transitions of cuprates, superconductor-insulator transitions in two dimension and metal-insulator transitions \cite{Guevara2003, Ziesche1999}. Quantum phase transition is characterized by a singularity in the ground state energy of the system as a function of an external parameter or coupling constant $\lambda$ \cite{Sachdev2001}. In addition, QPT is characterized by a diverging correlation length $\xi$ in the vicinity of the quantum critical point defined by the parameter value $\lambda_c$. The correlation length diverges as $\xi^{-1} \sim J |\lambda-\lambda_c|^{\nu}$, where $J$ is an inverse length scale and $\nu$ is a critical exponent. Quantum phase transitions in many body systems are accompanied by a significant change in the quantum correlations within the system. This led to a great interest in investigating the behaviour of quantum entanglement close the critical points of transitions, which may shed some light on the different properties of the ground state wave function as it goes through the transition critical point. On the other hand, these investigations may clarify the role entanglement plays in quantum phase transitions and how it is related to the different properties of that transition such as its order and controlling parameters. The increasing interest in studying the different properties of entangled states of complex systems motivated a huge amount of research in that area. One of the main consequences of these research is the consideration of entanglement as a physical resource which can be utilized to execute specific physical tasks in many body systems \cite{Bennett1996,Bennett1996-2}. Osborne and Nielsen have argued that the physical property responsible for long range quantum correlations accompanying quantum phase transitions in complex systems is entanglement, and that it becomes maximum at the critical point \cite{Tobias2002-2}. The renormalization group calculations demonstrated that quantum phase transitions have a universal character independent of the dynamical properties of the system and is only affected by specific global properties such as the symmetry of the system \cite {Cardy1996}. In order to test whether the entanglement would show the same universal properties in the similar systems, the pairwise entanglement was studied in the $XY$ spin model and its special case of the Ising model \cite{Osborne2002} where it was shown that the entanglement reaches a maximum value at the critical point of the phase transition in the Ising system. Also the entanglement was proved to obey a scaling behaviour in the vicinity of transition critical point for a class of one dimensional magnetic system, $XY$ model in a transverse magnetic field \cite{Osterloh2002}. Furthermore, the scaling properties of entanglement in $XXZ$ and $XY$ spin-$1/2$ chains near and at the transition critical point were investigated and the resemblance between the critical entanglement in spin system and entropy in conformal field theories was emphasized \cite{Vidal2003}. Several works have discussed the relation between entanglement and correlation functions and as a consequence the notion of localizable entanglement was introduced, which enabled the definition of entanglement correlation length bounded from above by entanglement assistance and from below by classical functions and diverges at a quantum phase transition \cite{Verstraete2004, Popp2005, Popescu1992, Sen(De)2003}. Quantum discord which measures the total amount of correlations in a quantum state and discerns it from the classical ones, first introduced by Olliver and Zurek \cite{Ollivier2001}, was used to study quantum phase transition in $XY$ and $XXZ$ spin systems \cite{Dillenschneider2008}. It was demonstrated that while the quantum correlations increases close to the critical points, the classical correlations decreases,in $XXZ$ model, and is monotonous, in Ising model, in the vicinity of the critical points.

\subsection{Dynamics of entanglement}
In addition to the interest in the static behavior of entanglement in many body systems, its dynamical behaviour has attracted great attention as well, where different aspects of this dynamics have been investigated recently. One of the most important aspects is the propagation of entanglement through a many body system starting from a specific part within the system. The speed of propagation of entanglement through the system depends on different conditions and parameters such as the initial set up of the system, impurities within the system, the coupling strength among the system constituents, and the external magnetic field \cite{Amico2004, Christandl2004, Hartmann2006}.

In most treatments, the system is prepared in an initial state described by an initial Hamiltonian $H_i$, then its time evolution is studied under the effect of different parameters, internal and external, which causes creation, decay, vanishing or just transfer of entanglement through the system. In many cases the system is abruptly changed from its initial state to another one causing sudden change in entanglement as well.

The creation of entanglement between different parts of a many body system rather than the transfer of entanglement through it was also investigated.
The creation of entanglement between the end spins of a spin-$1/2$ $XY$ chain was studied \cite{Glave2009}. A global time-dependent interaction between the nearest neighbour spins on the chain was applied to an initial separable state and the creation of entanglement between the end spins on the chain was tested. It was demonstrated that the amount of entanglement created dynamically was significantly larger than that created statically.

As heat can be extracted from a many body solid state system and be used to create heat, it was shown that entanglement can be extracted from a many body system by means of external probes and be used in quantum information processing \cite{DeChiara2006}. The idea is to scatter a pair of independent non-interacting particles simultaneously by an entangled many body solid state system (for instance solid state spin chain) or optical lattice with cold atoms where each incident particle interacts with a different entangled system particle. It was demonstrated that the entanglement was extracted from the many body system and transferred to the incident probes and the amount of entanglement between the probes pair is proportional to the entanglement within the many body system and vanishes for a disentangled system. Recently the time evolution of entanglement between an incident mobile particle and a static particle was investigated \cite{Ciccarello2009}. It was shown that the entanglement increases monotonically during the transient but then saturates to a steady state value. The results were general for any model of two particles, where it was demonstrated that the transient time depends only on the group velocity and the wave packet width for the incident quais-monochromatic particle and independent of the type and strength of the interaction. On the other hand, entanglement information extraction from spin-boson environment using non-interacting multi-qubit systems as a probe was considered \cite{Oxtoby2009}. The environment consists of a small number of quantum-coherent two-level fluctuators (TLFs) with a damping caused by independent bosonic baths. A special attention was devoted to the quantum correlations (entanglement) that build up in the probe as a result of the TLF-mediated interaction.

The macroscopic dynamical evolution of spin systems was demonstrated in what is known as Quantum domino dynamics. In this phenomenon, a one dimensional spin-$1/2$ system with nearest neighbour interaction in an external magnetic field is irradiated by a weak resonant transverse field \cite{Lee2005,Furman2006}. It was shown that a wave of spin flip can be created through the chain by an initial single spin flip. This can be utilized as a signal amplification of spin flipping magnetization.

\section{Dynamics of entanglement in one-dimensional spin systems}

\subsection{Effect of a time-dependent magnetic field on entanglement}

We consider a set of $N$ localized spin-$\frac{1}{2}$ particles coupled through exchange
interaction $J$ and subject to an external magnetic field of strength $h$. We investigate the
dynamics of entanglement in the systems in presence of a time-dependent magnetic fields.
The Hamiltonian for such a system is given by\cite{HuangZ2005}

\begin{equation}\label{initialHamiltonian}
H = -\frac{J}{2}(1+\gamma)\sum_{i=1}^N\sigma_i^x \sigma_{i+1}^x -
\frac{J}{2}(1-\gamma)\sum_{i=1}^N \sigma_i^y \sigma_{i+1}^y -
\sum_{i=1}^N h(t) \sigma_i^z \, ,
\end{equation}

\noindent where $J$ is the coupling constant, $h(t)$ is the time-dependent external
magnetic field, $\sigma^a$ are the Pauli matrices
($a = x,y,z$), $\gamma$ is the degree of anisotropy and $N$ is the number of sites.
We can set J=1 for convenience and use periodic boundary conditions.
Next we  transform the spin operators into fermionic operators. So that, the Hamiltonian assumes the following form
\begin{equation}\label{finalHamiltonian}
H=\sum_{p=1}^{N/2}{\alpha_p(t)[c_p^{+}c_p+c_{-p}^{+}c_{-p}]+
i\delta_p[c_p^{+}c_{-p}^{+}+c_{p}c_{-p}]+2h(t)}=\sum_{p=1}^{N/2}\tilde{H}_p.
\end{equation}
\noindent where, $\alpha_p(t)=-2cos\phi_p-2h(t)$,~~$\delta_p=2\gamma sin\phi_p$
and $\phi_p$=2$\pi$p/N~.
It is easy to show~$[\tilde{H}_p,\tilde{H}_q]=0~$, which means
the space of $\tilde{H}$ decomposes into noninteracting subspace,
each of four dimensions. No matter what $h(t)$ is, there will be no
 transitions among those subspaces.
 Using the following basis for the $\emph{p}$th subspace:
$(|0>; c_p^+c_{-p}^+|0>; c_p^+|0>; c_{-p}^+|0>)$,
we can explicitly get
\begin{equation}\label{subspaceHamiltonian}
\tilde{H}_p(t)=\scriptsize{
\left(\begin{array}{cccc}
2h(t)&-i\delta_p&0&0\\
i\delta_p&-4cos\phi_p-2h(t)&0&0\\
0&0&-2cos\phi_p&0~\\
0&0&0&-2cos\phi_p
\end{array}\right)}.
\end{equation}
\indent We only consider the systems which at time t=0 are in the thermal
equilibrium at temperature T. Let $\rho_p(t)$ be the density matrix of the
$\emph{p}$th subspace, we have~$\rho_p(0)=e^{-{\beta}\tilde{H}_p(0)}$,
where $\beta=1/kT$ and k is the Boltzmann's constant.
 Therefore, using Eq. {\ref{subspaceHamiltonian}}, we can have $\rho_p(0)$.
Let $U_p(t)$ be the time-evolution matrix in the $\emph{p}$th subspace, namely($\hbar=1$):
$i\frac{dU_p(t)}{dt}=U_p(t)\tilde{H}_p(t)~$,
with the boundary condition $U_p(0)=I$~. Now,
the Liouville equation of this system is
\begin{equation}
i\frac{d\rho(t)}{dt}=[H(t),\rho(t)]~.
\end{equation}
which can be decomposed into uncorrelated subspaces and solved exactly.
Thus, in the $\emph{p}th$ subspace, the solution of Liouville equation is
$\rho_p(t)=U_p(t)\rho_p(0)U_p(t)^\dagger~$.

\begin{figure}
\includegraphics[angle=270,width=0.8\textwidth]{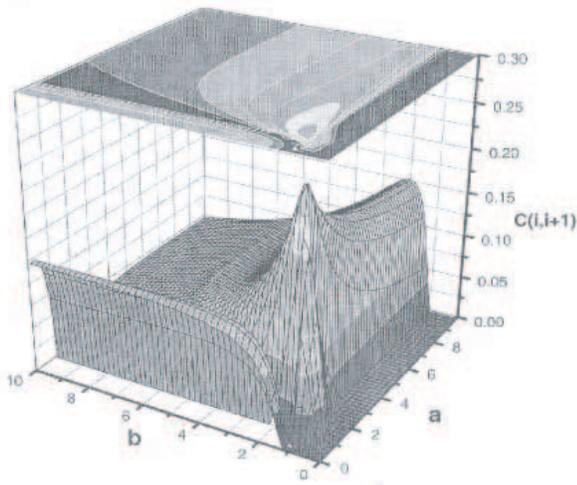}
\caption{\it Nearest-neighbor concurrence $C$ at zero temperature
as a function of the initial magnetic field $a$ for the step function case
with final field $b$.}
\label{Fig8}
\end{figure}

As a first step to investigate the dynamics of the entanglement we can take
the magnetic field to be a  step function then generalize it to other relevant
functional forms such as an oscillating one\cite{HuangZ2005}. Figure (\ref{Fig8}) shows the results for
nearest-neighbor concurrence $C(i,i+1)$ at temperature $T=0$ and $\gamma=1$
as a function of the initial magnetic field $a$ for the step function case
with final field $b$.
For $a < 1$ region, the concurrence increases very fast near $b = 1$ and reaches
a  limit $C(i,i+1) \sim 0.125$  when $~b\to\infty~$. It is surprising that
 the concurrence will not
 disappear when $b$ increases with $a < 1$. This indicates that the
 concurrence will not disappear as the final external magnetic field increase
 at infinite time. It shows that this model is not in agreement with the
 obvious physical intuition, since we expect that increasing the external
 magnetic field will destroy the spin-spin correlations functions and make
 the concurrence vanishes. The concurrence approaches
 maximum $C(i,i+1) \sim 0.258$
 at $(a=1.37,b=1.37)$,  and decreases
 rapidly as $a \neq b$. This indicates that the fluctuation of the external
 magnetic field near the equilibrium state will rapidly  destroy the entanglement.
 However, in the region where $a > 2.0$,
 the concurrence is close to zero when  $ b < 1.0$ and maximum close to $1$. Moreover,
it disappear in the limit of $b\to\infty$.\\

Now, let us examine the system size effect on the
entanglement with three different external magnetic fields
changing with time $t$\cite{HuangZ2006}:

\begin{equation}
h_{I}(t)=\left\{
\begin{array}{ll}
a &~~~~ t\le0\\
b + (a-b)e^{-Kt} &~~~~ t > 0
\end{array}
\right\},
\end{equation}
\begin{equation}
h_{II}(t)=\left\{
\begin{array}{ll}
a &~~~~ t\le0\\
a - a sin(K t) &~~~~ t > 0
\end{array}
\right\},
\end{equation}
\begin{equation}
h_{III}(t)=\left\{
\begin{array}{ll}
0 &~~~~ t\le0\\
a - a cos(K t) &~~~~ t > 0
\end{array}
\right\},
\end{equation}

where $a$, $b$ and $K$ are  varying parameters.

We have found that the entanglement fluctuates shortly after a disturbance
by an external magnetic
field when the system size is small.
For larger system size,
the entanglement  reaches a stable state for a long time before it
fluctuates. However, this fluctuation of entanglement disappears
when the system size goes to infinity. We also show that in
a periodic external magnetic field, the  nearest neighbor
entanglement displays a periodic structure with a period  related
to that of the magnetic field.
For the exponential external magnetic field,
by varying the constant $K$
we have found that as time evolves, $C(i,i+1)$ oscillates but it does not
reach its equilibrium value at $t \rightarrow \infty$. This confirms
the fact that the nonergodic
behavior of the concurrence  is a
general behavior for slowly changing magnetic field. For the
periodic magnetic field $h_{II}=a(1-sin{[-K t])}$ the nearest neighbor concurrence
is at maximum at $t=0$ for values of $a$ close to one, since the system exhibit
a quantum phase transition at $\lambda_c=J/h=1$, where in our calculations
we fixed $J=1$.
Moreover for the two
periodic $sin{[-K t]}$  and $cos{[-K t]}$ fields the nearest neighbor concurrence
displays a periodic structure according to the periods of their
respective magnetic fields\cite{HuangZ2006}.
\begin{figure}
\includegraphics[width=0.5\textwidth,height=0.5\textheight]{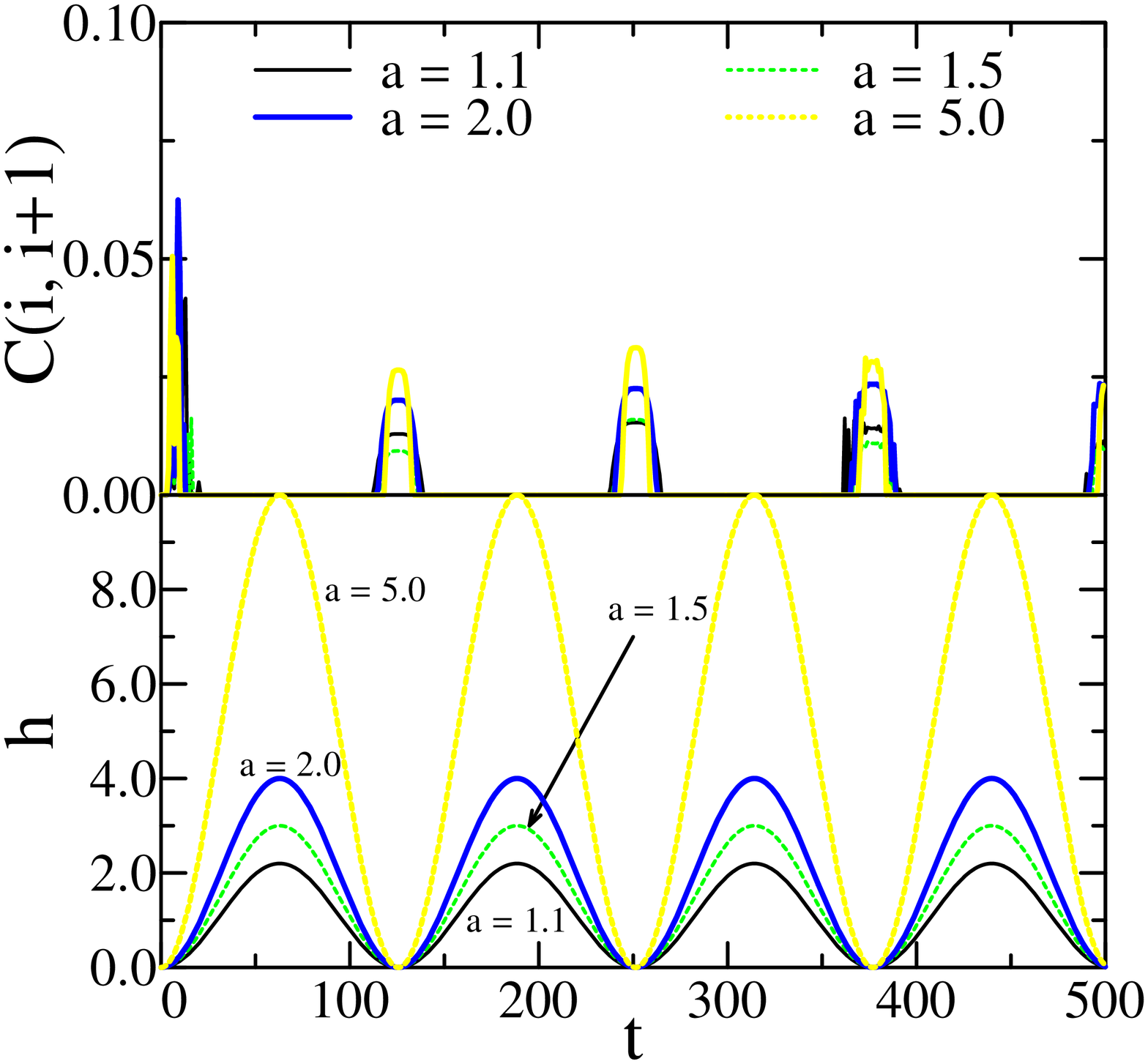}
\caption{\it The nearest neighbor concurrence $C(i,i+1)$ (upper panel) and the periodic
external magnetic field $h_{III} (t)=a(1-cos[Kt])$, see Eq. (14) in the text (lower panel)
for $K = 0.05$ with different values of $a$  as a
function of time $t$.}
\label{Fig9}
 \end{figure}
For the periodic external magnetic field $h_{III}(t)$, we show
in Figure (\ref{Fig9}) that the nearest neighbor concurrence  $C(i,i+1)$ is zero
at $t=0$ since the external magnetic field $h_{III}(t=0)=0$ and the spins
aligns along the $x$-direction: the total wave function is factorisable.
By increasing the external magnetic field  we see the appearance of
nearest neighbor concurrence but very small. This indicates that the concurrence can not be
produced without background external magnetic field in the Ising system.
However, as time evolves one can see the periodic structure of the
nearest neighbor concurrence according to the periodic structure
of the external magnetic field $h_{III}(t)$ \cite{HuangZ2006}.

\subsection{Decoherence in one dimesional spin system}

Recently, there has been a special interest in solid state systems
as they facilitate the fabrication of large integrated networks that would
be able to implement realistic quantum computing algorithms on a large scale.
On the other hand, the strong coupling between a solid state system and
its complex environment makes it a significantly challenging mission to achieve the
high coherence control required to manipulate the system. Decoherence
is considered as one of the main obstacles toward realizing
an effective quantum computing system \cite{Zurek1991,Bacon2000,Shevni2005,deSousa2003}.
The main effect of decoherence is
to randomize the relative phases of the possible states of the isolated
system as a result of coupling to the environment. By randomizing the
relative phases, the system loses all quantum interference effects and
may end up behaving classically.

As a system of special interest, there has been great efforts to study the
mechanism of electron phase decoherence and determine the
time scale for such process (the decoherence time), in solid
state quantum dots both theoretically \cite{Khaetskii2003, Elzerman2004, Florescu2006, Coish2004,  Shevni2005} and experimentally  \cite{Huttel2004, Tyryshkin2003, Abe2004, Johnson2005, Petta2005}. The main source of electron spin decoherence in a quantum dot is the inhomogeneous hyperfine
coupling between the electron spin and the nuclear spins.

In order to study the decoherence of a two state quantum system as a result of coupling to a spin bath, we examined the time evolution of a single spin coupled by exchange interaction to an environment of interacting spin bath
modelled by the XY-Hamiltonian.
The Hamiltonian for such system is given by\cite{Huang2006}
 \begin{equation}\label{impurityhamiltonian}
H = -\frac{1+\gamma}{2} \sum_{i=1}^N J_{i,i+1} \sigma_i^x \sigma_{i+1}^x -
\frac{1-\gamma}{2} \sum_{i=1}^N J_{i,i+1} \sigma_i^y \sigma_{i+1}^y -
\sum_{i=1}^N h_{i} \sigma_i^z \, ,
\end{equation}

\noindent where $J_{i,i+1}$ is the exchange interaction between
sites $i$ and $i+1$, $h_{i}$ is the strength of the external magnetic
field on site $i$, $\sigma^a$ are the Pauli matrices ($a = x,y,z$),
$\gamma$ is the degree of anisotropy and $N$ is the number of sites.
We consider the centered spin on the $l_{th}$ site as the single spin quantum system
and the rest of the chain as its environment, where in this case $l = (N+1)/2$.
The single spin directly interacts with its nearest neighbor spins through exchange interaction $J_{l-1,l}=J_{l,l+1}=J'$.
We assume exchange interactions between spins in the environment are uniform,
and simply set it as $J=1$. The centered spin is considered as inhomogeneously
coupled to all the spins in the environment by being directly coupled to its
nearest neighbors and indirectly to all other spins in the chain through its
nearest neighbors.

\begin{figure}
\subfigure[]{\includegraphics[width=0.48\textwidth,height=0.3\textheight]{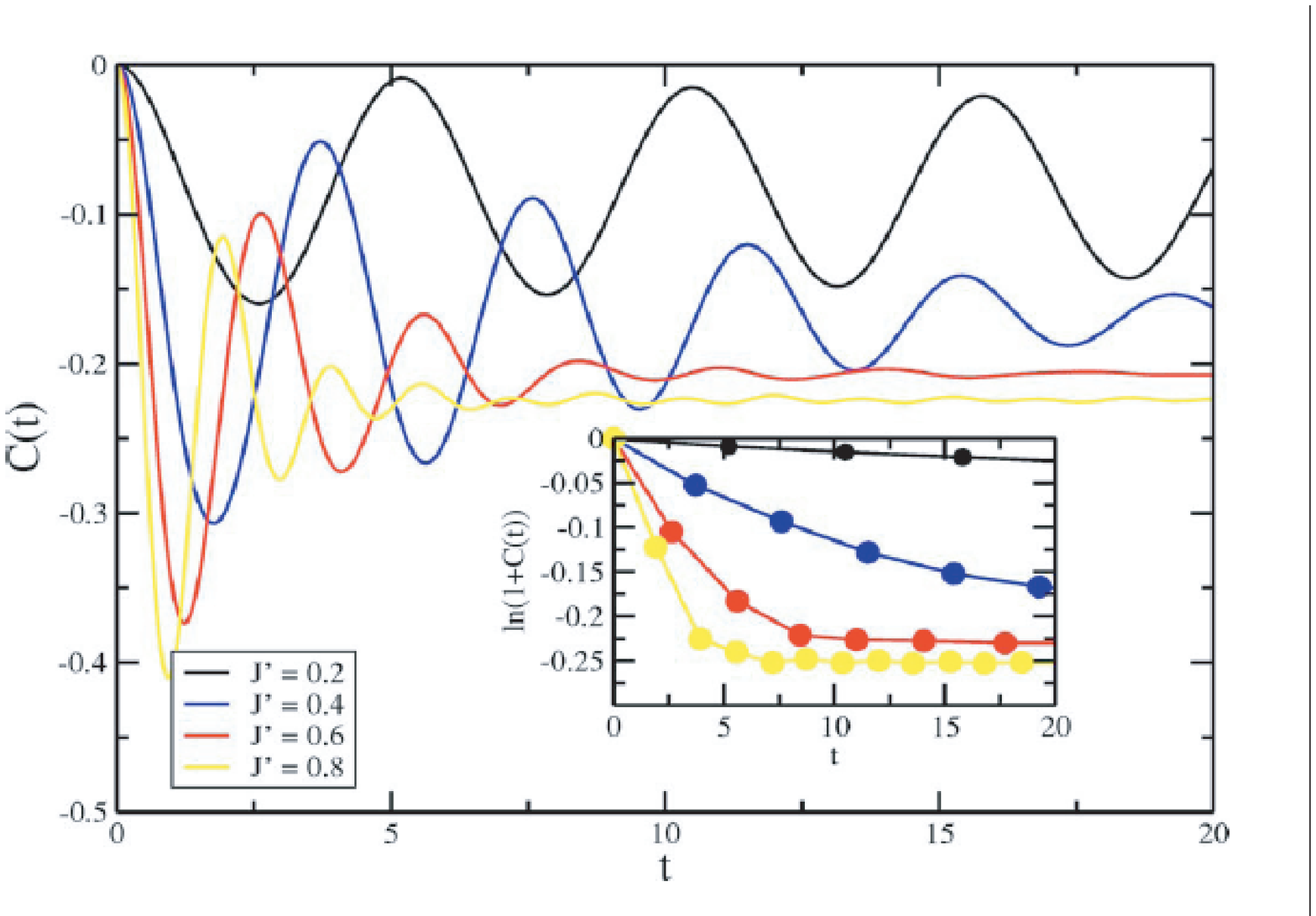}}\quad
\subfigure[]{\includegraphics[width=0.48\textwidth,height=0.3\textheight]{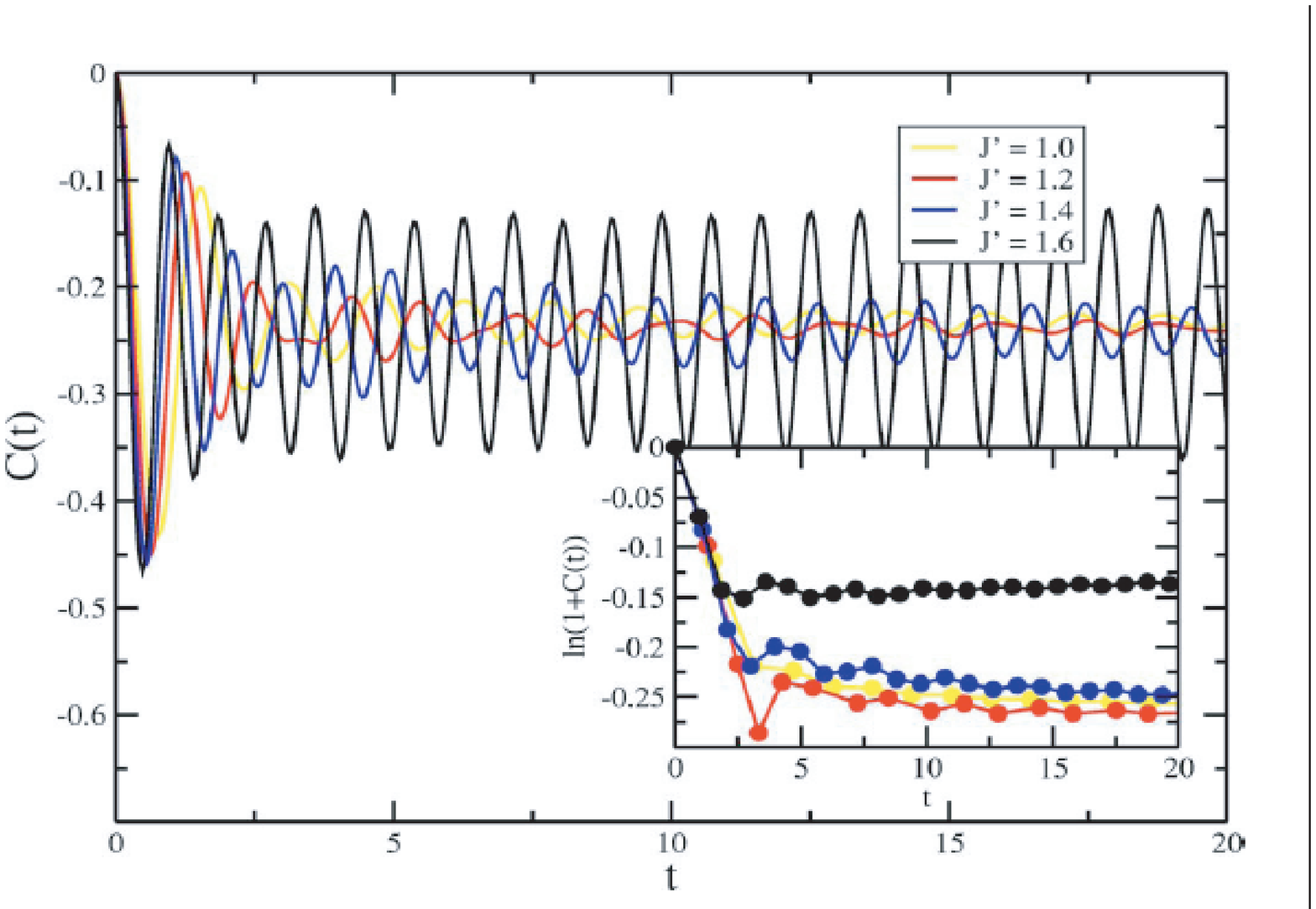}}
\caption{\it The spin correlation function $C(t)$ of centered spin for N=501, h=0.5, and $\gamma$ = 1.0 versus time t for different values of the coupling (a) $J'\leq J$; (b) $J' \geq J$ at zero temperature. The decay profile for each case is shown in the inner
panel.}
\label{JCP1}
\end{figure}

\begin{figure}
\includegraphics[width=0.5\textwidth,height=0.3\textheight]{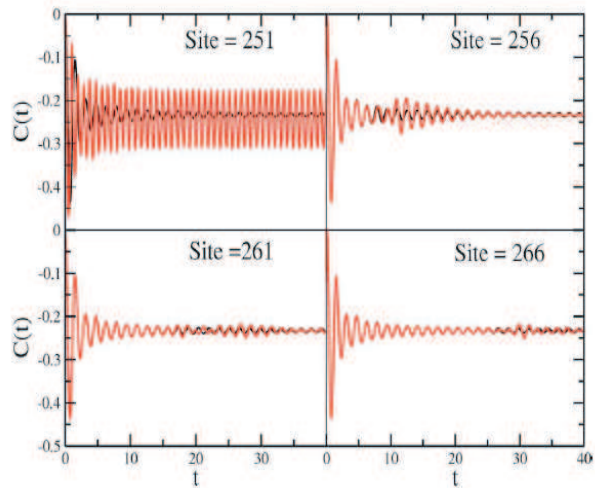}
\caption{\it The spin correlation function $C(t)$ vs. time t with $J'=1.0$ (black) and
$J'=1.5$ (red) for the centered spin (upper left) and the spin at site 256 (upper right), 261 (lower left), and 266 (lower right) in the environment. N=501, h=0.5, and $\gamma$ =1.0.}
\label{JCP2}
\end{figure}

By evaluating the spin correlator $C(t)$ of the single spin,the $j_{th}$ site\cite{Huang2006}
\begin{equation}
C_j(t)=\rho_j^z(t,\beta)-\rho_j^z(0,\beta),
\end{equation}
we observed that the decay rate
of the spin oscillations strongly depends on the relative magnitude of the
exchange coupling between the single spin and its nearest neighbor $J'$ and
coupling among the spins in the environment $J$.
The decoherence time varies significantly
based on the relative couplings magnitudes of $J$ and $J'$.
The decay rate law has a Gaussian profile when the  two exchange couplings are of the same order $J' \sim J$ but converts to exponential and
then a power law as we move to the regimes of $J' > J $ and $J' < J $ as shown in Fig. \ref{JCP1}.
We also showed that the spin
oscillations propagate from the single spin to the environmental
spins with a certain speed as depicted in Fig. \ref{JCP2}.
Moreover,  the amount of
saturated decoherence induced into the spin state depends on this relative magnitude
and approaches maximum value for a relative magnitude of unity. Our results suggests that setting the interaction within the environment in such a way that its magnitude is much higher or lower than the interaction with the single spin may reduce the decay rate of the spin state. The reason behind this phenomenon could be that the variation in the
coupling strength along the chain at one point (where the single spin exits) blocks the
propagation of decoherence along the chain by reducing the entanglement among
the spins within the environment which reduces its decoherence effect on the
single spin in return\cite{Huang2006}.
This result might be applicable in general to similar cases of
a centered quantum system coupled inhomogeneously to an interacting environment with
large degrees of freedom.

\subsection{An exact treatment of the system with step time-dependent coupling and magnetic field}

The obvious demand in quantum computation for a controllable mechanism to couple the qubits, led to one of the most interesting proposals in that regard which is to introduce a time-dependent exchange interaction between the two valence spins on a doubled quantum dot system as the coupling mechanism \cite{Loss1998, Burkard1999}. The coupling can be pulsed over definite intervals resulting a swap gate which can be achieved by raising and lowering the potential barrier between the two dots through controllable gate voltage. The ground state of the two-coupled electrons is a spin singlet, which is a highly entangled spin state.

There has been many studies focusing on the entanglement at zero and finite temperature for isotropic and anisotropic Heisenberg spin chains in presence and absence of an external magnetic field \cite{Wang2004, Asoudeh2006, ZhangJ2009, Rossignoli2005, Abdalla2008, Sadiek2009, Wichterich2009, Sodano2010}. Particularly, the dynamics of thermal entanglement has been studied in an $XY$ spin chain considering a constant nearest neighbor exchange interaction, in presence of a time varying magnetic field represented by a step, exponential and sinusoidal functions of time which we have discussed above \cite{HuangZ2005,HuangZ2006}.

Recently, The dynamics of entanglement in a one dimensional Ising spin chain at zero temperature was investigated numerically where the number of spins was seven at most \cite{Furman2008}. The generation and transportation of the entanglement through the chain under the effect of an external magnetic field and irradiated by a weak resonant field were studied. It was shown that the remote entanglement between the spins is generated and transported though only nearest neighbor coupling was considered. Latter the anisotropic $XY$ model for a small number of spins, with a time dependent nearest neighbor coupling at zero temperature was studied too \cite{Alkurtass2011}. The time-dependent spin-spin coupling was represented by a dc part and a sinusoidal ac part. It was found that there is an entanglement resonance through the chain whenever the ac coupling frequency is matching the Zeeman splitting.

Here we investigate the time evolution of quantum entanglement in an infinite one dimensional $XY$ spin chain system coupled through nearest neighbor interaction under the effect of a time varying magnetic field $h(t)$ at zero and finite temperature. We consider a time-dependent nearest neighbor Heisenberg coupling $J(t)$ between the spins on the chain. We discuss a general solution for the problem for any time dependence form of the coupling and magnetic field and present an exact solution for a particular case of practical interest, namely a step function form for both the coupling and the magnetic field. We focused on the dynamics of entanglement between any two spins in the chain and its asymptotic behavior under the interplay of the time-dependent coupling and magnetic field. Moreover, we investigated the persistence of quantum effects specially close to critical points of the system as it evolves in time and as its temperature increases.

The Hamiltonian for the $XY$ model of a one dimensional lattice with $N$ sites in a time-dependent external magnetic field $h(t)$ with a time-dependent coupling $J(t)$ between the nearest neighbor spins on the chain is given by

\begin{equation}
H=-\frac{J(t)}{2} (1+\gamma) \sum_{i=1}^{N} \sigma_{i}^{x} \sigma_{i+1}^{x}-\frac{J(t)}{2}(1-\gamma)\sum_{i=1}^{N} \sigma_{i}^{y} \sigma_{i+1}^{y}- \sum_{i=1}^{N} h(t) \sigma_{i}^{z}\, ,
\label{eq:H}
\end{equation}
where $\sigma_{i}$'s are the Pauli matrices and $\gamma$ is the anisotropy parameter.

Following the standard procedure to treat the Hamiltonian (\ref{eq:H}), we transform the Hamiltonian into the form \cite{Lieb1961}

\begin{equation}
H=\sum_{p=1}^{N/2} \tilde{H}_{p}\, ,
\label{eq:Hsum}\end{equation}
with $\tilde{H}_{p}$ given by
\begin{equation}
\tilde{H}_{p}=\alpha_{p}(t) [c_{p}^{\dagger} c_{p}+c_{-p}^{\dagger} c_{-p}]+i J(t) \delta_{p} [c_{p}^{\dagger} c_{-p}^{\dagger}+c_{p} c_{-p}]+2 h(t)\, ,
\label{eq:Hp}\end{equation}
where $\alpha_{p}(t)=-2 J(t) \cos \phi_{p} - 2 h(t)$ and  $\delta_{p}=2 \gamma \sin \phi_{p}$.

Writing the matrix representation of $\tilde{H}_{p}$ in the basis $\{ \left|0\right\rangle, c_{p}^{\dagger}c_{-p}^{\dagger}\left|0\right\rangle, c_{p}^{\dagger}\left|0\right\rangle, c_{-p}^{\dagger}\left|0\right\rangle \}$ we obtain
\begin{equation}
\tilde{H}_{p}=\left(\begin{array} {cccc}
2 h(t) & -i J(t)\delta_{p} & 0 & 0\\
i J(t) \delta_{p} & -4 J(t)\cos \phi_{p}-2 h(t) & 0 & 0\\
0 & 0 & -2 J(t)\cos \phi_{p} & 0\\
0 & 0 & 0 & -2 J(t)\cos \phi_{p}\\
\end{array}\right)\, .
\label{eq:Hmatrix}\end{equation}

Initially the system is assumed to be in a thermal equilibrium state and therefore its initial density matrix is given by
\begin{equation}
\rho_{p}(0)=e^{-\beta \tilde{H}_{p}(0)}\, ,
\label{eq:rho0}\end{equation}
where $\beta=1/k T$, $k$ is Boltzmann constant and $T$ is the temperature.

Since the Hamiltonian is decomposable we can find the density matrix at any time $t$, $\rho_{p}(t)$, for the $p$th subspace by solving Liouville equation given by
\begin{equation}
i \dot{\rho}_{p}(t)=[H_p(t),\rho_{p}(t)] \, ,
\label{eq:Liouville}
\end{equation}
which gives
\begin{equation}
\rho_{p}(t)=U_{p}(t) \rho_{p}(0) U_{p}^{\dagger}(t) \, .
\label{eq:UrhoU}
\end{equation}
where $U_{p}(t)$ is time evolution matrix which can be obtained by solving the equation
\begin{equation}
i \dot{U}_{p}(t)=U_{p}(t) \tilde{H}_{p}(t) \, .
\label{eq:Udot}
\end{equation}
Since $\tilde{H}_{p}$ is block diagonal $U_{p}$ should take the form

\begin{equation}
U_{p}(t)=\left(\begin{array}{cccc}
U_{11}^{p} & U_{12}^{p} & 0 & 0\\
U_{21}^{p} & U_{22}^{p} & 0 & 0\\
0 & 0 & U_{33}^{p} & 0\\
0 & 0 & 0 & U_{44}^{p}\\
\end{array}\right)\, .
\label{eq:U}
\end{equation}
Fortunately, eq. (\ref{eq:Udot}) may have an exact solution for a time-dependent step function form for both exchange coupling and the magnetic field which we adopt in this work. Other time-dependent function forms will be considered in a future work where other techniques can be applied. The coupling and magnetic field are represented respectively by

\begin{equation}
J(t) = J_0 + (J_1 - J_0) \theta(t) \, ,
\label{eq:stepJ}\end{equation}

\begin{equation}
h(t) = h_0 + (h_1 - h_0) \theta(t) \, ,
\label{eq:steph}\end{equation}
where $\theta(t)$ is the usual mathematical step function.
With this set up, the matrix elements of $U_p$ can be evaluated. The reduced density matrix of any two spins is evaluated in terms of the magnetization defined by
\begin{equation}
M=\frac{1}{N}\sum_{j=1}^{N}(S_{j}^{z})=\frac{1}{N}\sum_{p=1}^{1/N}M_p \:,
\label{eq:Mdef}
\end{equation}
and the spin-spin correlation functions defined by
\begin{equation}
S^{x}_{l,m}=\left\langle S^{x}_{l} S^{x}_{m} \right\rangle, \;\;\;S^{y}_{l,m}=\left\langle S^{y}_{l} S^{y}_{m} \right\rangle, \;\;\; S^{z}_{l,m}=\left\langle S^{z}_{l} S^{z}_{m} \right\rangle\, ,
\label{eq:Sdef}\end{equation}
Using the obtained density matrix elements one can evaluate the entanglement between any pair of spins using the Wootters method \cite{Wootters1998}.
\subsubsection{Transverse Ising Model}

The completely anisotropic $XY$ model, Ising model, is obtained by setting $\gamma=1$ in the Hamiltonian (\ref{eq:H}).
Defining a dimensionless coupling parameter $\lambda=J/h$, the ground state of the Ising model is characterized by a quantum phase transition that takes place at $\lambda$ close to the critical value $\lambda_c = 1$ \cite{Osborne2002}. The order parameter is the magnetization $\langle \sigma^x \rangle$ which differs from zero for $\lambda \geq \lambda_c$ and zero otherwise. The ground state of the system is paramagnetic when $\lambda \rightarrow 0$ where the spins get aligned in the magnetic field direction, the $z$ direction. For the other extreme case when $\lambda \rightarrow \infty$ the ground state is ferromagnetic and the spins are all aligned in the $x$ direction.

\begin{figure}[htbp]
\begin{minipage}[c]{\textwidth}
 \centering
   \subfigure{\label{PRA1_fig:2a}\includegraphics[width=7cm]{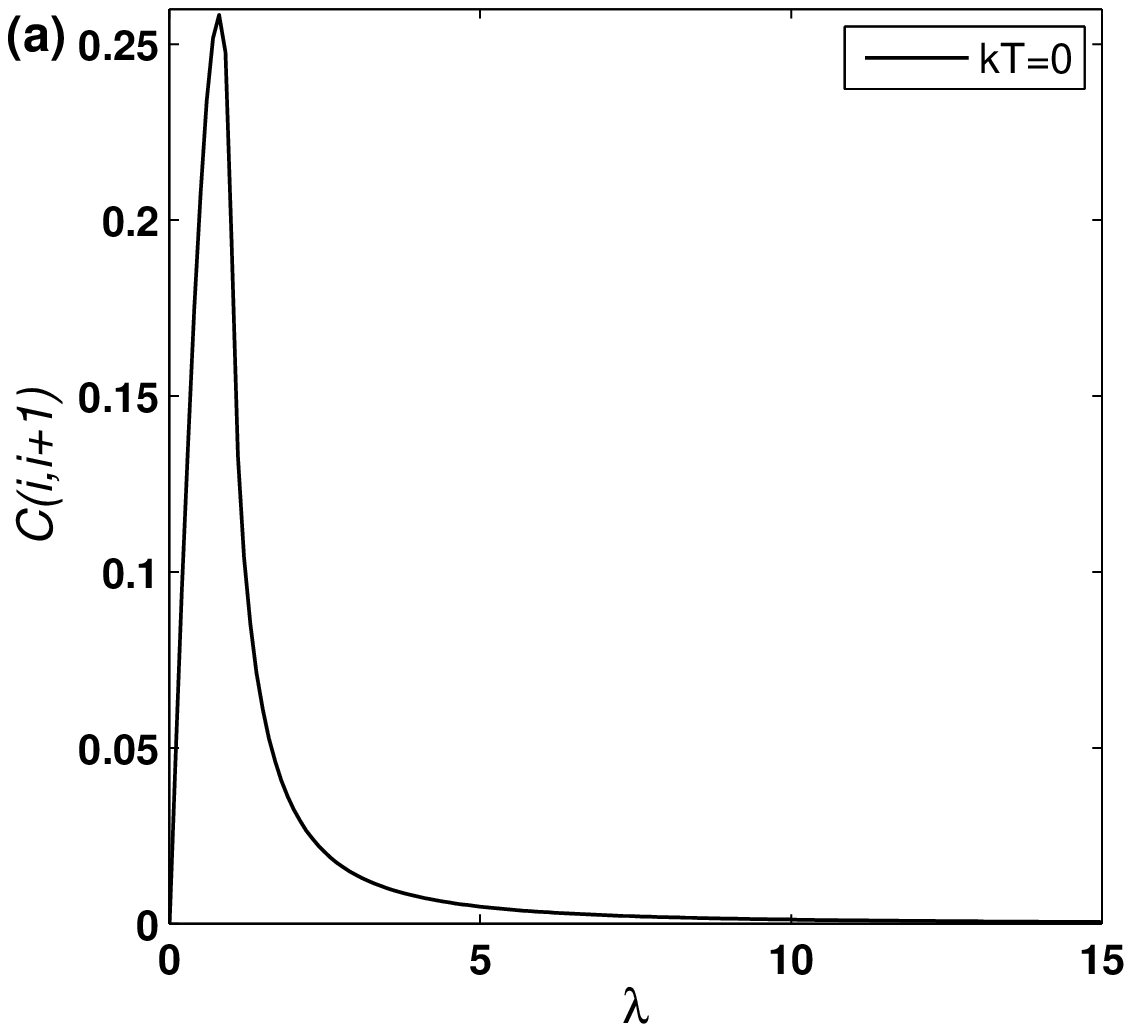}}\quad
   \subfigure{\label{PRA1_fig:2b}\includegraphics[width=7cm]{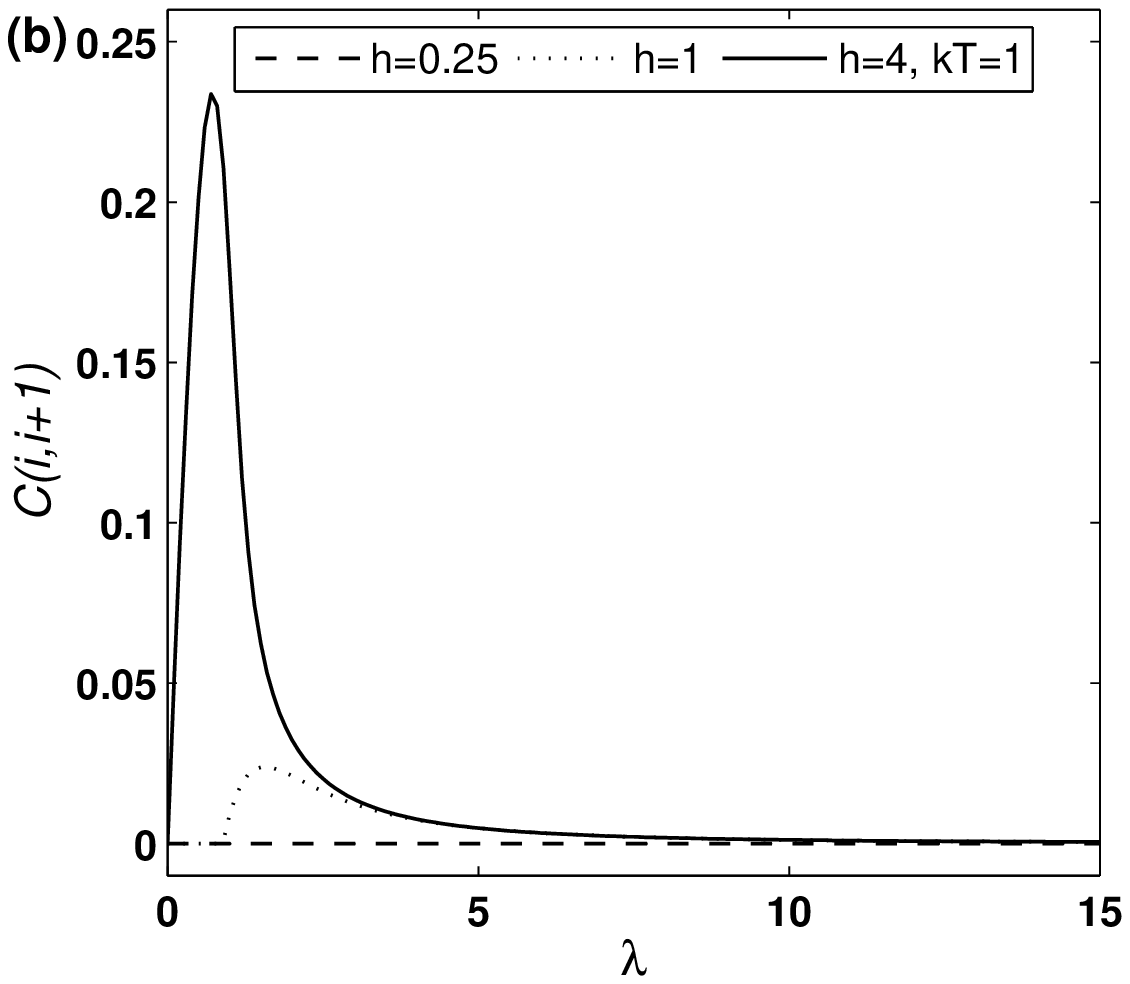}}\\
   \subfigure{\label{PRA1_fig:2c}\includegraphics[width=7cm]{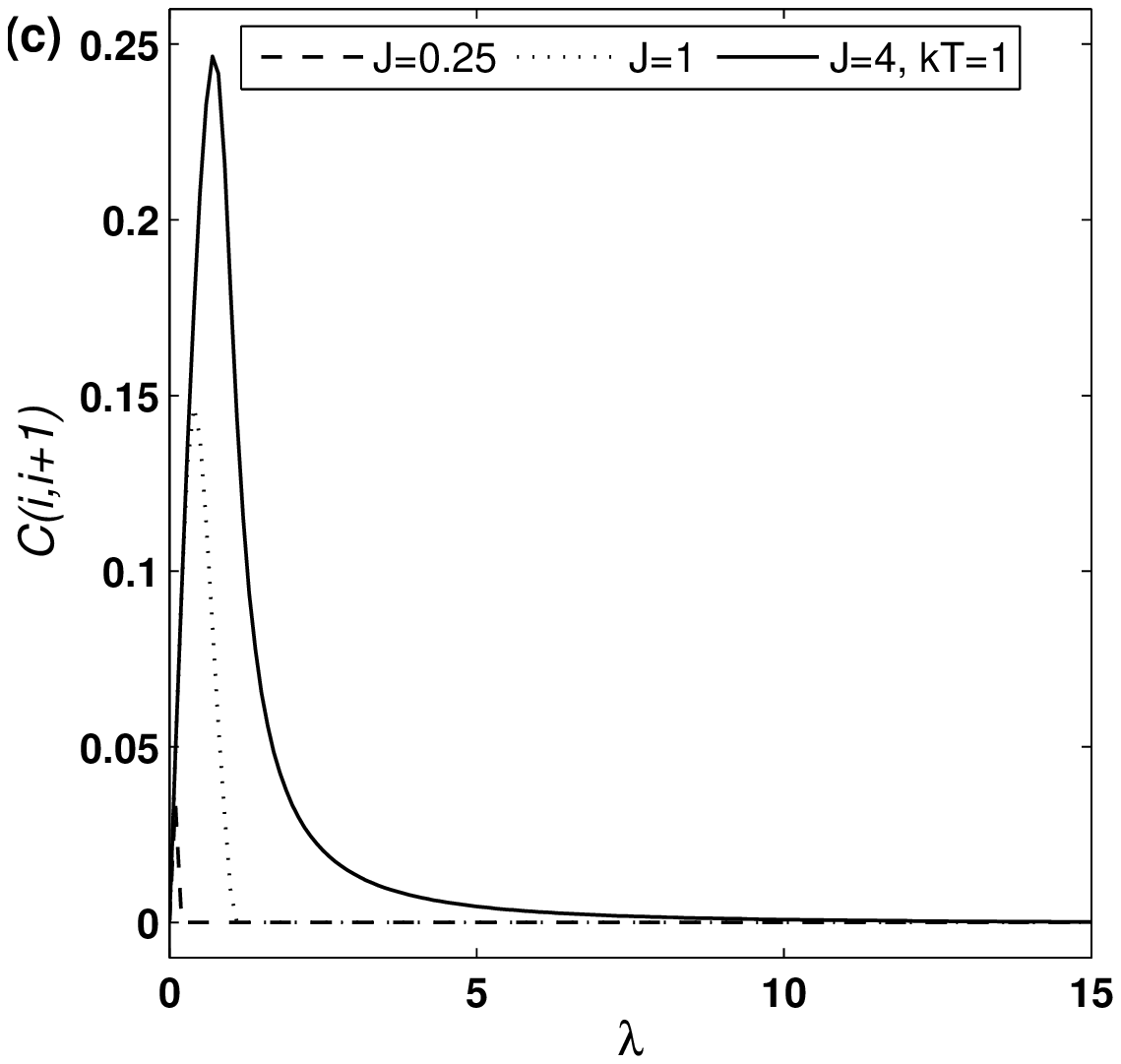}}\quad
   \subfigure{\label{PRA1_fig:2d}\includegraphics[width=7cm]{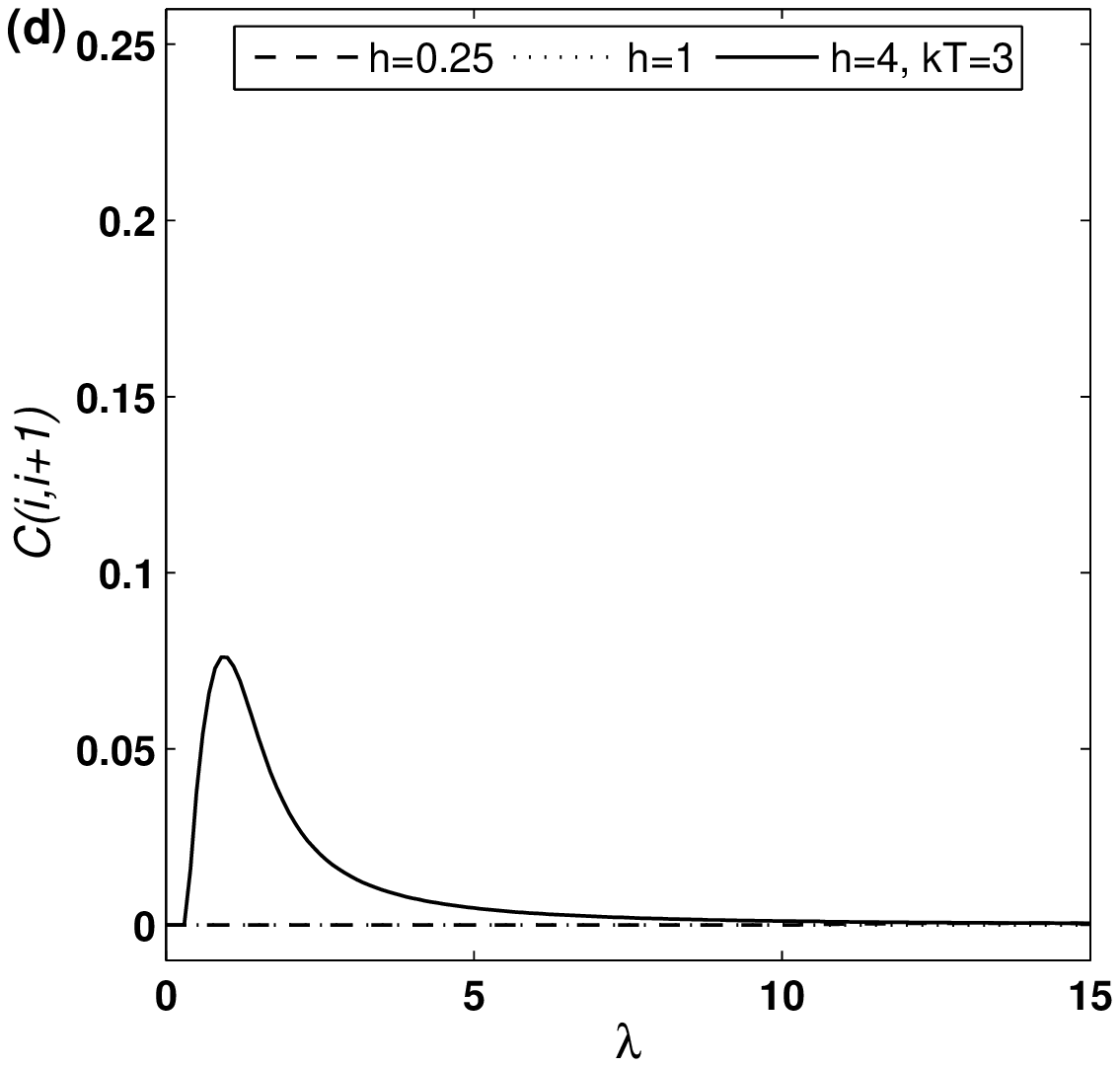}}
   \caption{{\protect\footnotesize $C(i,i+1)$ as a function of $\lambda$ for $h=h_{0}=h_{1}$ and $J=J_{0}=J_{1}$ at (a) $kT=0$ with any combination of $J$ and $h$; (b) $kT=1$ with $h_{0}=h_{1}=0.25, 1, 4$; (c) $kT=1$ with $J_{0}=J_{1}=0.25, 1, 4$; (d) $kT=3$ with $h_{0}=h_{1}=0.25, 1, 4$.}}
 \label{PRA1_fig:2}
 \end{minipage}
\end{figure}

We explored the dynamics of the nearest-neighbor concurrence $C(i,i+1)$ at zero temperature while the coupling parameter (and or the magnetic field) is a step function in time. We found that the concurrence $C(i,i+1)$ shows a nonergodic behavior. This behavior follows from the nonergodic properties of the magnetization and the spin-spin correlation functions as reported by previous studies \cite{HuangZ2005, Mazur1969, Barouch1970}. At higher temperatures the nonergodic behavior of the system sustains but with reduced magnitude of the asymptotic concurrence (as $t \rightarrow \infty$).

We studied the behavior of the nearest neighbor concurrence $C(i,i+1)$ as a function of $\lambda$ for different values of $J$ and $h$ at different temperatures.
As can be seen in Fig.~\ref{PRA1_fig:2a}, the behavior of $C(i,i+1)$ at zero temperature depends only on the ratio $J/h$ (i.e. $\lambda$) rather than their individual values. Studying entanglement at non-zero temperatures shows that the maximum value of $C(i,i+1)$ decreases as the temperature increases. Furthermore, $C(i,i+1)$ shows a dependence on the individual values of $J$ and $h$, not only their ratio as illustrated in Fig.~\ref{PRA1_fig:2b}, \ref{PRA1_fig:2c} and \ref{PRA1_fig:2d}.
\begin{figure}[htbp]
\begin{minipage}[c]{\textwidth}
 \centering
   \subfigure{\label{PRA1_fig:4a}\includegraphics[width=8cm]{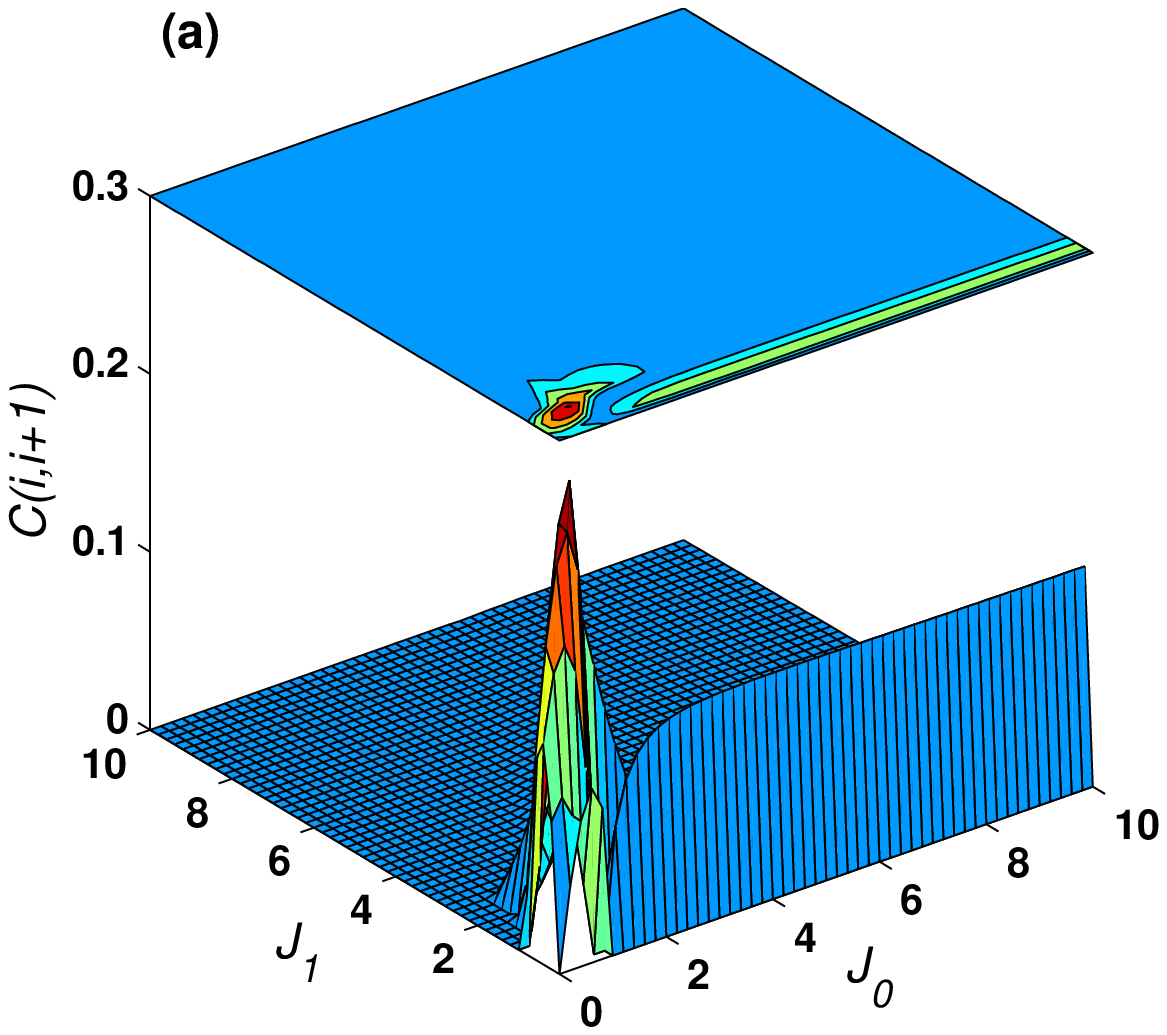}}\quad
    \subfigure{\label{PRA1_fig:4b}\includegraphics[width=8cm]{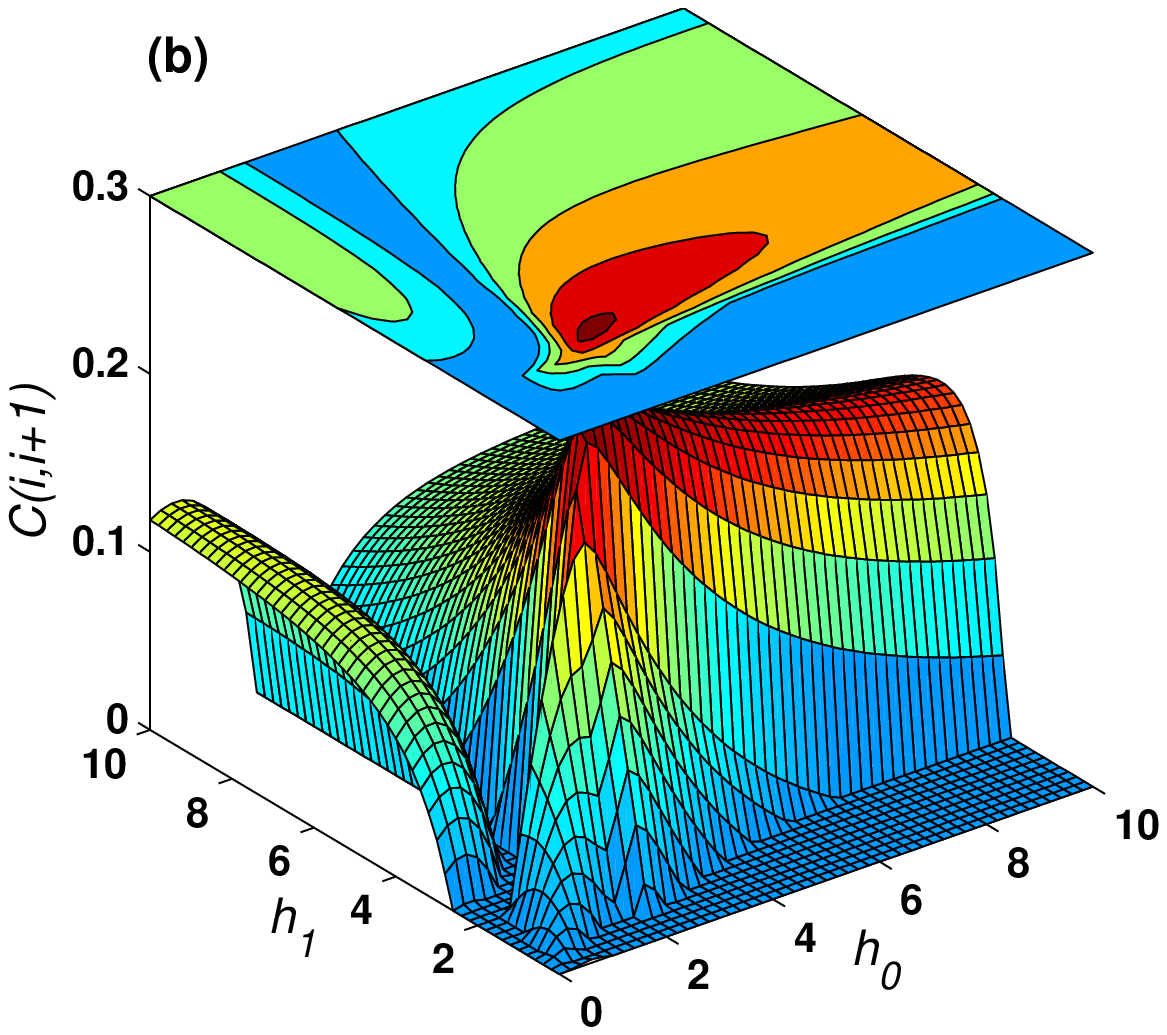}}
     \caption{{\protect\footnotesize The asymptotic behavior of $C(i,i+1)$ as a function of (a) $J_{0}$ and $J_{1}$ with $h_{0}=h_{1}=1$; (b) $h_{0}$ and $h_{1}$ with $J_{0}=J_{1}=2$, where $h_{0}$, $h_{1}$ and $J_{0}$ are in units of $J_{1}$, $kT=0$ and $\gamma=1$.}}
 \label{PRA1_fig:4}
 \end{minipage}
\end{figure}
We investigated the dependence of the asymptotic behavior (as $t \rightarrow \infty$) of the nearest neighbor concurrence on the magnetic field and coupling parameters $h_0$, $h_1$, $J_0$ and $J_1$ at zero temperature. In Fig.~\ref{PRA1_fig:4a} we present a 3-dimensional plot for the concurrence versus $J_0$ and $J_1$ where we set the magnetic field at $h_0=h_1=1$, while Fig.~\ref{PRA1_fig:4b} shows the asymptotic behavior of the nearest neighbor concurrence as a function of $h_{0}$ and $h_{1}$, while fixing the coupling parameter at $J_{0}=J_{1}=2$. In both cases as can be noticed the entanglement reaches its maximum value close to the critical value $\lambda_c=1$. Also as can concluded the behaviour of the asymptotic concurrence is much more sensitive to the change in the magnetic field parameters compared to the coupling ones.

\begin{figure}[htbp]
\begin{minipage}[c]{\textwidth}
 \centering
   \subfigure{\label{PRA1_fig:5a}\includegraphics[width=8cm]{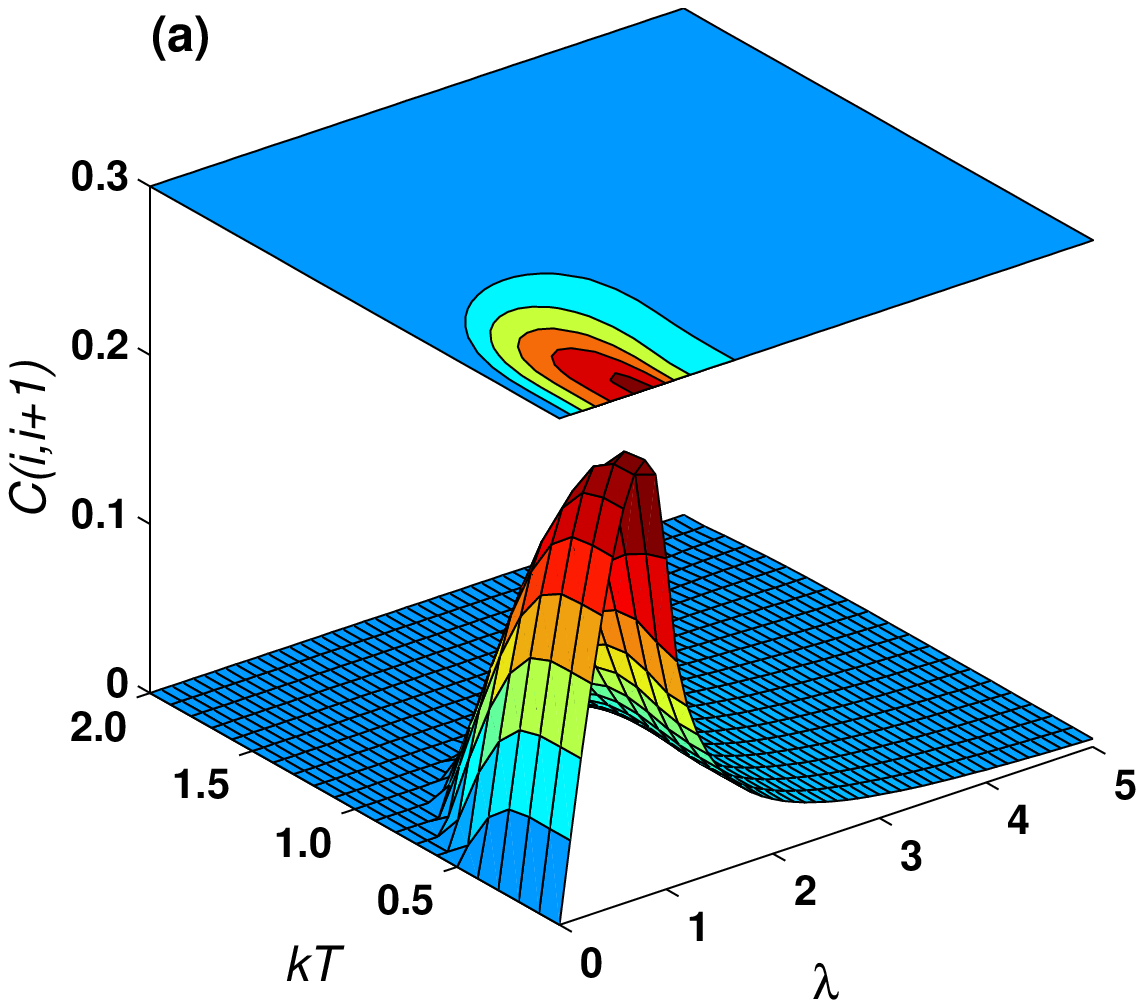}}\quad
   \subfigure{\label{PRA1_fig:5b}\includegraphics[width=8cm]{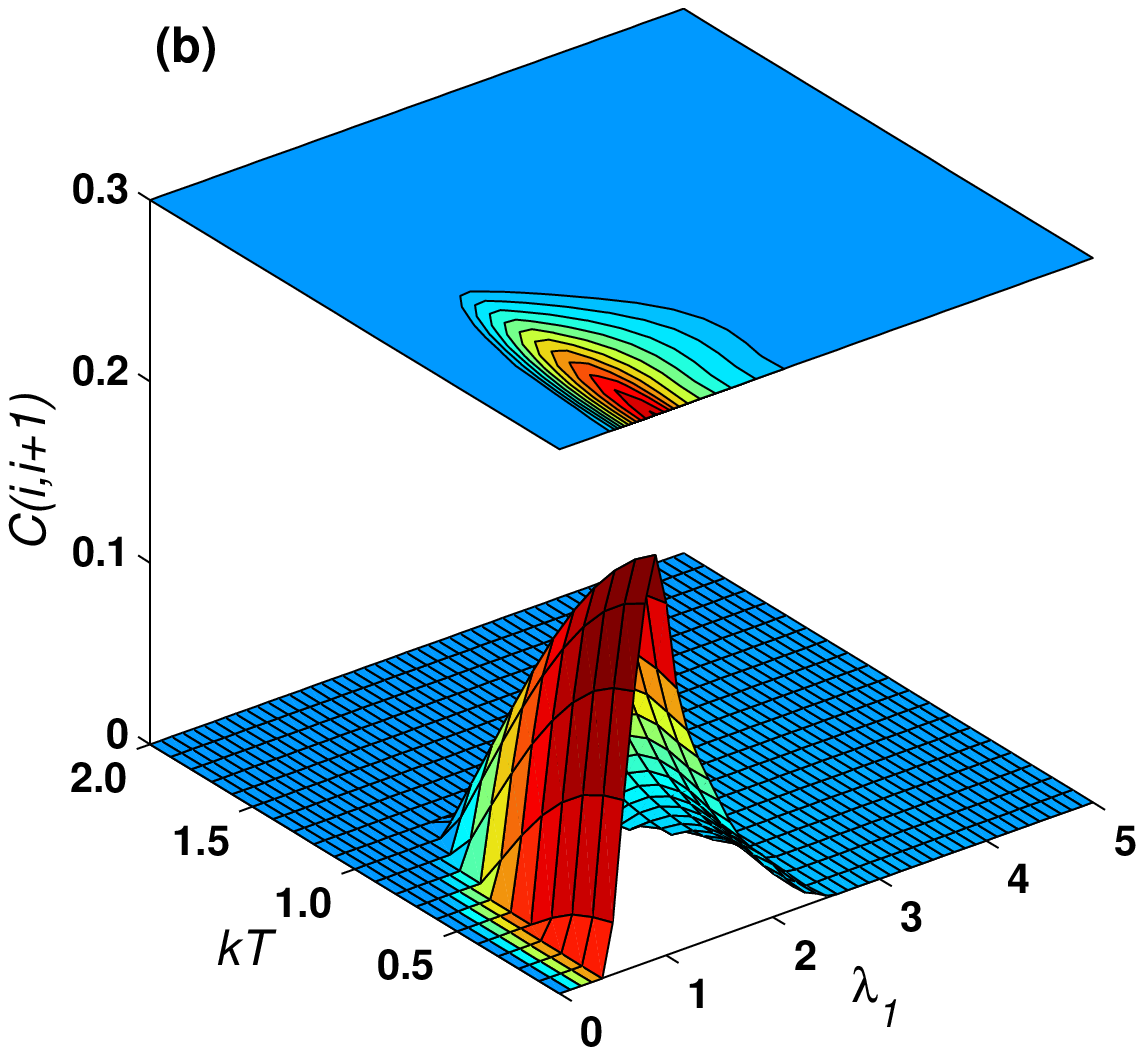}}
      \caption{{\protect\footnotesize The asymptotic behavior of $C(i,i+1)$ as a function of (a) $\lambda$ and $kT$, in units of $J_1$, with $\gamma=1$, $h_{0}=h_{1}=1$ and $J_{0}=J_{1}$; (b) $\lambda_{1}$ and $kT$ with $\gamma=1$, $h_{0}=h_{1}=1$ and $J_{0}=1$.}}
 \label{PRA1_fig:5}
 \end{minipage}
\end{figure}
There has been great interest in investigating the effect of temperature on the quantum entanglement and the critical behavior of many body systems and particularly spin systems \cite{Osborne2002, Sachdev2001, Sondhi1997, Arnesen2001, Gunlycke2001}. Osborne and Nielsen have studied the persistence of quantum effects in the thermal state of the transverse Ising model as temperature increases \cite{Osborne2002}. Here we investigate the persistence of quantum effects under both temperature and time evolution of the system in presence of the time-dependent coupling and magnetic field. Interestingly, the time evolution of entanglement shows a very similar profile to that manifested in the static case, i.e. the system evolves in time preserving its quantum character in the vicinity of the critical point and $kT=0$ under the time varying coupling. Studying this behavior at different values of $J_0$ and $h_0$ shows that the threshold temperature, at which $C(i,i+1)$ vanishes, increases as $\lambda_{0}$ increases.
\subsubsection{Partially Anisotropic XY Model}
We now turn to the partially anisotropic system where $\gamma=0.5$. First, we studied the time evolution of nearest neighbor concurrence for this model and it showed a non-ergodic behavior, similar to the isotropic case, which also follows from the non-ergodic behavior of the spin correlation functions and magnetization of the system. Nevertheless, the equilibrium time in this case is much longer than the isotropic.
\begin{figure}[htbp]
\begin{minipage}[c]{\textwidth}
 \centering
   \subfigure{\label{PRA1_fig:8a}\includegraphics[width=8cm]{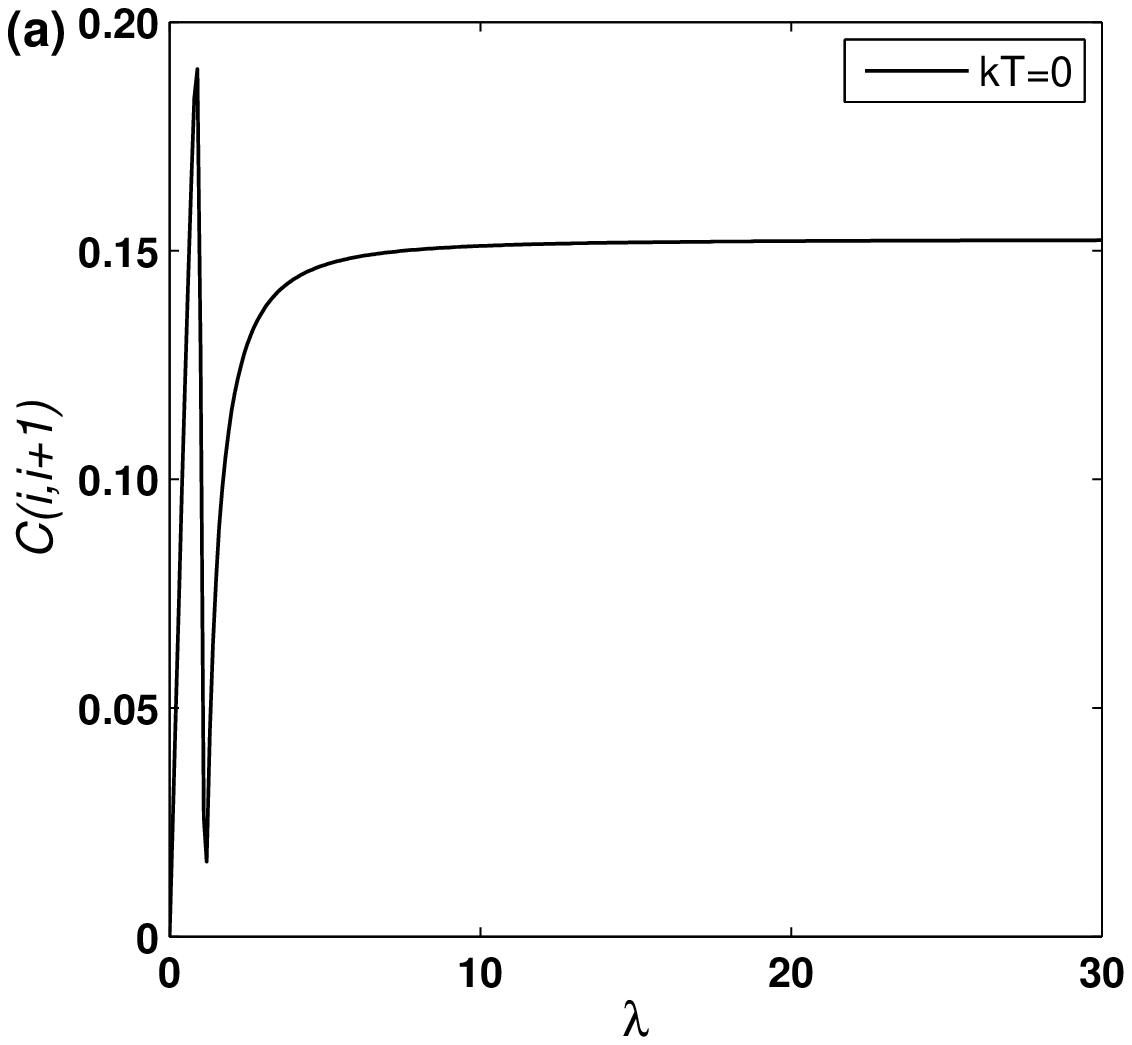}}\quad
   \subfigure{\label{PRA1_fig:8b}\includegraphics[width=8cm]{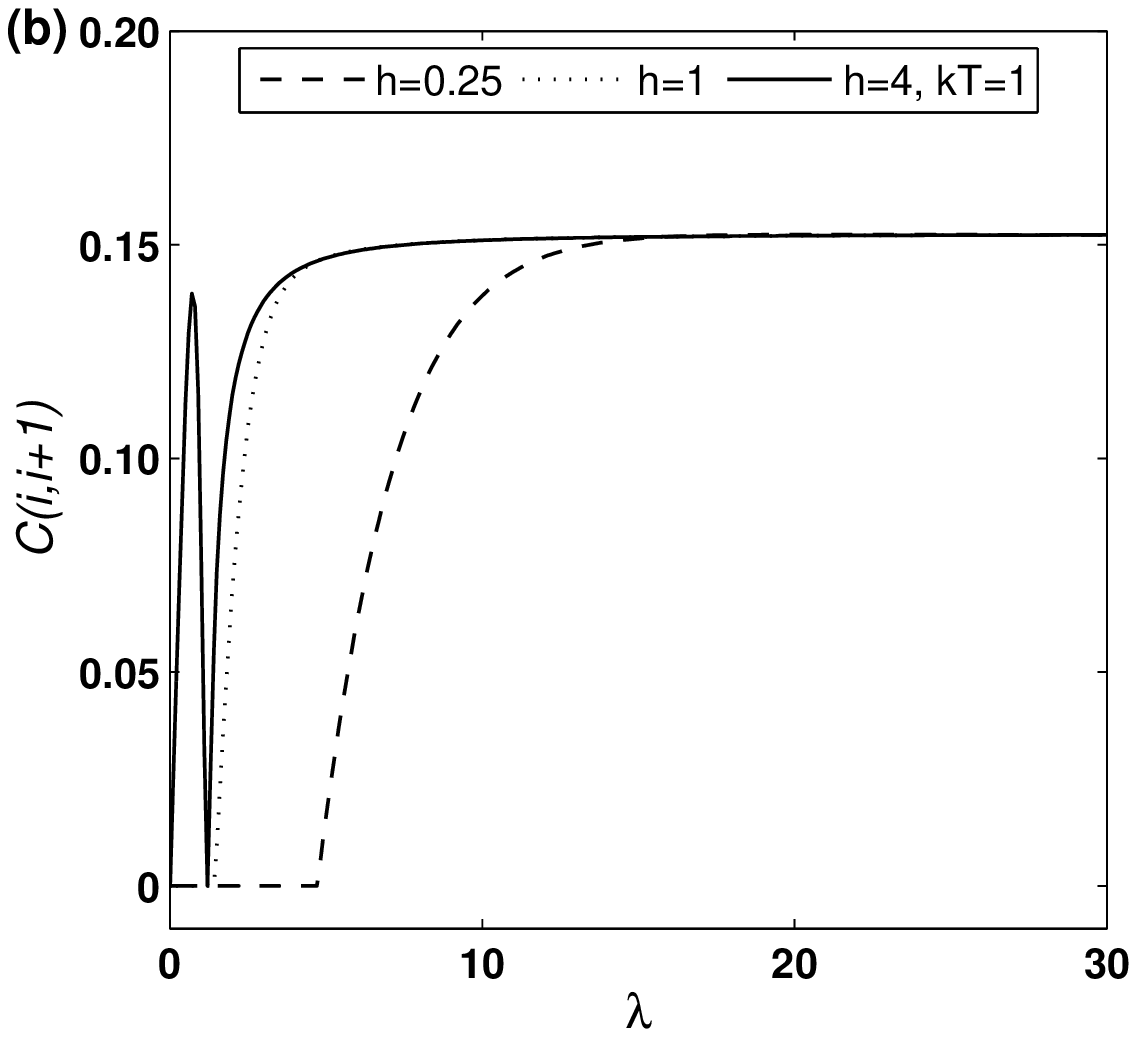}}\\
   \subfigure{\label{PRA1_fig:8c}\includegraphics[width=8cm]{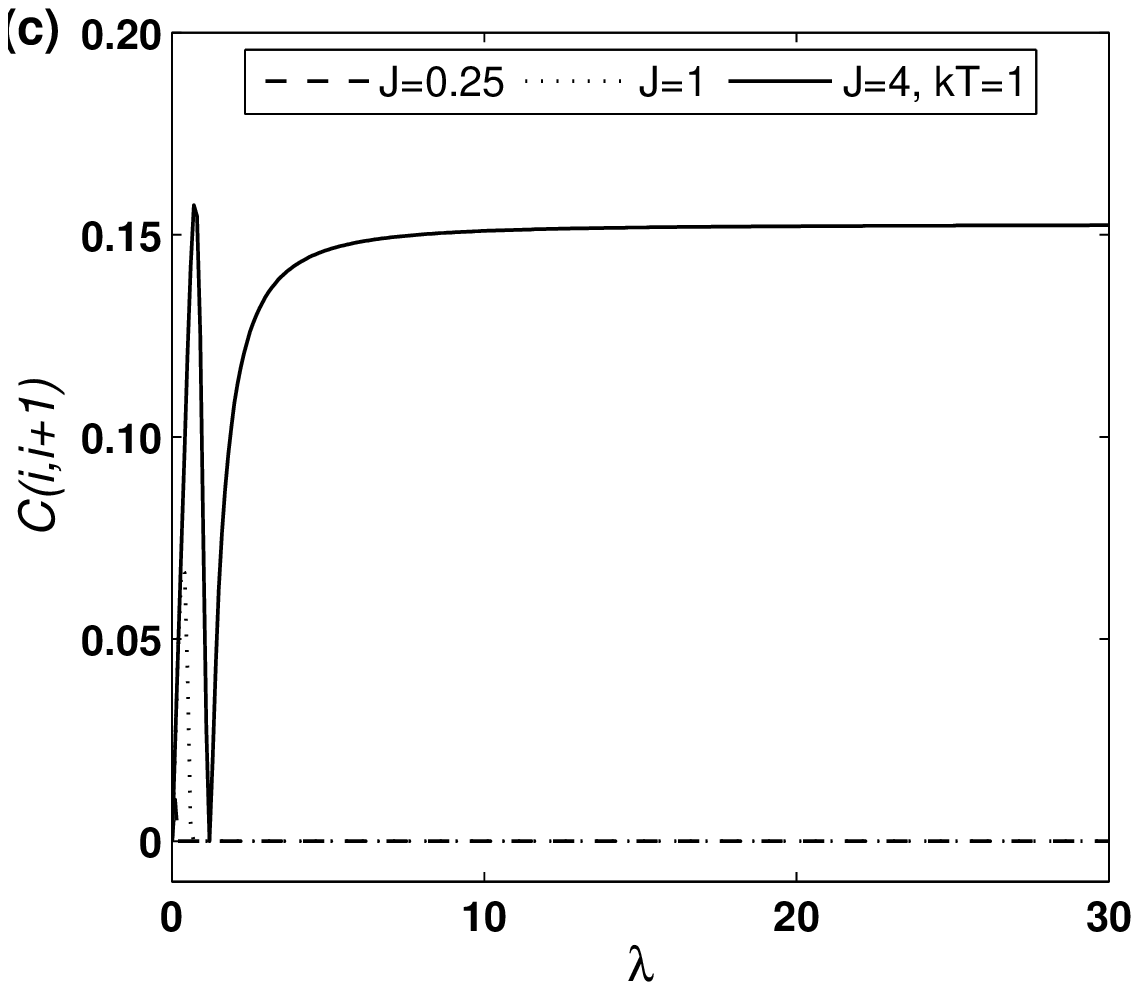}}\quad
   \subfigure{\label{PRA1_fig:8d}\includegraphics[width=8cm]{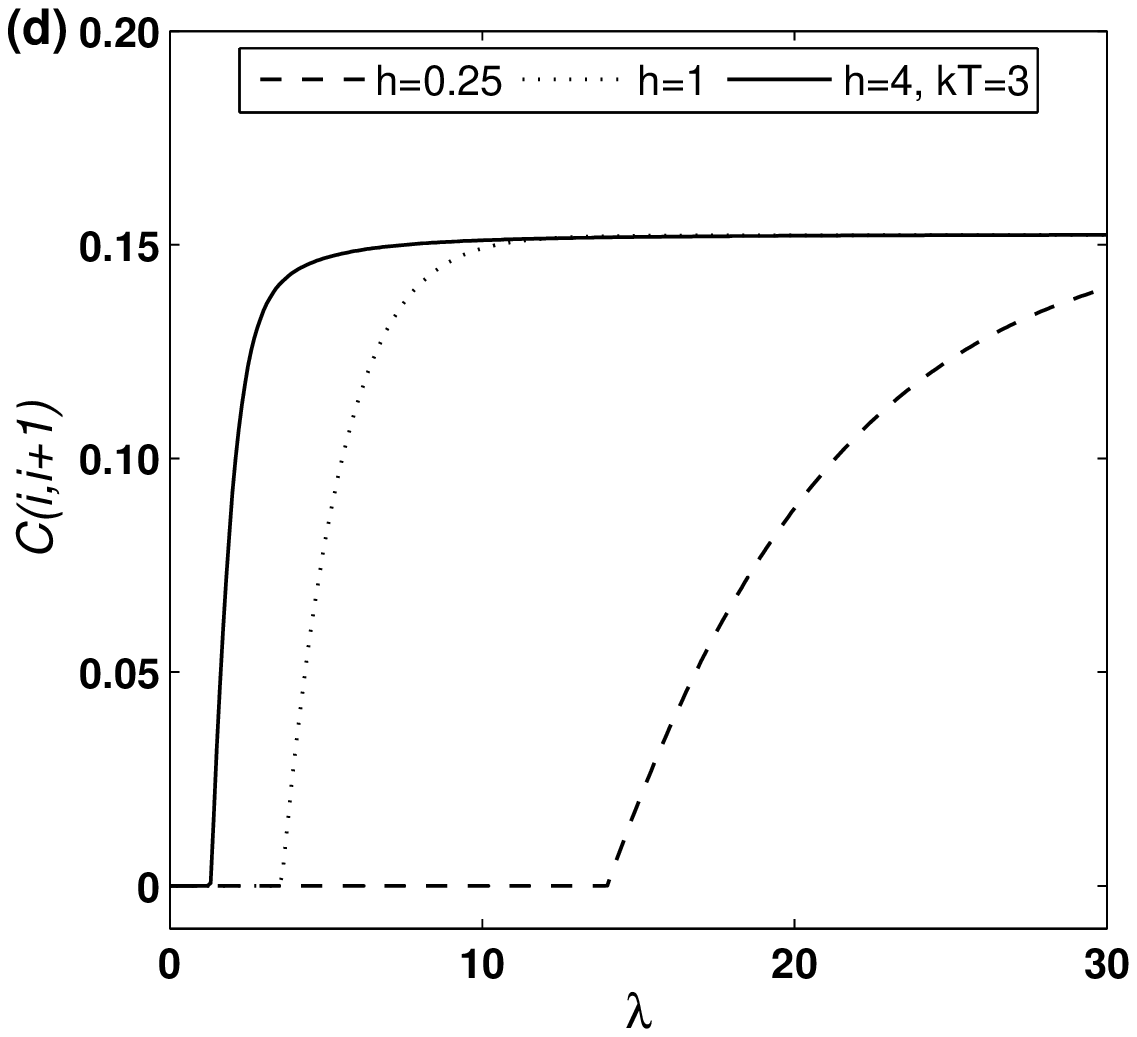}}
   \caption{{\protect\footnotesize The asymptotic behavior of $C(i,i+1)$ with $\gamma=0.5$ as a function of $\lambda$ when $h_{0}=h_{1}$ and $J_{0}=J_{1}$ at (a) $kT=0$ with any combination of constant $J$ and $h$; (b) $kT=1$ with $h_{0}=h_{1}=0.25, 1, 4$; (c) $kT=1$ with $J_{0}=J_{1}=0.25, 1, 4$; (d) $kT=3$ with $h_{0}=h_{1}=0.25, 1, 4$.}}
 \label{PRA1_fig:8}
 \end{minipage}
\end{figure}
We have investigated the behaviour of the nearest neighbour concurrence $C(i,i+1)$ as a function of $\lambda$ for different values of $J$ and $h$ and at different temperatures as shown in Fig.~\ref{PRA1_fig:8}. We first studied the zero temperature case at different constant values of $J$ and $h$. For this particular case, $C(i,i+1)$ depends only on the ratio of $J$ and $h$, similar to the isotropic case, rather than their individual values as depicted in Fig.~\ref{PRA1_fig:8a}. Interestingly, the concurrence shows a complicated critical behavior in the vicinity of $\lambda=1$, where it reaches a maximum value first and immediately drops to a minimum (very small) value before raising again to its equilibrium value. The raising of the concurrence from zero as $J$ increases, for $\lambda < 1$, is expected as in that case part of the spins which were originally aligned in the $z$-direction change directions into the $x$ and $y$-directions. The sudden drop of the concurrence in the vicinity of $\lambda=1$, where $\lambda$ is slightly larger than 1, suggests that significant fluctuations is taking place and the effect of $J_x$ is dominating over both $J_y$ and $h$ which aligns most of the spins into the x-direction leading to a reduced entanglement value. Studying the thermal concurrence in Fig.~\ref{PRA1_fig:8b}, \ref{PRA1_fig:8c} and \ref{PRA1_fig:8d} we note that the asymptotic value of $C(i,i+1)$ is not affected as the temperature increases. However the critical behavior of the entanglement in the vicinity of $\lambda=1$ changes considerably as the temperature is raised and the other parameters are varied. The effect of higher temperature is shown in Fig.~\ref{PRA1_fig:8d} where the critical behavior of the entanglement disappears completely at all values of $h$ and $J$ which confirms that the thermal excitations destroy the critical behavior due to suppression of quantum effects.
\begin{figure}[htbp]
\begin{minipage}[c]{\textwidth}
 \centering
   \subfigure{\label{PRA1_fig:10a}\includegraphics[width=8cm]{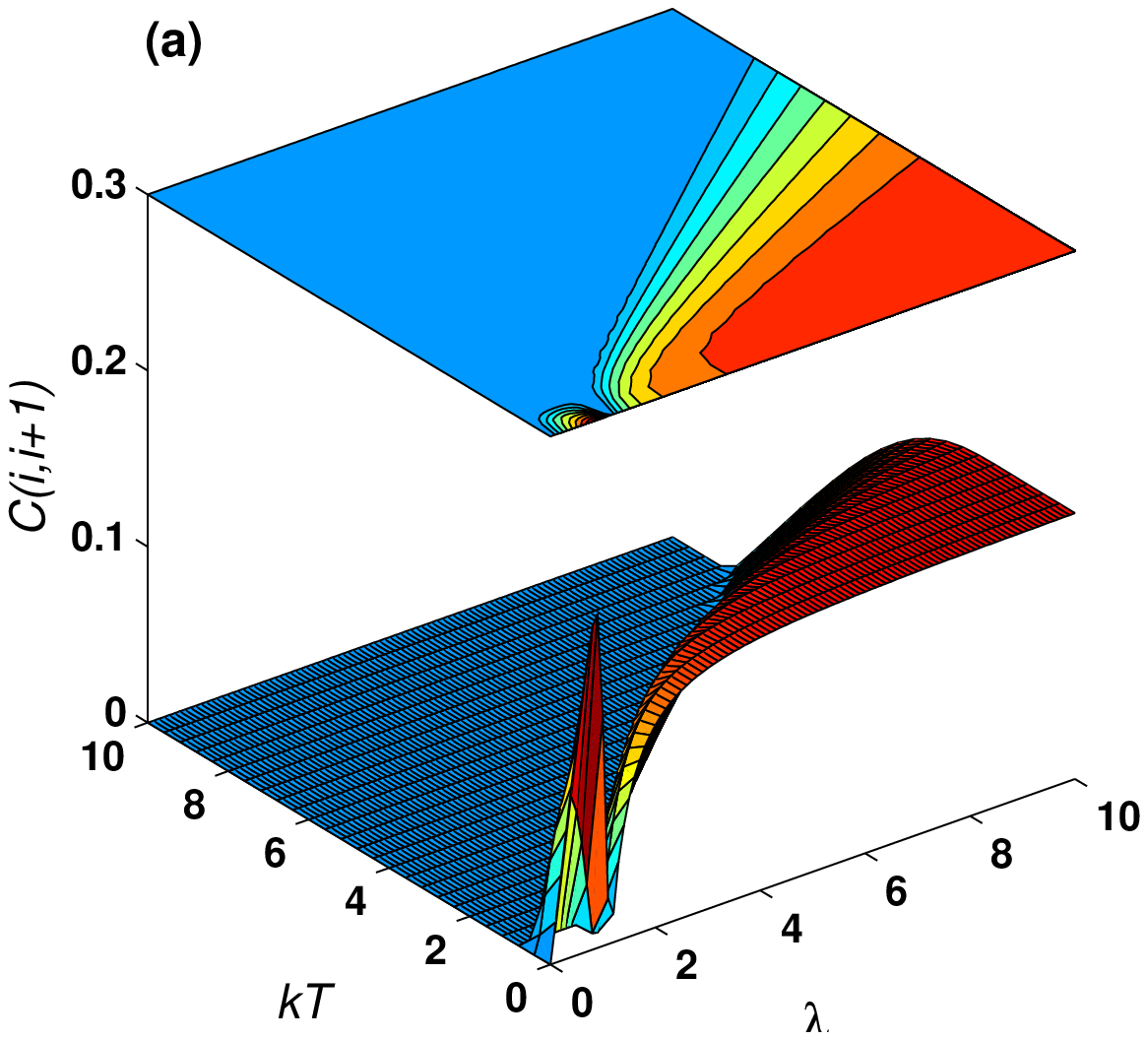}}\quad
   \subfigure{\label{PRA1_fig:10b}\includegraphics[width=8cm]{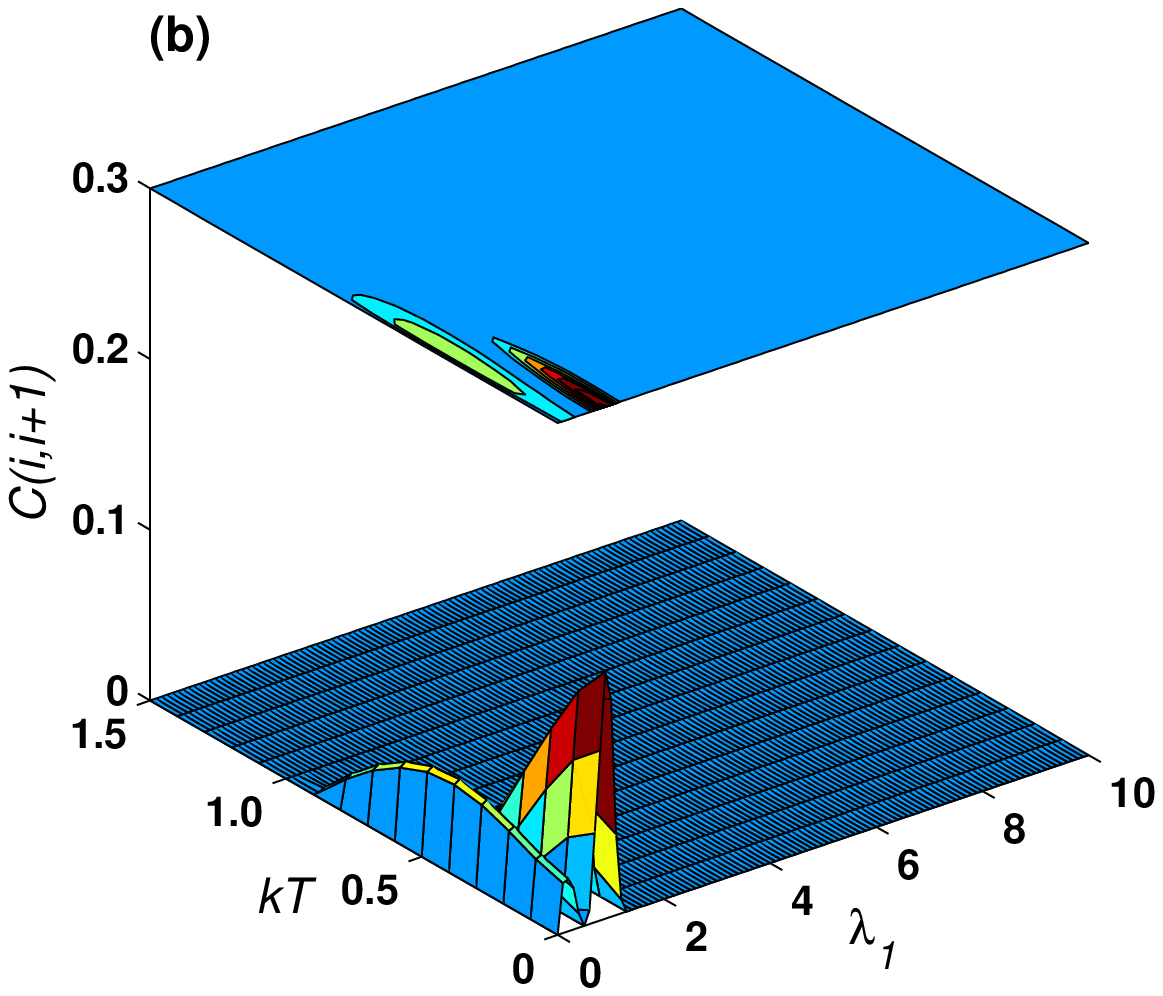}}
      \caption{{\protect\footnotesize The asymptotic behavior of $C(i,i+1)$ as a function of (a) $\lambda$ and $kT$, in units of $J_{1}$, with $\gamma=0.5$, $h_{0}=h_{1}=1$ and $J_{0}=J_{1}$; (b) $\lambda_{1}$ and $kT$ with $\gamma=0.5$, $h_{0}=h_{1}=1$ and $J_{0}=1$.}}
 \label{PRA1_fig:10}
 \end{minipage}
 \end{figure}
The persistence of quantum effects as temperature increases and time elapses in the partially anisotropic case is examined and presented in Fig.~\ref{PRA1_fig:10}. As demonstrated, the concurrence shows the expected behavior as a function of $\lambda$ and decays as the temperature increases. As shown, the threshold temperature where the concurrence vanishes is determined by the value of $\lambda$, it increases as $\lambda$ increases. The asymptotic behavior of the concurrence as a function $\lambda_1$ and $kT$ is illustrated. Clearly the non-zero concurrence shows up at small values of $kT$ and $\lambda_1$. The concurrence has two peaks versus $\lambda_1$ but as the temperature increases, the second peak disappears.
\subsubsection{Isotropic XY Model}
Now we consider the isotropic system where $\gamma=0$ (i.e. $J_{x}=J_{y}$).
We started with the dynamics of nearest neighbor concurrence, we found that $C(i,i+1)$ takes a constant value that does not depend on the final value of the coupling $J_{1}$ and magnetic field $h_{1}$. This follows from the dependence of the spin correlation functions and the magnetization on the initial state only as well.
This is because the initial coupling parameters $J_x$ and $J_y$, which are equal, force the spins to be equally aligned into the $x$- and $y$-directions, apart from those in the $z$-direction, causing a finite concurrence. Increasing the coupling parameters strength would not change that distribution or the associated concurrence at constant magnetic field.
\begin{figure}[htbp]
\begin{minipage}[c]{\textwidth}
 \centering
   \subfigure{\label{PRA1_fig:14a}\includegraphics[width=6cm]{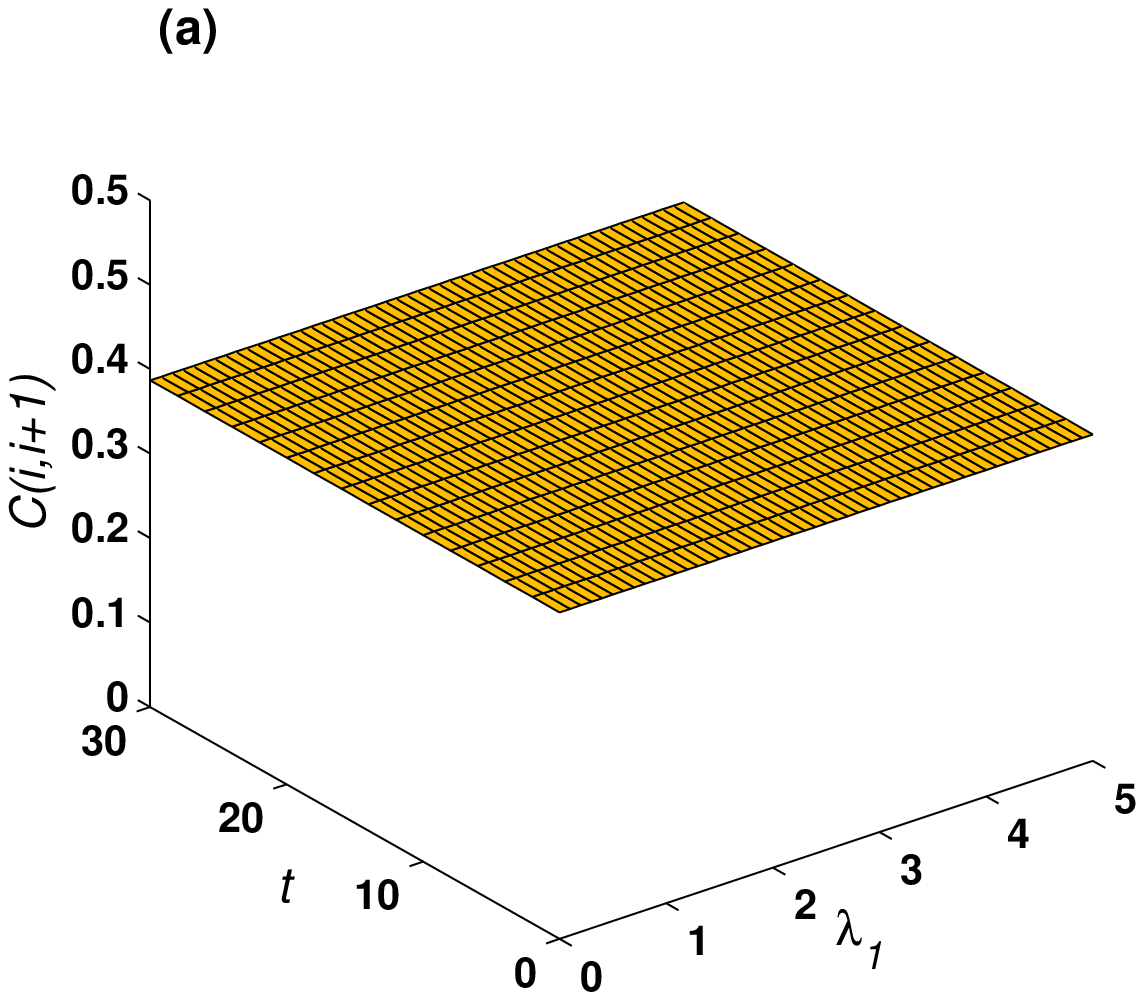}}\quad
   \subfigure{\label{PRA1_fig:14b}\includegraphics[width=6cm]{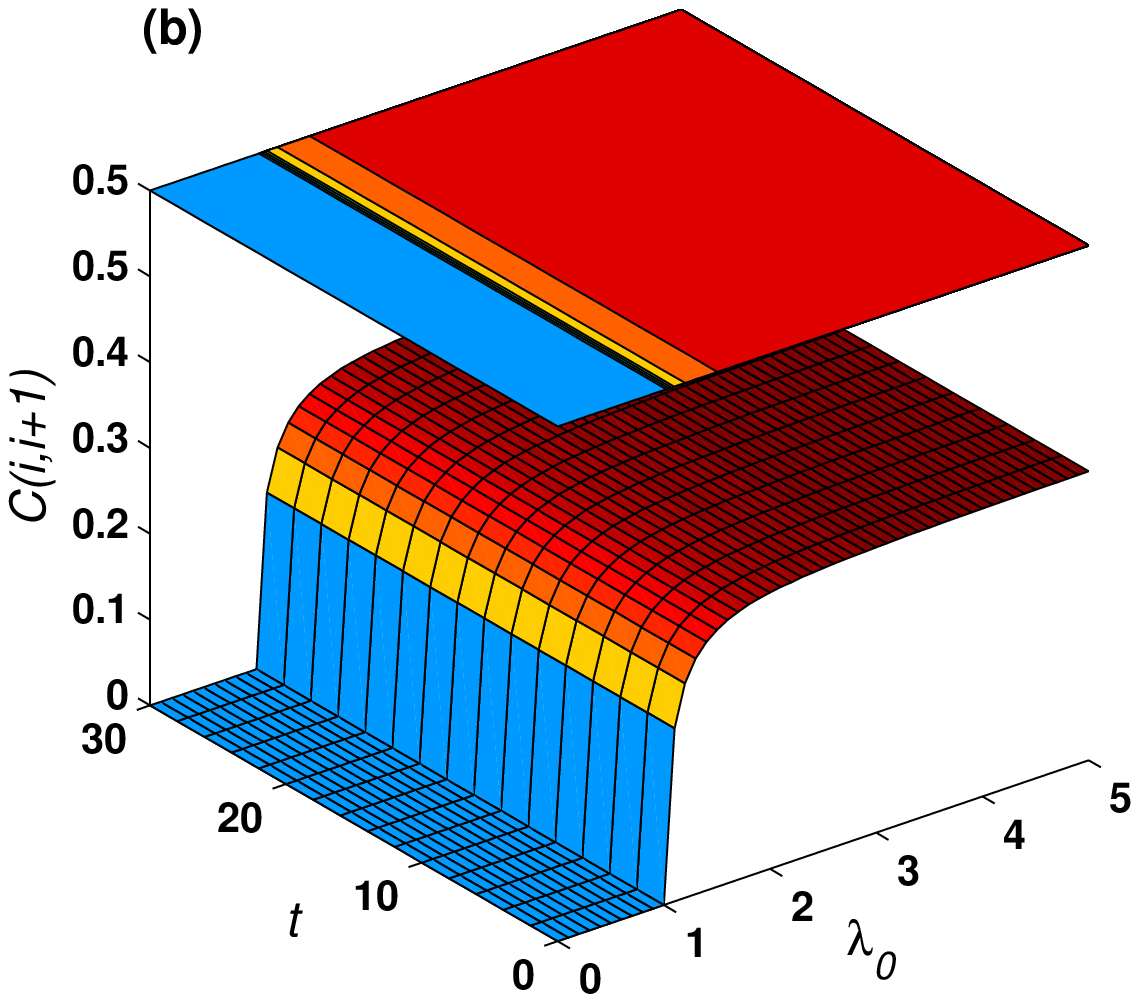}}\\
   \caption{{\protect\footnotesize (a) $C(i,i+1)$ as a function of $\lambda_{1}$ and $t$, in units of $J_{1}^{-1}$, at $kT=0$ with $\gamma=0$, $h_{0}=h_{1}=1$ and $J_{0}=5$; (b) $C(i,i+1)$ as a function of $\lambda_{0}$ and $t$ at $kT=0$ with $\gamma=0$, $h_{0}=h_{1}=1$ and $J_{1}=5$.}}
 \label{PRA1_fig:14}
 \end{minipage}
\end{figure}
The time evolution of nearest neighbor concurrence as a function of the time-dependent coupling is explored in Fig.~\ref{PRA1_fig:14}. Clearly, $C(i,i+1)$ is independent of $\lambda_{1}$. Studying $C(i,i+1)$ as a function of $\lambda_{0}$ and $t$ with $h_{0}=h_{1}=1$ at $kT=0$ for various values of $J_{1}$, we noticed that the results are independent of $J_{1}$. Again, as can be noticed when $J_{0}<h_{0}$ the magnetic field dominates and $C(i,i+1)$ vanishes. While for $J_{0} \geq h_{0}$, $C(i,i+1)$ has a finite value, as discussed above.
\begin{figure}[htbp]
\begin{minipage}[c]{\textwidth}
 \centering
   \subfigure{\label{PRA1_fig:17a}\includegraphics[width=6.5cm]{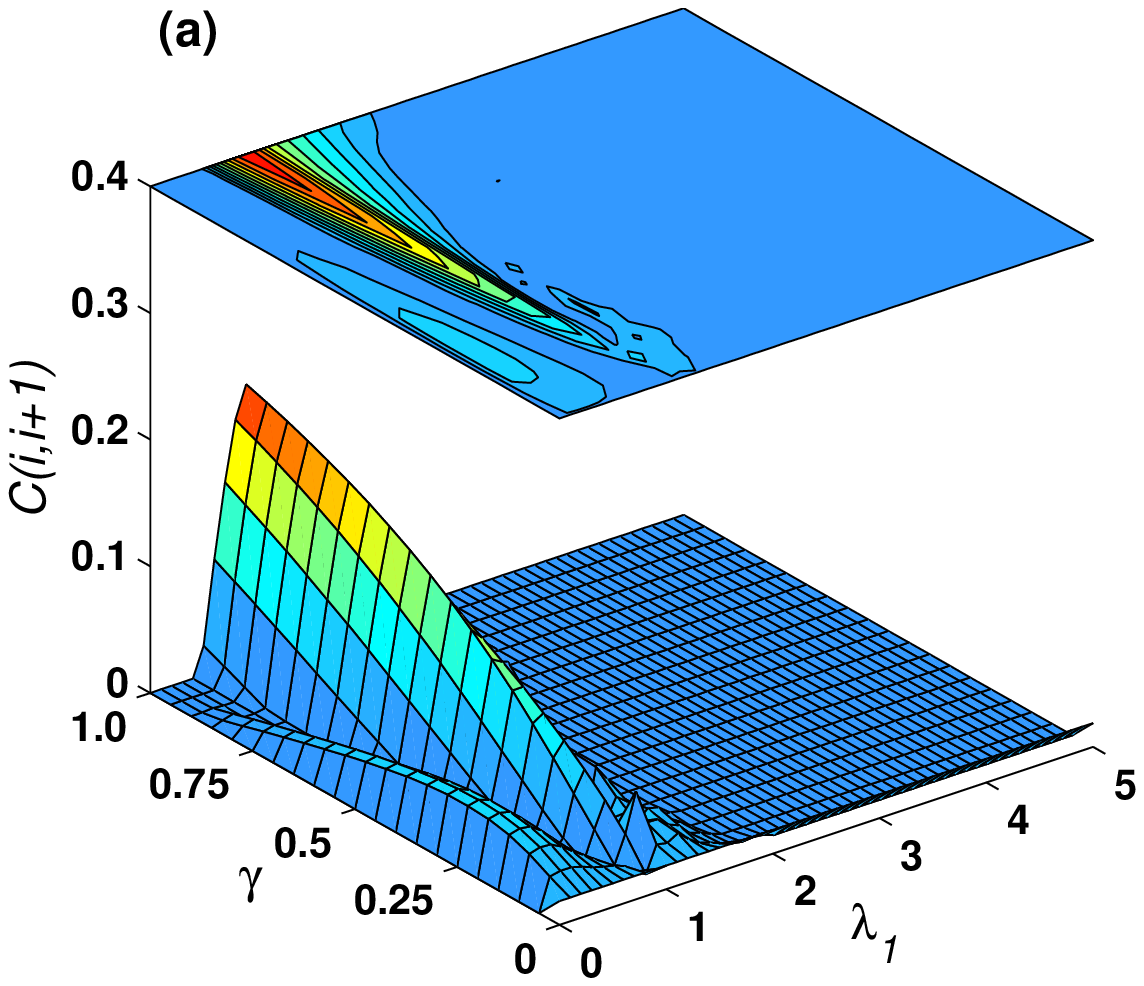}}\quad
   \subfigure{\label{PRA1_fig:17b}\includegraphics[width=6.5cm]{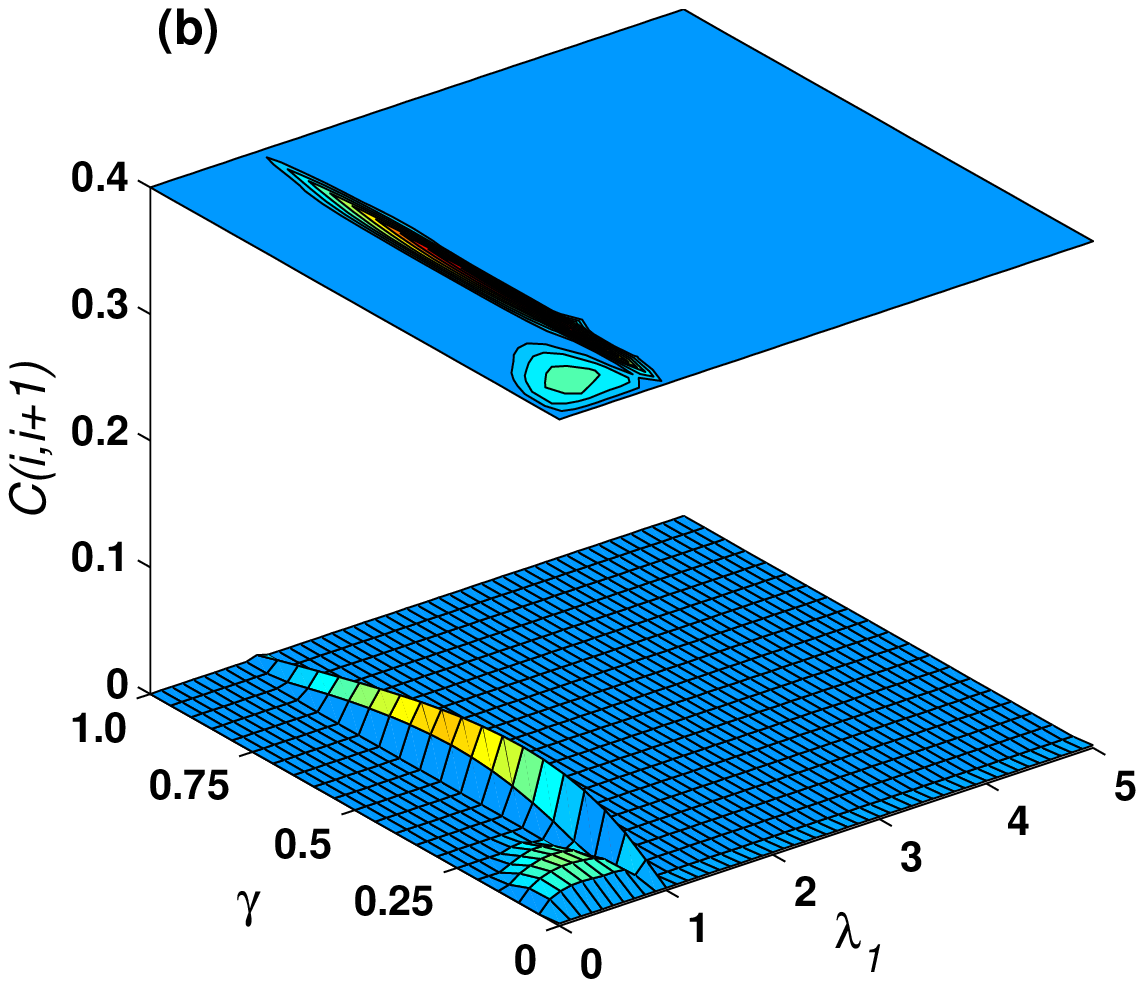}}\\
   \caption{{\protect\footnotesize (a) The asymptotic behavior of $C(i,i+1)$ as a function of $\lambda_{1}$ and $\gamma$ with $h_{0}=h_{1}=1$ and $J_{0}= 1$ at $kT=0$; (b) The asymptotic behavior of $C(i,i+2)$ as a function of $\lambda_{1}$ and $\gamma$ with $h_{0}=h_{1}=1$ and $J_{0}= 1$ at $kT=0$.}}
 \label{PRA1_fig:17}
\end{minipage}
\end{figure}
Finally we explored the asymptotic behavior of the nearest neighbor and next nearest neighbor concurrence in the $\lambda$-$\gamma$ phase space of the one dimensional $XY$ spin system under the effect of a time-dependent coupling $J(t)$ as shown in Fig.~\ref{PRA1_fig:17}. As one can notice, the non-vanishing concurrences appear in the vicinity of $\lambda=1$ or lower and vanishes for higher values. One interesting feature is that the maximum achievable nearest neighbor concurrence takes place at $\gamma=1$, i.e. in a completely anisotropic system, while the maximum next nearest neighbor concurrence is achievable in a partially anisotropic system, where $\gamma \approx 0.3$.
\subsection{Time evolution of the driven spin system in an external time-dependent magnetic field}
We investigate the time evolution of quantum entanglement in a one dimensional $XY$ spin chain system coupled through nearest neighbor interaction under the effect of an external magnetic field at zero and finite temperature. We consider both time-dependent nearest neighbor Heisenberg coupling $J(t)$ between the spins on the chain and magnetic field $h(t)$, where the function forms are exponential, periodic and hyperbolic in time. Particularly we focus on the concurrence as a measure of entanglement between any two adjacent spins on the chain and its dynamical behavior under the effect of the time-dependent coupling and magnetic fields. We apply both analytical and numerical approaches to tackle the problem. The Hamiltonian of the system is given by

\begin{equation}
H=-\frac{J(t)}{2} (1+\gamma) \sum_{i=1}^{N} \sigma_{i}^{x} \sigma_{i+1}^{x}-\frac{J(t)}{2}(1-\gamma)\sum_{i=1}^{N} \sigma_{i}^{y} \sigma_{i+1}^{y}- \sum_{i=1}^{N} h(t) \sigma_{i}^{z}\, ,
\label{eq:H2}
\end{equation}
where $\sigma_{i}$'s are the Pauli matrices and $\gamma$ is the anisotropy parameter.
\subsubsection{Numerical and exact solutions}
Following the standard procedure to treat the Hamiltonian (\ref{eq:H2}) \cite{Lieb1961}, we again obtain the Hamiltonian in the form
\begin{equation}
H=\sum_{p=1}^{N/2} \tilde{H}_{p}\, ,
\label{eq:Hsum2}\end{equation}
with $\tilde{H}_{p}$ given by
\begin{equation}
\tilde{H}_{p}=\alpha_{p}(t) [c_{p}^{\dagger} c_{p}+c_{-p}^{\dagger} c_{-p}]+i J(t) \delta_{p} [c_{p}^{\dagger} c_{-p}^{\dagger}+c_{p} c_{-p}]+2 h(t)\, ,
\label{eq:Hp2}\end{equation}
where $\alpha_{p}(t)=-2 J(t) \cos \phi_{p} - 2 h(t)$ and  $\delta_{p}=2 \gamma \sin \phi_{p}$.
Initially the system is assumed to be in a thermal equilibrium state and therefore its initial density matrix is given by
\begin{equation}
\rho_{p}(0)=\frac{e^{-\beta \tilde{H}_{p}(0)}}{Z}\, , \hspace{0.5cm}  Z=Tr(e^{-\beta \tilde{H}_{p}(0)})\, ,
\label{eq:rho02}\end{equation}
where $\beta=1/k T$, $k$ is Boltzmann constant and $T$ is the temperature.

Since the Hamiltonian is decomposable we can find the density matrix at any time $t$, $\rho_{p}(t)$, for the $p$th subspace by solving Liouville equation, in the Heisenberg representation by following the same steps we applied in Eqs. \ref{eq:Liouville}-\ref{eq:U}.

To study the effect of a time-varying coupling parameter $J(t)$ we consider the following forms
\begin{eqnarray}
J_{exp}(t)&=&J_{1}+\left(J_{0}-J_{1}\right) e^{-K t} \, ,\\
J_{cos}(t)&=&J_{0}-J_{0} \cos\left(K t\right)\, ,\quad\quad\\
J_{sin}(t)&=&J_{0}-J_{0} \sin\left(K t\right)\, ,\quad\quad\\
J_{tanh}(t)&=&J_{0}+\frac{J_1-J_0}{2} \left[\tanh\left(K (t-\frac{5}{2})\right)+1\right]\,.
\end{eqnarray}

Note that Eq.~(\ref{eq:Udot}), in the current case, gives two systems of coupled differential equations with variable coefficients. Such systems can only be solved numerically which we carried out in this work. Nevertheless, an exact solution of the system can be obtained using a general time-dependent coupling $J(t)$ and a magnetic field in the following form:
\beq
J(t)=\lambda \: h(t)
\label{eq:proportional}
\eeq
where $\lambda$ is a constant. Using (\ref{eq:Udot}) and (\ref{eq:proportional}) we obtain the non-vanishing matrix elements
\beq
i \left(\begin{array}{cc}
\dot{u}_{11} & \dot{u}_{12}\\
\dot{u}_{21} & \dot{u}_{22}\\
\end{array}\right)=	\left(\begin{array}{cc}
u_{11} & u_{12}\\
u_{21} & u_{22}\\
\end{array}\right) \left(\begin{array}{cc}
\frac{2}{\lambda} & -i \delta_p\\
i \delta_p & -4 \cos \phi_p - \frac{2}{\lambda}\\
\end{array}\right) J(t)\, ,
\label{eq:U2dot}
\eeq
and
\beq
i \: \dot{u}_{33}=-2 \: \cos \phi_p \: J(t) \: u_{33}\, , \, \, \, \, \, u_{44}=u_{33}\, .
\eeq

These system of coupled differential equations can be exactly solved to yield
\beq
u_{11}=\cos^2 \theta e^{-i \lambda_1\int^t_0{J(t') dt'}} + \sin^2 \theta e^{-i \lambda_2\int^t_0{J(t') dt'}}\, ,
\eeq
\beq
u_{12}=-i\sin \theta\cos \theta \left\{e^{-i \lambda_1\int^t_0{J(t') dt'}} - e^{-i \lambda_2\int^t_0{J(t') dt'}}\right\}\, ,
\eeq
\beq
u_{21}=-u_{12}\, ,
\eeq
\beq
u_{22}=\sin^2 \theta e^{-i \lambda_1\int^t_0{J(t') dt'}} + \cos^2 \theta e^{-i \lambda_2\int^t_0{J(t') dt'}}\, ,
\eeq
\beq
u_{33}=u_{44}=e^{2i \cos\phi_p \int^t_0{J(t') dt'}} \, ,
\eeq
where
\beq
\sin\theta=\sqrt{\frac{\sqrt{\delta_p^2+(2\cos\phi_p+\frac{2}{\lambda})}-(2\cos\phi_p+\frac{2}{\lambda})}{2\sqrt{\delta_p^2+(2\cos\phi_p+\frac{2}{\lambda})}}}\, ,
\eeq
\beq
\cos\theta=\sqrt{\frac{\sqrt{\delta_p^2+(2\cos\phi_p+\frac{2}{\lambda})}+(2\cos\phi_p+\frac{2}{\lambda})}{2\sqrt{\delta_p^2+(2\cos\phi_p+\frac{2}{\lambda})}}}\, .
\eeq
Where the angles $\phi$ and $\theta$ were found to be
\beq
\phi=(n+1) \pi, \: \:\:\: \tan{2\theta} = \frac{\delta_p}{2\cos\phi_p+\frac{2}{\lambda}}\: ,
\eeq
where $n=0,\pm1,\pm2,\ldots$, therefore
\beq
\sin{2\theta} = \frac{\delta_p}{\sqrt{\delta_p^2+(2\cos\phi_p+\frac{2}{\lambda})}}, \, \:\:\: \cos{2\theta} = \frac{2\cos\phi_p+\frac{2}{\lambda}}{\sqrt{\delta_p^2+(2\cos\phi_p+\frac{2}{\lambda})}}\,.
\eeq
As usual, upon obtaining the matrix elements of the evolution operators, either numerically or analytically,  one can evaluate the matrix elements of the two-spins density operator with the help of the magnetization and the spin-spin correlation funtions of the system and finally produce the concurrence using Wootters method \cite {Wootters1998}.
\subsubsection{Constant magnetic field and time-varying coupling}
First we studied the dynamics of the nearest neighbor concurrence $C(i,i+1)$ for the completely anisotropic system, $\gamma=1$, when the coupling parameter is $J_{exp}$ and the magnetic field is a constant using the numerical solution.
\begin{figure}[htbp]
\begin{minipage}[c]{\textwidth}
 \centering
   \subfigure{\label{fig:expK01}\includegraphics[width=7cm]{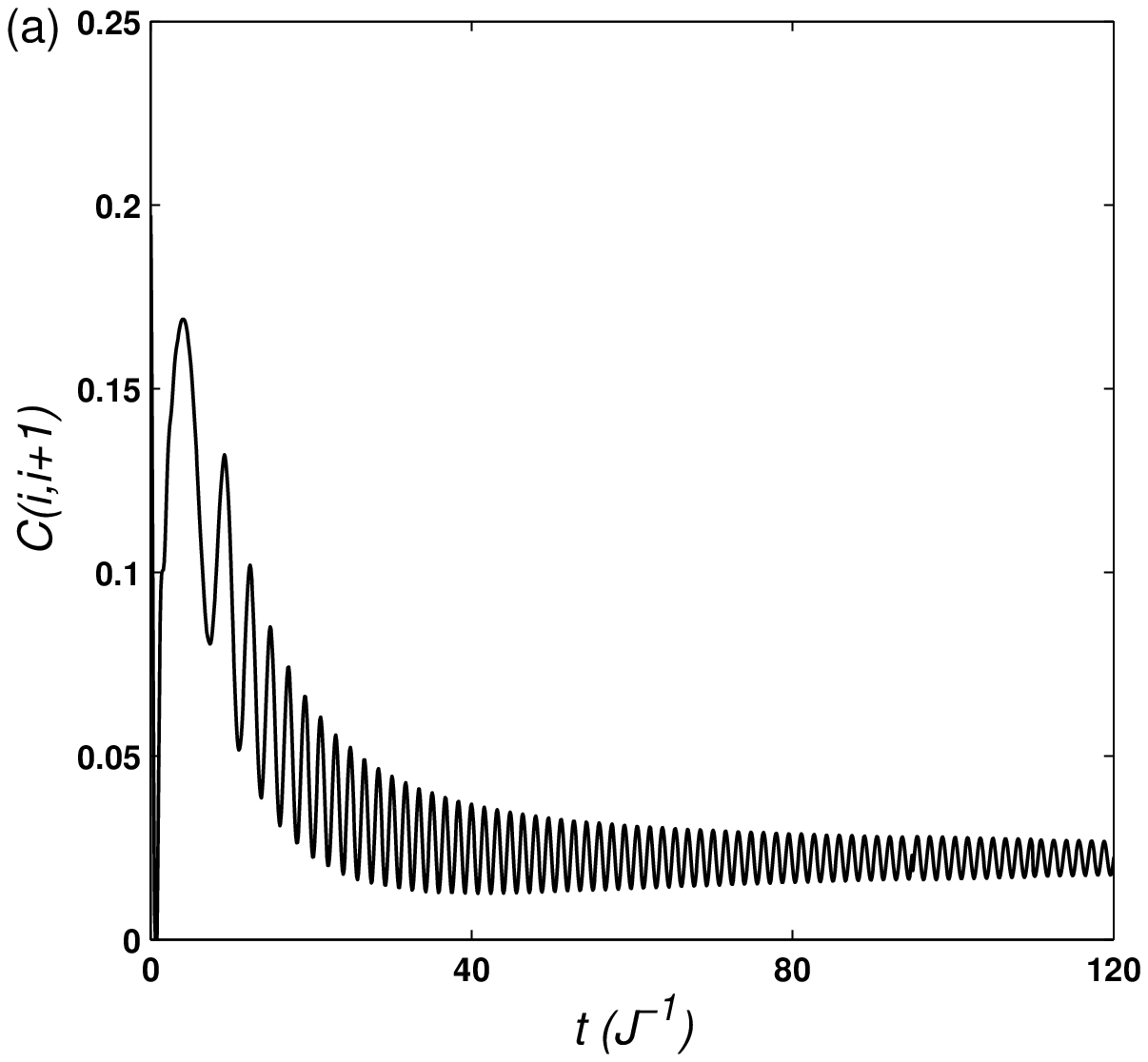}}\quad
   \subfigure{\label{fig:expK10}\includegraphics[width=7cm]{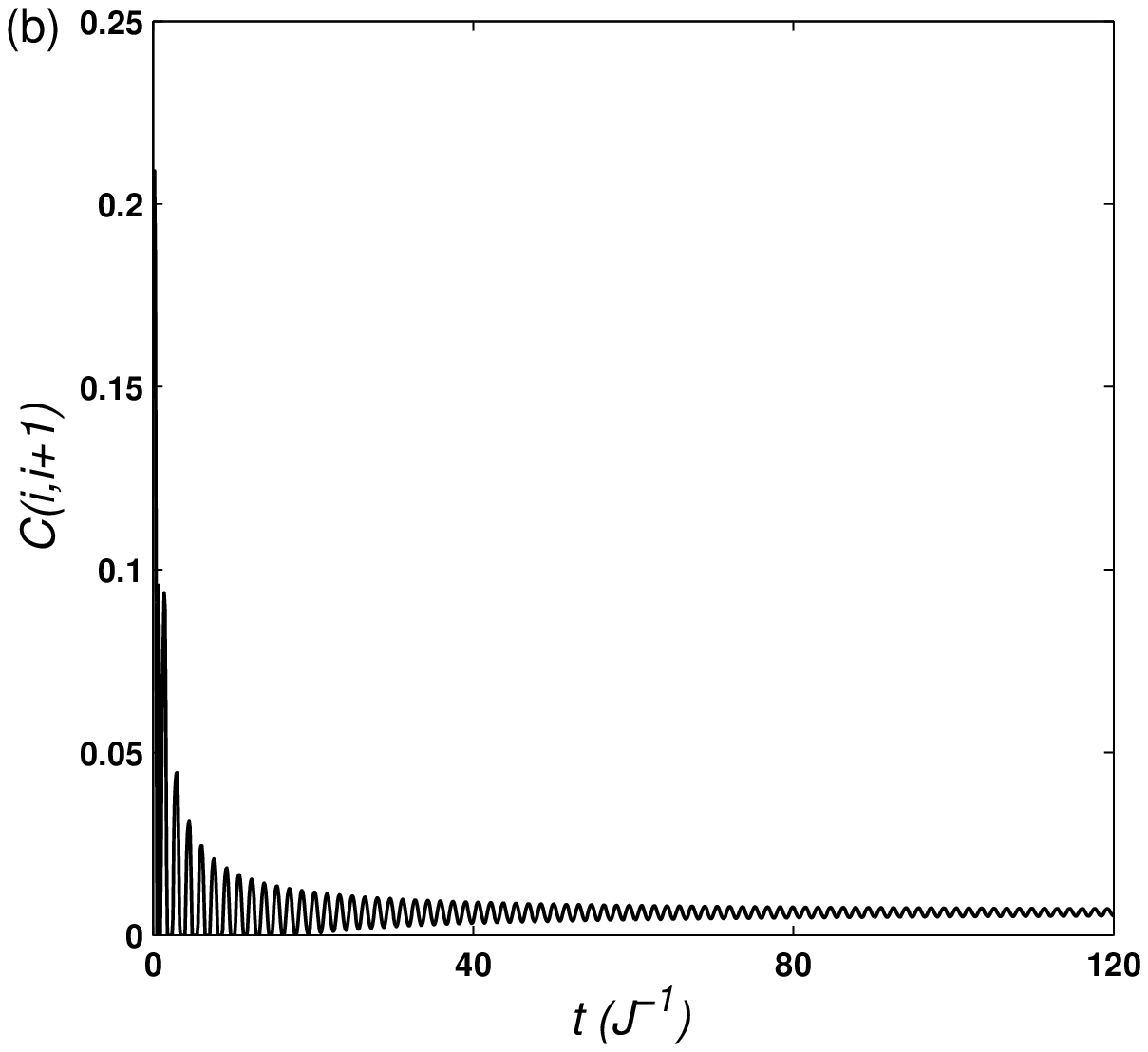}}
   \caption{{\protect\footnotesize $C(i,i+1)$ as a function of $t$ with $J_{0}=0.5, J_{1}=2, h=1, N=1000$ at $kT=0$ and (a) $J= J_{exp}, K=0.1$ (b) $J= J_{exp}, K=10$ ; (c) $J= J_{tanh}, K=0.1$ ; (d) $J= J_{tanh}, K=10$.}}
  \label{fig:Ks}
 \end{minipage}
\end{figure}
As can be noticed in Fig.~\ref{fig:Ks}, the asymptotic value of the concurrence depends on $K$ in addition to the coupling parameter and magnetic field. The larger the transition constant is, the lower is the asymptotic value of the entanglement and the more rapid decay is. This result demonstrates the non-ergodic behavior of the system, where the asymptotic value of the entanglement is different from the one obtained under constant coupling $J_{1}$.

\begin{figure}[htbp]
 \centering
	\includegraphics[width=14cm]{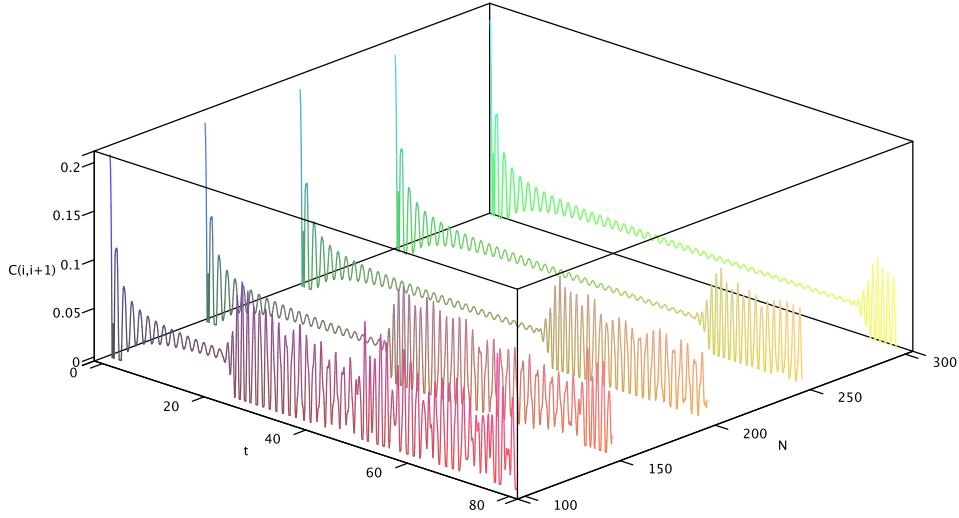}
   \caption{{\protect\footnotesize $C(i,i+1)$ as a function of $t$ (units of $J^{-1}$) with $J= J_{exp}, J_{0}=0.5, J_{1}=2, h=1, K=1000$ at $kT=0$ and $N$ varies from $100$ to $300$.}}
  \label{fig:Ns}
\end{figure}

\begin{figure}[tbph]
\begin{minipage}[c]{\textwidth}
 \centering
   \subfigure{\label{fig:Jcosa}\includegraphics[width=6.5cm]{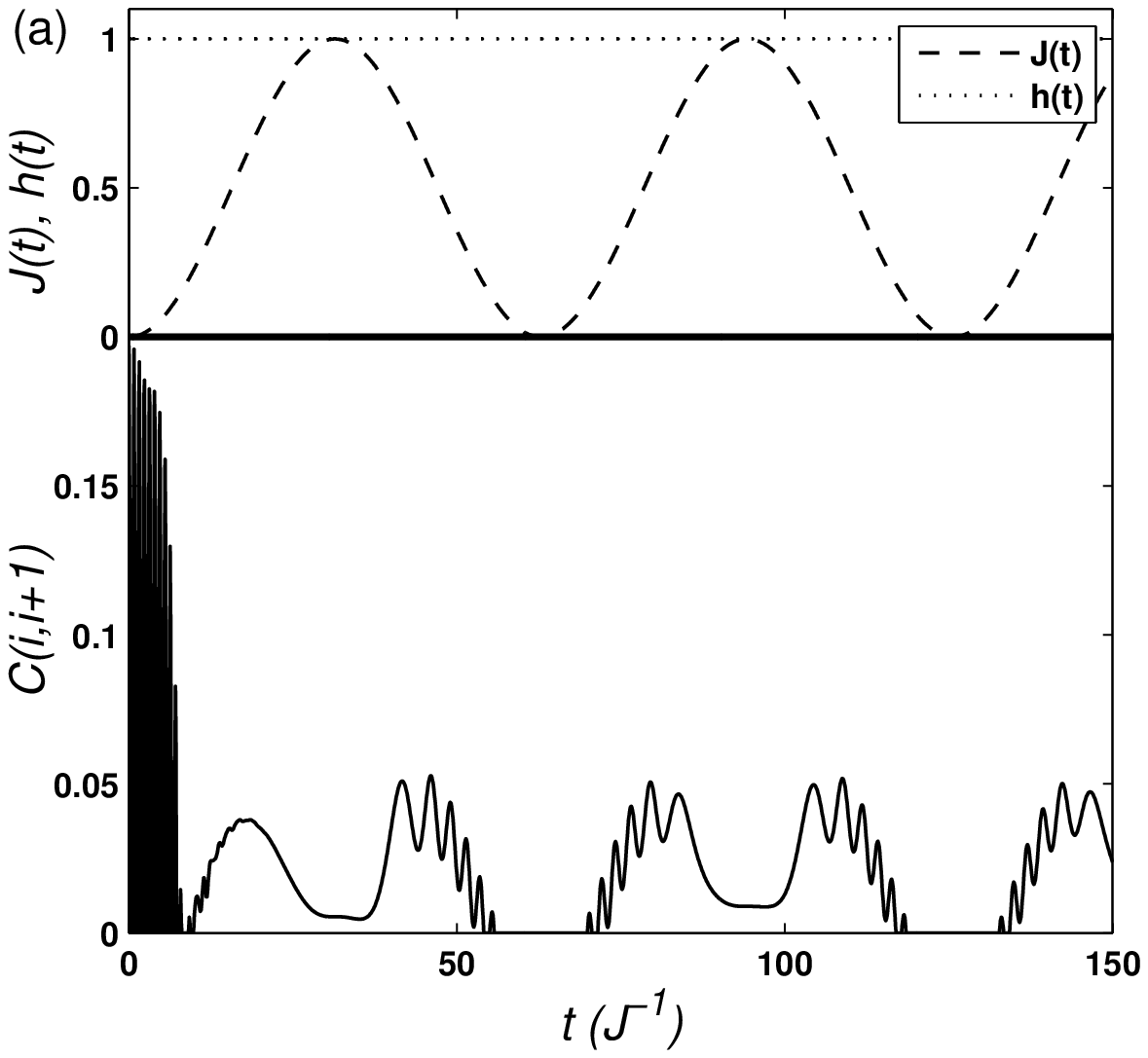}}
   \subfigure{\label{fig:Jcosb}\includegraphics[width=6.5cm]{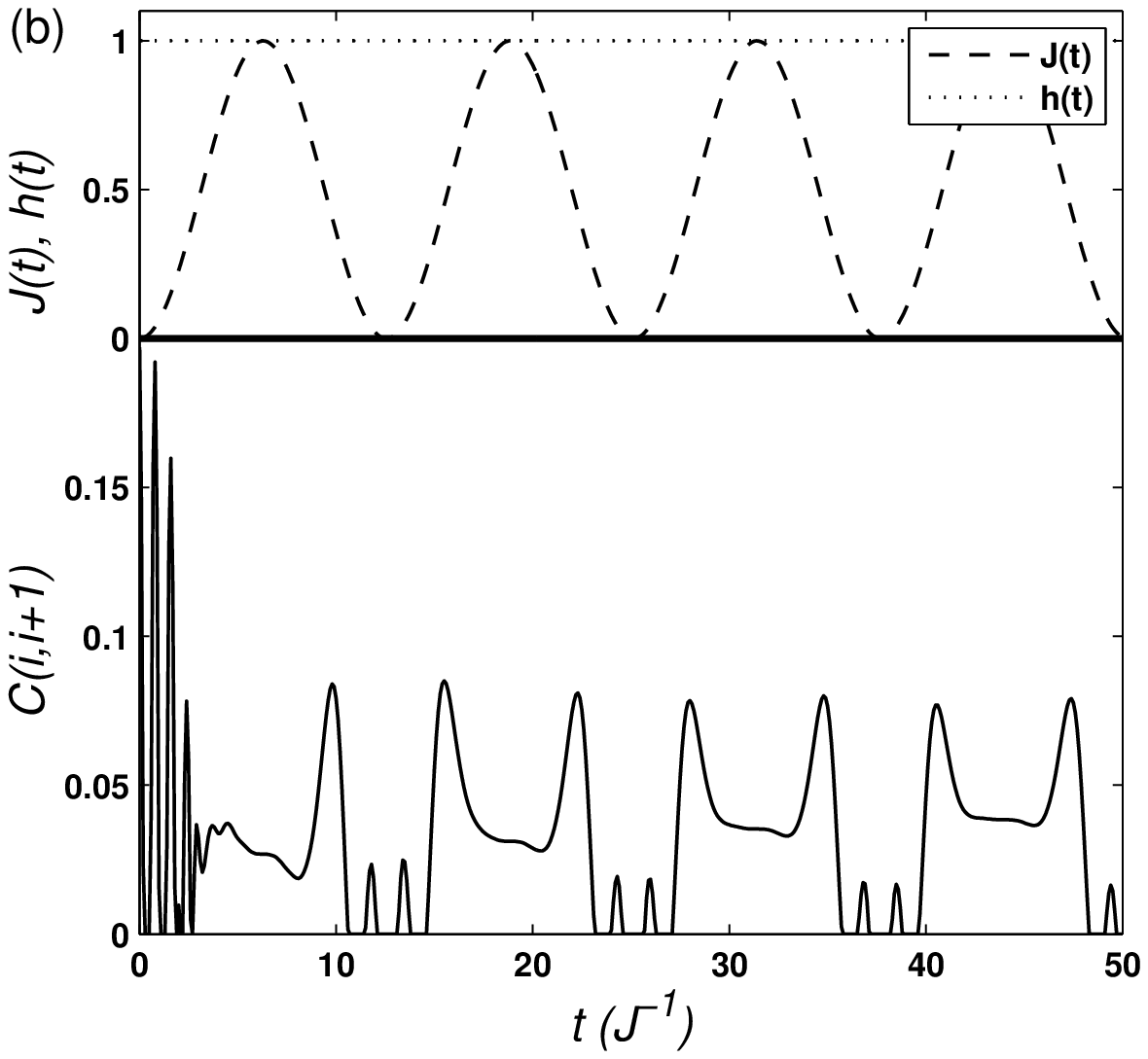}}
   \caption{{\protect\footnotesize Dynamics of nearest neighbor concurrence with $\gamma=1$ for $J_{cos}$ where $J_{0}=0.5, h=1$ at $kT=0$ and (a) $K=0.1$ ; (b) $K=0.5$.}}
   \label{fig:Jcos}
   \end{minipage}
\end{figure}

We have examined the effect of the system size $N$ on the dynamics of the concurrence, as depicted in Fig.~\ref{fig:Ns}. We note that for all values of $N$ the concurrence reaches an approximately constant value but then starts oscillating after some critical time $t_c$, that increases as $N$ increases, which means that the oscillation will disappear as we approach an infinite one-dimensional system. Such oscillations are caused by the spin-wave packet propagation \cite{HuangZ2006}.
We next study the dynamics of the nearest neighbor concurrence when the coupling parameter is $J_{cos}$ with different values of $K$, i.e. different frequencies, which is shown in Fig.~\ref{fig:Jcos}. We first note that $C(i,i+1)$ shows a periodic behavior with the same period of $J(t)$. We have shown in our previous work \cite{Sadiek2010} that for the considered system at zero temperature the concurrence depends only on the ratio $J/h$. When $J \approx h$ the concurrence has a maximum value. While when $J>>h$ or $J<<h$ the concurrence vanishes. In Fig.~\ref{fig:Jcos}, one can see that when $J=J_{max}$, $C(i,i+1)$ decreases because large values of $J$ destroy the entanglement, while $C(i,i+1)$ reaches a maximum value when $J=J_{0}=0.5$. As $J(t)$ vanishes, $C(i,i+1)$ decreases because of the magnetic field domination.

\subsubsection{Time-dependent magnetic field and coupling}
\begin{figure}[htbp]
\begin{minipage}[c]{\textwidth}
 \centering
  	\subfigure{\label{fig:bothcos}\includegraphics[width=7cm]{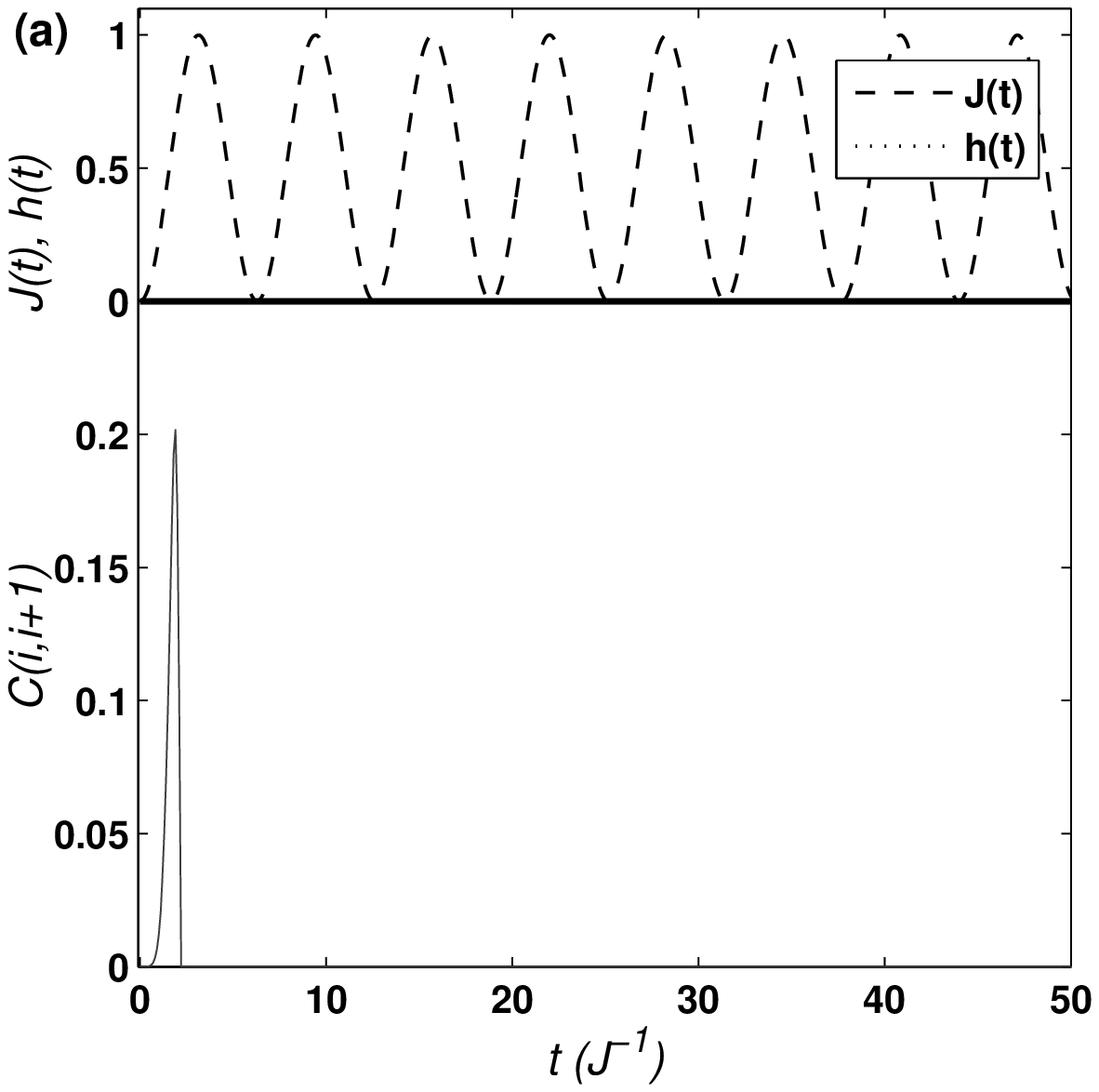}}
  	\subfigure{\label{fig:bothsin}\includegraphics[width=7cm]{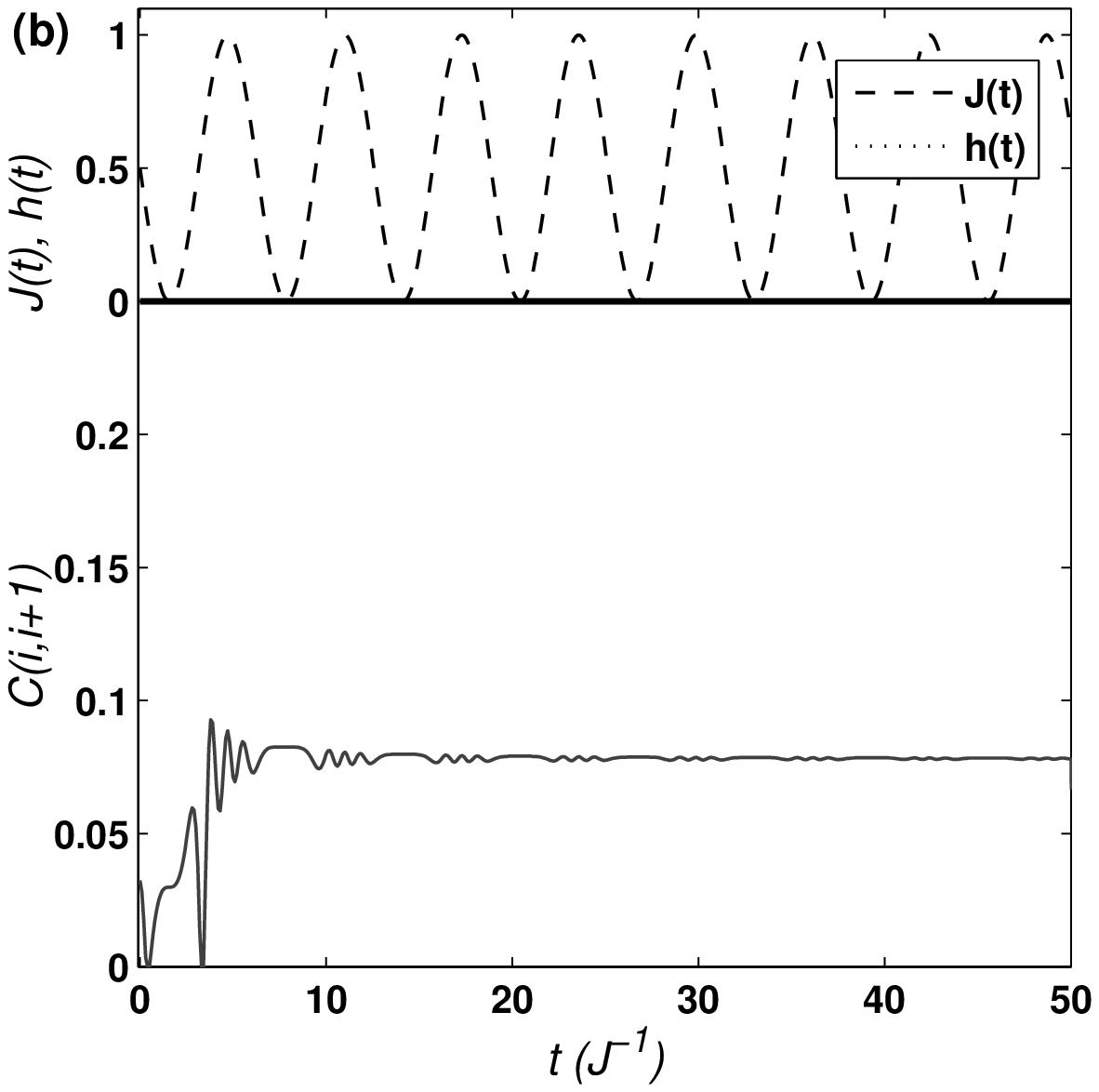}}
		\caption{{\protect\footnotesize Dynamics of nearest neighbor concurrence with $\gamma=1$ at $kT=0$ with $J_{0}=h_{0}=0.5,K=1$ for (a) $J_{cos}$ and $h_{cos}$ ; (b) $J_{sin}$ and $h_{sin}$.}}
  \label{fig:both2}
\end{minipage}
\end{figure}

\begin{figure}[htbp]
\begin{minipage}[c]{\textwidth}
 \centering
  	\label{fig:JexpLvar}\includegraphics[width=7cm]{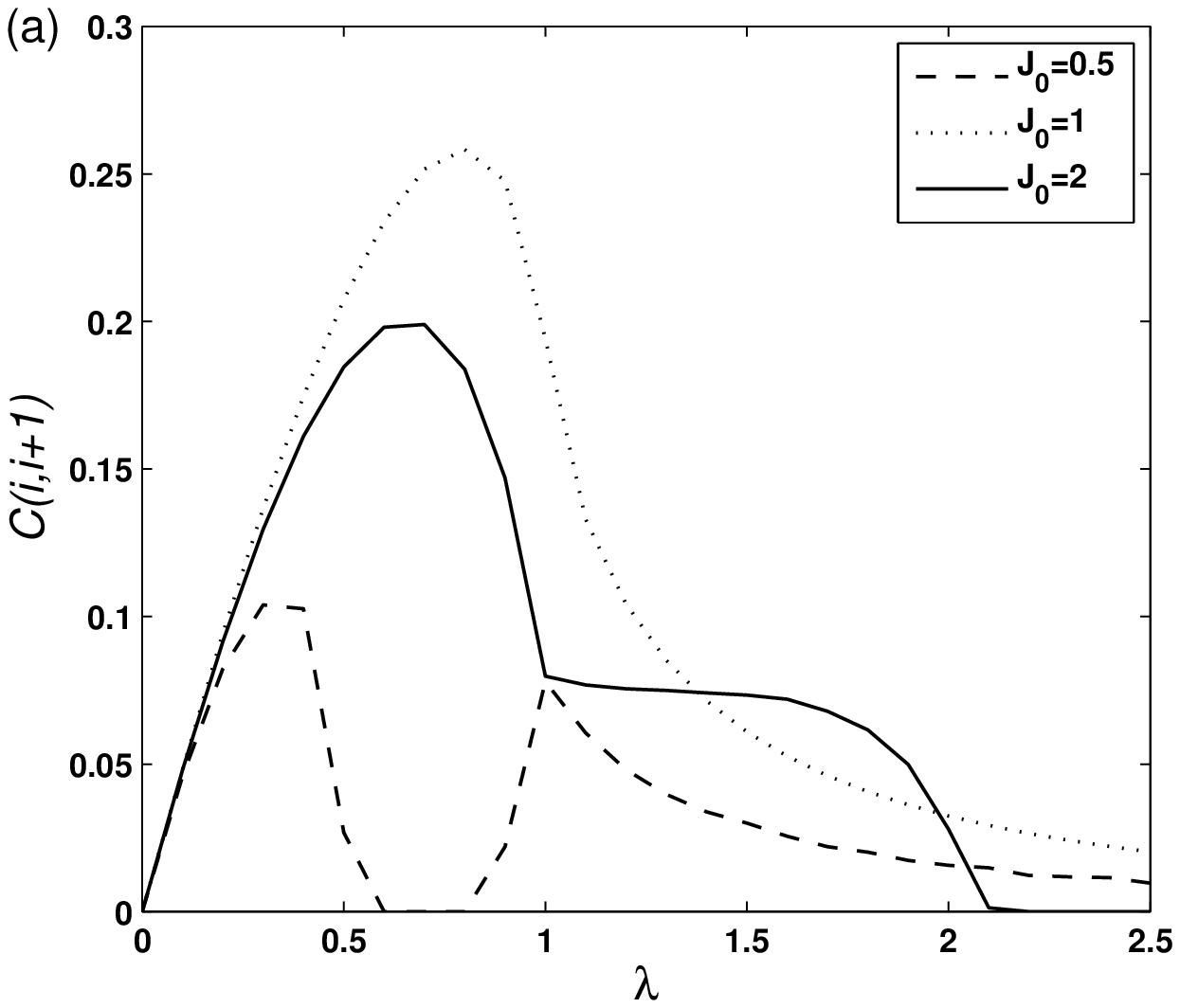}
  	\label{fig:JLvarkTnon0}\includegraphics[width=7cm]{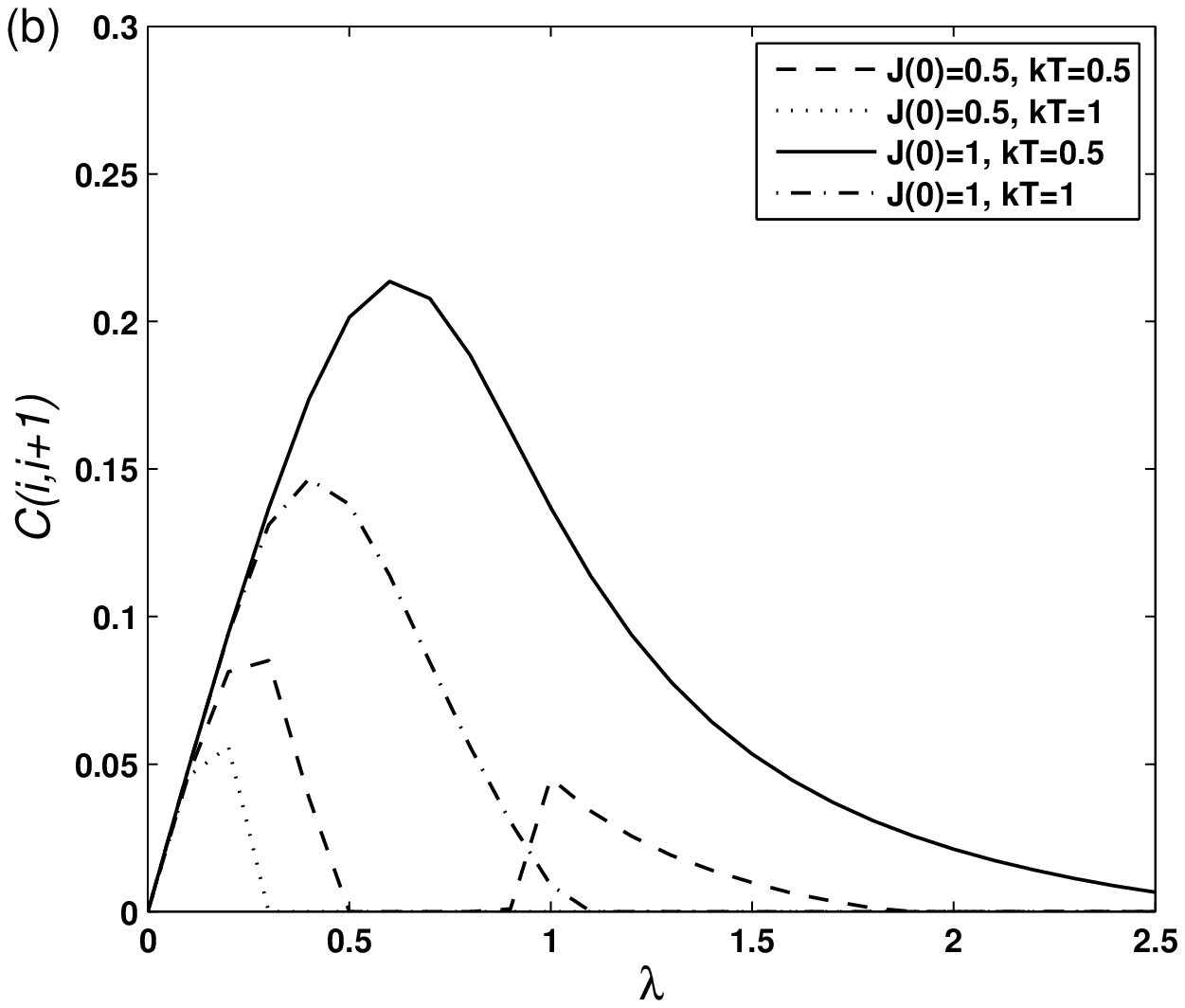}
		\caption{{\protect\footnotesize The behavior asymptotic value of $C(i,i+1)$ as a function of $\lambda$ with $\gamma=1$ at (a) $kT=0$ ; (b) $kT=0.5, 1$.}}
  \label{fig:Lvar}
\end{minipage}
\end{figure}

\begin{figure}[htbp]
\begin{minipage}[c]{\textwidth}
 \centering
  	\label{fig:Gamma05}\includegraphics[width=7cm]{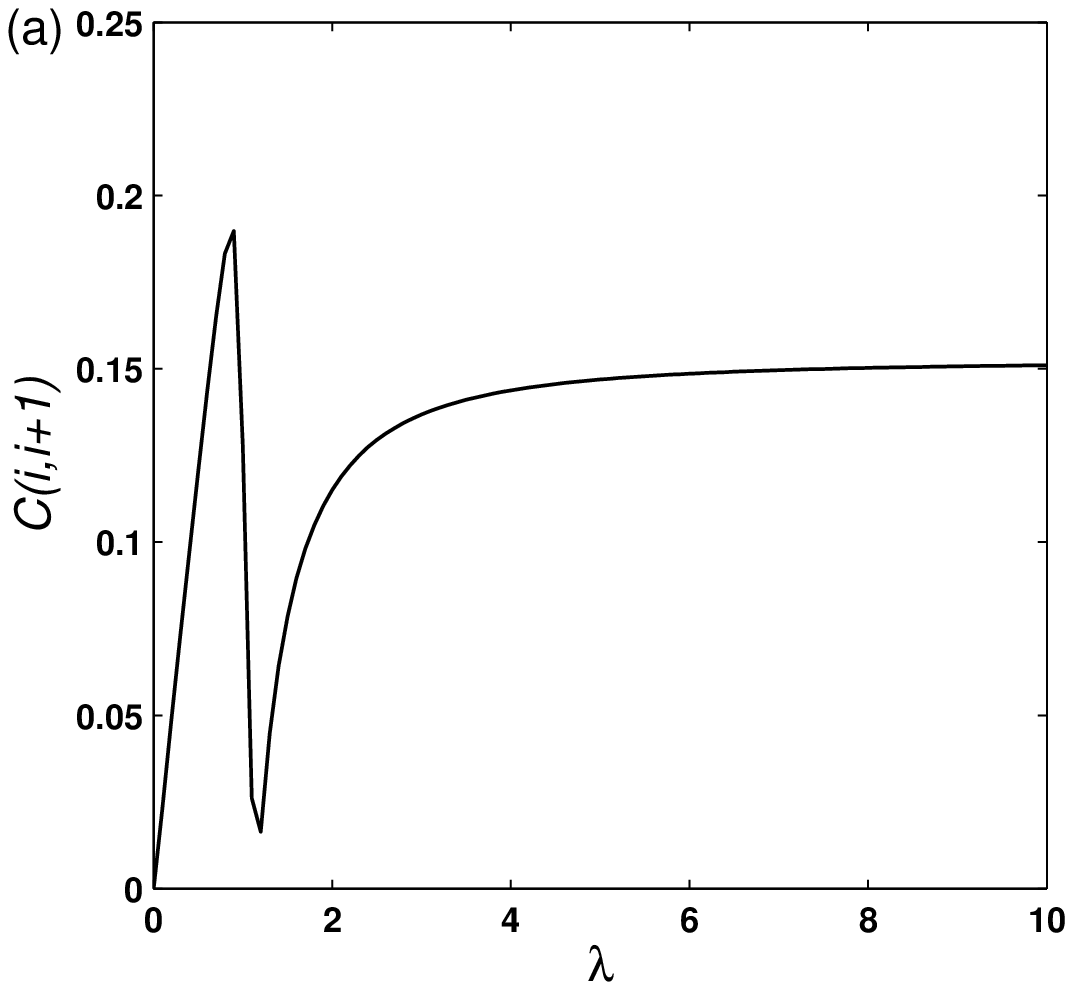}
  	\label{fig:Gamma0}\includegraphics[width=7cm]{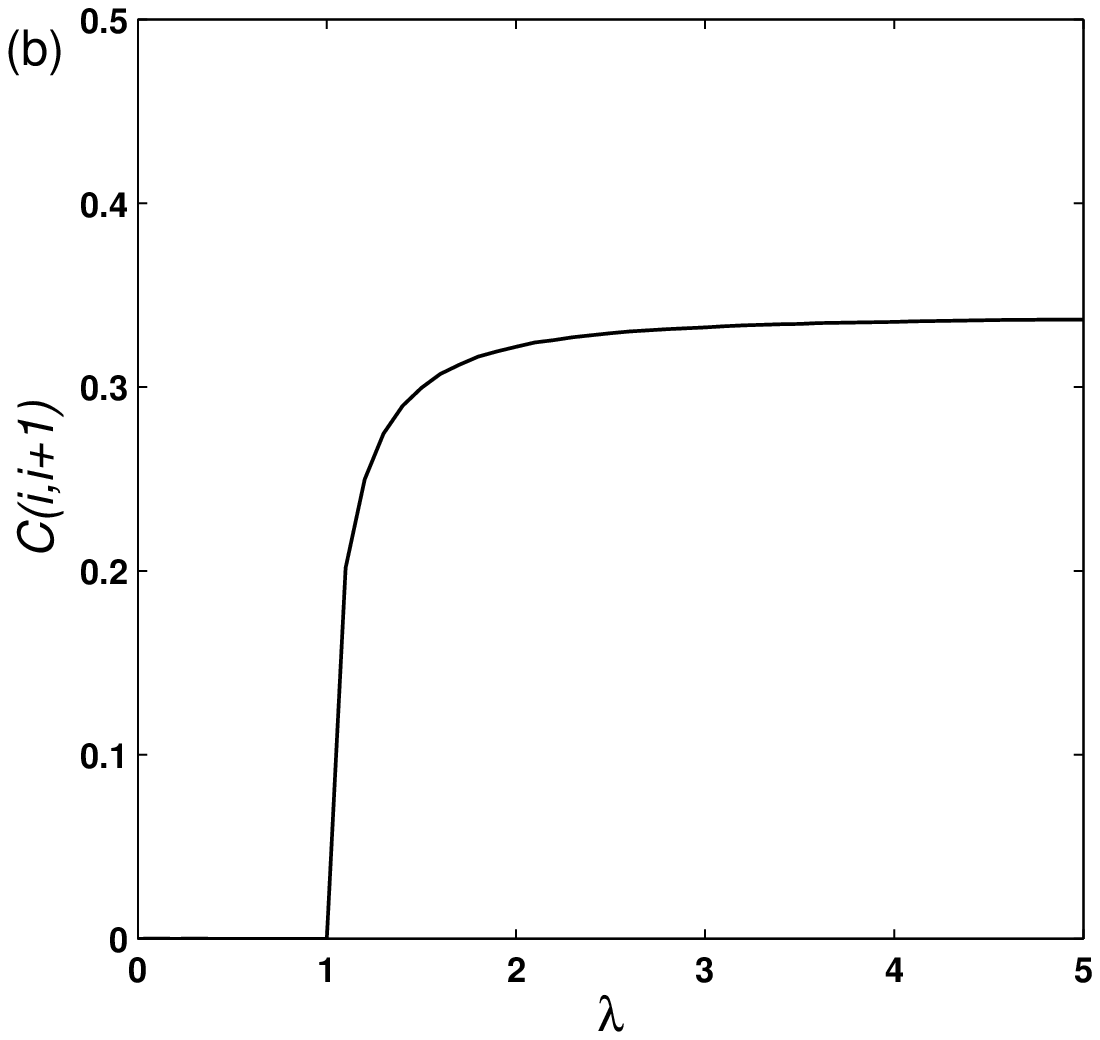}
		\caption{{\protect\footnotesize The behavior asymptotic value of $C(i,i+1)$ as a function of $\lambda$ at $kT=0$ with (a) $\gamma=0.5$ ; (b) $\gamma=0$.}}
  \label{fig:Gamma050}
\end{minipage}
\end{figure}

Here we use the exact solution to study the concurrence for four different forms of coupling parameter $J_{exp}, J_{tanh}, J_{cos}$ and $J_{sin}$ when $J(t)=\lambda h(t)$ where $\lambda$ is a constant. We have compared the exact solution results with the numerical ones and they have shown coincidence.
Figure~\ref{fig:both2} show the dynamics of $C(i,i+1)$ when $h(t)=J(t)=J_{cos}$ and $h(t)=J(t)=J_{sin}$ respectively, where $J_{0}=0.5$ and $K=1$. Interestingly, the concurrence in this case does not show a periodic behavior as it did when $h(t)=1$, i.e. for a constant magnetic field, in Fig.~\ref{fig:Jcos}.
In Fig.~\ref{fig:Lvar}(a) we study the behavior of the asymptotic value of $C(i,i+1)$ as a function of $\lambda$ at different values of the parameters $J_0, J_1$ and $K$ where $J(t)=\lambda h(t)$. Obviously, the asymptotic value of $C(i,i+1)$ depends only on the initial conditions not on the form or behavior of $J(t)$ at $t > 0$. This result demonstrates the sensitivity of the concurrence evolution to its initial value. Testing the concurrence at non-zero temperatures demonstrates that it maintains the same profile but with reduced value with increasing temperature as can be concluded from Fig.~\ref{fig:Lvar}(b). Also the critical value of $\lambda$ at which the concurrence vanishes decreases with increasing temperature as can be observed, which is expected as thermal fluctuations destroy the entanglement.
Finally, in Fig.~\ref{fig:Gamma050} we study the partially anisotropic system, $\gamma=0.5$, and the isotropic system $\gamma=0$ with $J(0)=1$. We note that the behavior of $C(i,i+1)$ in this case is similar to the case of constant coupling parameter studied previously \cite{Sadiek2010}. We also note that the behavior depends only on the initial coupling $J(0)$ and not on the form of $J(t)$ where different forms of $J(t) $have been tested. It is interesting to notice that the results of Figs.~\ref{fig:both2}, \ref{fig:Lvar} and \ref{fig:Gamma050} confirm one of the main results of the previous works \cite{HuangZ2005, HuangZ2006, Sadiek2010} namely that the dynamic behavior of the spin system, including entanglement, depends only on the parameter $\lambda=J/h$ not the individual values of $h$ and $J$ for any degree of anisotropy of the system.
In these previous works both the coupling and magnetic field were considered time-independent, while in this work we have assumed $J(t)=\lambda h(t)$ where $h(t)$ can take any time-dependent form. This explains why the asymptotic value of the concurrence depends only on the initial value of the parameters regardless of their function form. Furthermore in the previous works, it was demonstrated that for finite temperatures the concurrence turns to be dependent not only on $\lambda$ but on the individual values of $h$ and $J$, while according to Fig.~\ref{fig:Lvar}(b) even at finite temperatures the concurrence still depends only on $\lambda$ where $J(t)=\lambda J(t)$ for any form of $h(t)$.

\section{Dynamics of entanglement in two-dimensional spin systems}
\subsection{An exact treatment of two-dimensional transverse Ising model in a triangular lattice}
Entanglement close to quantum phase transitions was originally analyzed by Osborne
and Nielsen \cite{Osborne2002}, and Osterloh et al. \cite{Osterloh2002}
for the Ising model in one dimension.  We studied before a set of localized
spins coupled through exchange interaction and subject to an external
magnetic filed \cite{Osenda2003,HuangZ2004,HuangZ2005,HuangZ2006}. We demonstrated
for such a class of one-dimensional magnetic systems, that
entanglement can be controlled and tuned by varying the
anisotropy parameter in the Hamiltonian and by
introducing impurities into the systems.
In particular, under certain conditions, the entanglement is
zero up to a critical point $\lambda_c$, where a quantum phase transition occurs,
and is different from zero above $\lambda_c$ \cite{Kais2007}.

In two and higher dimensions nearly all calculations for spin systems  were
obtained by means of numerical simulations \cite{Sandvik1991}. The concurrence
and localizable entanglement in two-dimensional quantum XY and XXZ models were
considered using quantum Monte Carlo \cite{Syljuasen2003,Syljuasen2004}. The results of these calculations
were qualitatively similar to the one-dimensional case, but entanglement is much smaller
in magnitude. Moreover, the maximum in the concurrence occurs at a position closer to the
critical point than in the one-dimensional case \cite{Amico2008}.

In this section, we introduce how to use the Trace Minimization Algorithm \cite{Sameh1982,Sameh2000}
to  carry out an exact calculation of entanglement
in a 19-site two dimensional transverse Ising model.
We classify the ground state properties according to its
entanglement for certain class on two-dimensional magnetic systems
and demonstrate  that entanglement can be controlled and tuned by varying the
parameter $\lambda=h/J$ in the Hamiltonian and by introducing impurities into the
systems.  We discuss the relationship of entanglement and quantum phase
transition, and the effects of impurities on the
entanglement.

\subsubsection{Trace minimization algorithm}
Diagonalizing a $2^{19}$
by $2^{19}$ Hamiltonian matrix and partially tracing its density matrix
is a numerically difficult task. We propose to compute the  entanglement of formation, first by applying the trace
minimization algorithm (Tracemin) \cite{Sameh1982,Sameh2000} to obtain the eigenvalues and eigenvectors of the constructed Hamiltonian. Then, we use these eigenpairs and new techniques detailed in the appendix to build partially traced density matrix.

The trace minimization algorithm was developed for computing a few of the smallest
eigenvalues and the corresponding eigenvectors of the large sparse generalized eigenvalue problem
\begin{equation}
AU=BU\Sigma,
\end{equation}
where matrices $A,B \in \mathbb{C}^{n\times n}$ are Hermitian
positive definite, $U=[u_1, ..., u_p] \in
\mathbb{C}^{n\times p}$ and $\Sigma \in \mathbb{R}^{p\times p}$ is a
diagonal matrix. The main idea of Tracemin is that minimizing $Tr(X^* AX)$, subject
to the constraints $X^* BX=I$, is equivalent to finding the
eigenvectors $U$ corresponding to the p smallest eigenvalues. This
consequence of Courant-Fischer Theorem can be restated as
\begin{equation}
\min_{X^* BX=I} Tr(X^* AX)=Tr(U^* AU)=\sum_{i=1}^p\lambda_i,
\end{equation}
where $I$ is the identity matrix. The following steps constitute a
single iteration of the Tracemin algorithm:\\
\begin{tabular}{l l}
$\bullet\quad G=X_{k}^* BX_{k}$ & (compute $G$)\\
$\bullet\quad G=V\Omega V^*$ & (compute the spectral decomposition of $G$)\\
$\bullet\quad H=\tilde{Q}^* A \tilde{Q}$ & (compute $H$, where $\tilde{Q}=X_k V\Omega^{-1/2}$)\\
$\bullet\quad H=W\Omega W^*$ &  (compute the spectral decomposition of $H$)\\
$\bullet\quad \bar{X}_k = \tilde{Q}W$ & (now $\bar{X}_k^*
A\bar{X}_k=\Lambda$ and $\bar{X}_k^* B\bar{X}_k = I$)\\
$\bullet\quad \bar{X}_{k+1}=\bar{X}_k -D$ & ( $D$ is determined s.t. $Tr(X_{k+1}^* AX_{k+1})<Tr(X_k^* AX_k)$ ).\\
\end{tabular}\\
In order to find the optimal update $D$ in the last step, we enforce the
natural constraint $\bar{X}_k^* BD=0$, and obtain
\begin{equation}
\label{tracemin_system_for_D}
\begin{pmatrix}
A & B\bar{X}_k \\
X_k^* B & 0
\end{pmatrix}
\begin{pmatrix}
D\\
L
\end{pmatrix}
=
\begin{pmatrix}
A\bar{X}_k\\
0
\end{pmatrix}
\end{equation}.

Considering the orthogonal projector $P=B\bar{X}_k (X_k^* B^2
X_k)^{-1} \bar{X}_k^*B$ and letting $D=(I-P)\bar{D}$, the linear system \eqref{tracemin_system_for_D} can be rewritten in the following form
\begin{equation}
\label{tracemin_projected_system}
(I-P)A(I-P)\bar{D}=(I-P)A\bar{X}_k.
\end{equation}
Notice that the Conjugate Gradient method can be
used to solve \eqref{tracemin_projected_system}, since it can be shown that the residual and
search directions $\text{r},\text{p} \in \text{Range}(P)^{\perp}$. Also, notice that the linear system \eqref{tracemin_projected_system} need to be solved only to a fixed relative precision at every iteration of Tracemin.

A reduced density matrix, built from the ground state which is obtained by Tracemin,
is usually constructed as follows:
diagonalize the system Hamiltonian $H(\lambda)$, retrieve
the ground state $|\Psi>$ as a function of $\lambda=h/J$, build the
density matrix $\rho=|\Psi><\Psi|$, and trace out contributions of all the other
spins in density matrix to get reduced density matrix
by $\rho(i,j)=\sum_{p}<u_{i}(A)|<v_{p}(B)|\rho|u_{j}(A)>|v_{p}(B)>$,
where ${u_{i}(A)}$ and ${v_{p}(B)}$ are bases of subspaces
$\epsilon(A)$ and $\epsilon(B)$. That includes creating a
$2^{19} \times 2^{19}$ density
matrix $\rho$
followed by permutations of rows, columns and some basic arithmetic
operations on the elements of $\rho$.
Instead of operating on a huge matrix, we pick up only certain
elements from $|\Psi>$, performing basic algebra to build a reduced density matrix
directly.

\subsubsection{General forms of matrix representation of the Hamiltonian }

By studying the  patterns of $\sum_{<i,j>}I\otimes\cdots
\sigma_{i}^{x}\otimes\cdots \sigma_{j}^{x}\otimes\cdots I$ and
$\sum_{i}I\otimes\cdots \sigma_{i}^z\otimes\cdots I$, one founds the
following rules.

\paragraph{$\sum_{i} \sigma_{i}^z$ for N spins}

\begin{figure}
\centering
\includegraphics[scale=0.5]{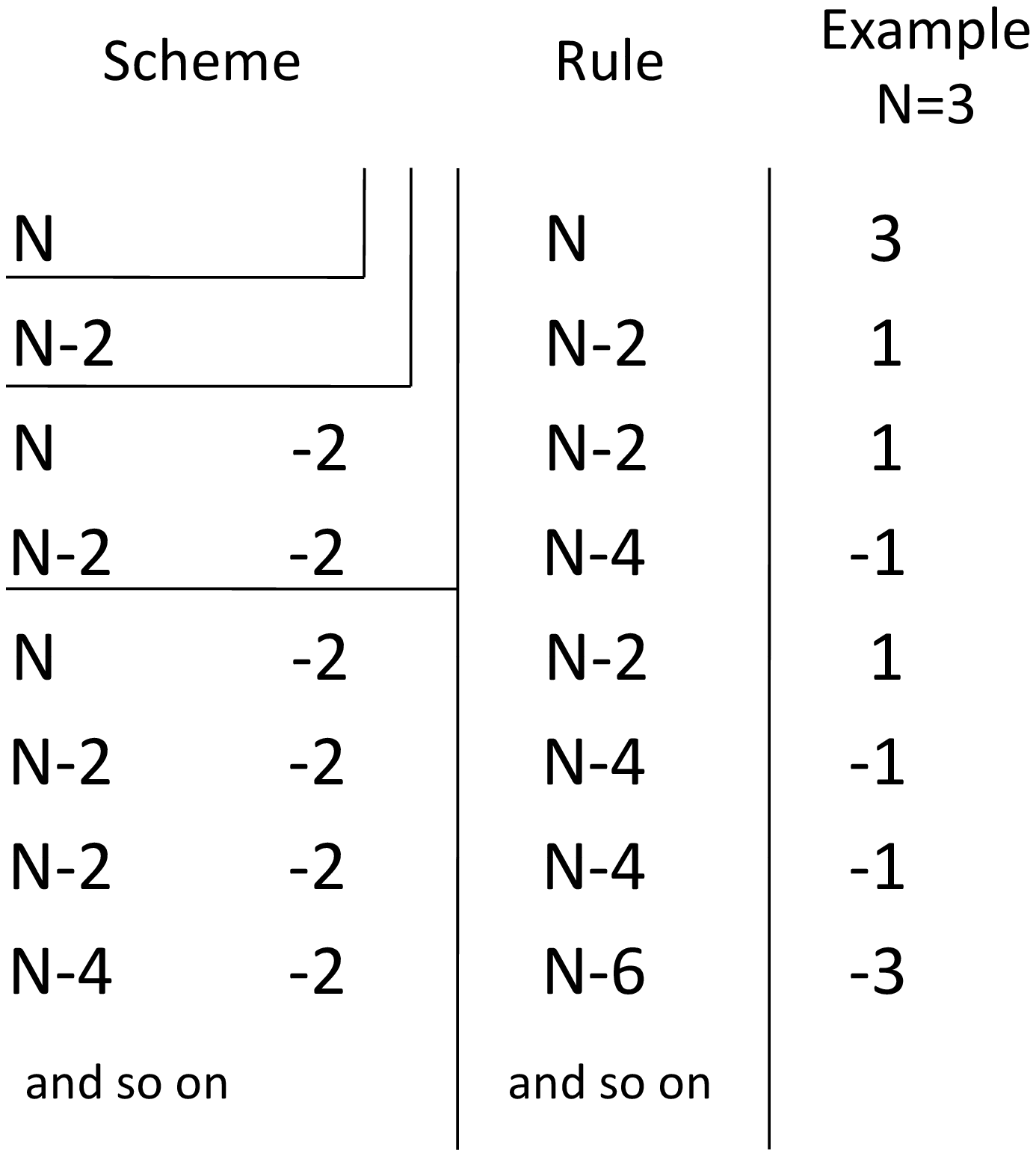}
\caption{Diagonal elements of $\sum_{i} \sigma_{i}^z$ for N spins.}
\label{fig10}
\end{figure}

The matrix is $2^N$ by $2^N$; it has only $2^N$ diagonal elements.
Elements follow the rules shown in Fig. \ref{fig10}.

If one stores these numbers in a vector, and  initializes $v=(N)$,
then the new v is the concatenation of the original v and the original v with 2
subtracted from each of its elements. We repeat this N times, i.e.,
\begin{equation}
v=\left(
\begin{array}{c}
v \\
v-2 \\
\end{array}
\right);
\end{equation}
\begin{eqnarray}
&v&=\left(
\begin{array}{c}
N
\end{array}
\right),\\
\Rightarrow &v&=\left(
\begin{array}{c}
N \\
N-2 \\
\end{array}
\right),\\
\Rightarrow &v&=\left(
\begin{array}{c}
N \\
N-2 \\
N-2 \\
N-4 \\
\end{array}
\right).
\end{eqnarray}

\paragraph{$\sum_{<i,j>}I\otimes\cdots \sigma_{i}^{x}\otimes\cdots
\sigma_{j}^{x}\otimes\cdots I$ for N spins}

\begin{figure}
\centering{}
\includegraphics[scale=0.4]{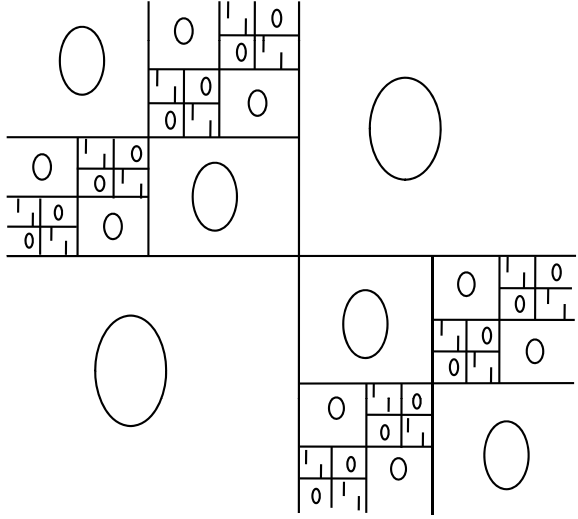}
\caption{Scheme of matrix $I\otimes \sigma_{3}^{x}\otimes
\sigma_{2}^{x}\otimes I\otimes I$}
\label{fig11}
\end{figure}

Since $\left( \begin{array}{ccc}
1 & 0\\
0 & 1\\
\end{array} \right)$ \& $\left( \begin{array}{ccc}
0 & 1\\
1 & 0\\
\end{array} \right)$
exclude each other, for matrix $I\otimes\cdots
\sigma_{i}^{x}\otimes\cdots \sigma_{j}^{x}\otimes\cdots I$, every
row/column contains only one ``1'', then the matrix owns $2^N$
``1''s and only ``1'' in it. If we know the position of ``1''s,
it turns out that we can set a $2^N$ by 1
array ``col'' to store the column position of ``1''s corresponding
to the 1st $\rightarrow$ $2^N$th rows. In fact, the non-zero elements
can be located by the properties stated below. For clarity, let us number N spins in the reverse order
as: N-1, N-2, \dots, 0, instead of 1, 2, \dots, N. The string of non-zero elements starts from the
first row at: $1+2^i+2^j$; with string length $2^j$; and number of such
strings $2^{N-j-1}$. For example, Fig. \ref{fig11} shows these rules  for a scheme of $I\otimes
\sigma_{3}^{x} \otimes \sigma_{2}^{x} \otimes I \otimes I$.

Again, because of the exclusion, the positions of non-zero element
``1'' of $I\otimes\cdots \sigma_{i}^{x}\otimes\cdots
\sigma_{j}^{x}\otimes\cdots I$ are different from those of
$I\otimes\cdots \sigma_{l}^{x}\otimes\cdots
\sigma_{m}^{x}\otimes\cdots I$. So $\sum_{<i,j>}I\otimes\cdots
\sigma_{i}^{x}\otimes\cdots \sigma_{j}^{x}\otimes\cdots I$ is a
$2^N$ by $2^N$ matrix with only 1 and 0.

After storing array ``col'', we repeat the algorithm
for all the nearest pairs $<i,j>$, and concatenate ``col''s to
position matrix ``c'' of $\sum_{<i,j>}I\otimes\cdots
\sigma_{i}^{x}\otimes\cdots \sigma_{j}^{x}\otimes\cdots I$. In the next
section we apply these rules to our problem.

\subsubsection{Specialized matrix multiplication}
Using the diagonal elements array ``v'' of $\sum_{i} \sigma_{i}^z$ and
position matrix of non-zero elements ``c'' for
$\sum_{<i,j>}I\otimes\cdots \sigma_{i}^{x}\otimes\cdots
\sigma_{j}^{x}\otimes\cdots I$, we can generate matrix H, representing the Hamiltonian. However, we
only need to compute the result of the matrix-vector multiplication H*Y in order to run Tracemin, which is the advantage of Tracemin, and consequently do not need to explicitly obtain H. Since matrix-vector multiplication is repeated many times throughout iterations, we propose an efficient implementation to speedup its computation
specifically for Hamiltonian of Ising model (and XY by
adding one term).

For simplicity, first let Y in H*Y be a vector and $J=h=1$ (in general
Y is a tall matrix and $J\neq h\neq1$). Then
\begin{eqnarray}
&&H*Y \nonumber\\
&=&\sum_{<i,j>}\sigma_i^x \sigma_j^x *Y+\sum_i
\sigma_i^z
*Y\nonumber\\
   &=&\left(
      \begin{array}{cccccc}
        1 &   & 1 &   & 1 &  \\
          & 1 &   & 1 &   & 1\\
          & \ldots&   & \ldots &   &  \\
          &   &   &   &   &  \\
          & \ldots&   &\ldots&   &  \\
        1 &   &   &   & 1 & 1\\
      \end{array}
    \right)*
    \left(
      \begin{array}{c}
        Y(1) \\
        Y(2) \\
        \vdots \\
        \vdots \\
        Y(2^N) \\
      \end{array}
    \right)+
    \left(
      \begin{array}{ccccc}
        v(1)  &   &   &   &   \\
          & v(2)  &   &   &   \\
          &   & \ddots&   &   \\
          &   &   & \ddots&   \\
          &   &   &   & v(2^N)\\
      \end{array}
    \right)*
        \left(
      \begin{array}{c}
        Y(1) \\
        Y(2) \\
        \vdots \\
        \vdots \\
        Y(2^N) \\
      \end{array}
    \right)\nonumber\\
   &=&\left(
      \begin{array}{c}
        Y(c(1,1))+Y(c(1,2))+\ldots+Y(c(1,\#ofpairs) \\
        \vdots \\
        Y(c(k,1))+Y(c(k,2))+\ldots+Y(c(k,\#ofpairs) \\
        \vdots \\
        Y(c(2^N,1))+Y(c(2^N,2))+\ldots+Y(c(2^N,\#ofpairs)\\
      \end{array}
    \right)+
    \left(
      \begin{array}{c}
        v(1)*Y(1) \\
        v(2)*Y(2) \\
        \vdots \\
        \vdots \\
        v(2^N)*Y(2^N)
      \end{array}
    \right),
\end{eqnarray}
where p\# stands for the number of pairs.

\begin{figure}
\centering{}
\includegraphics[scale=0.6]{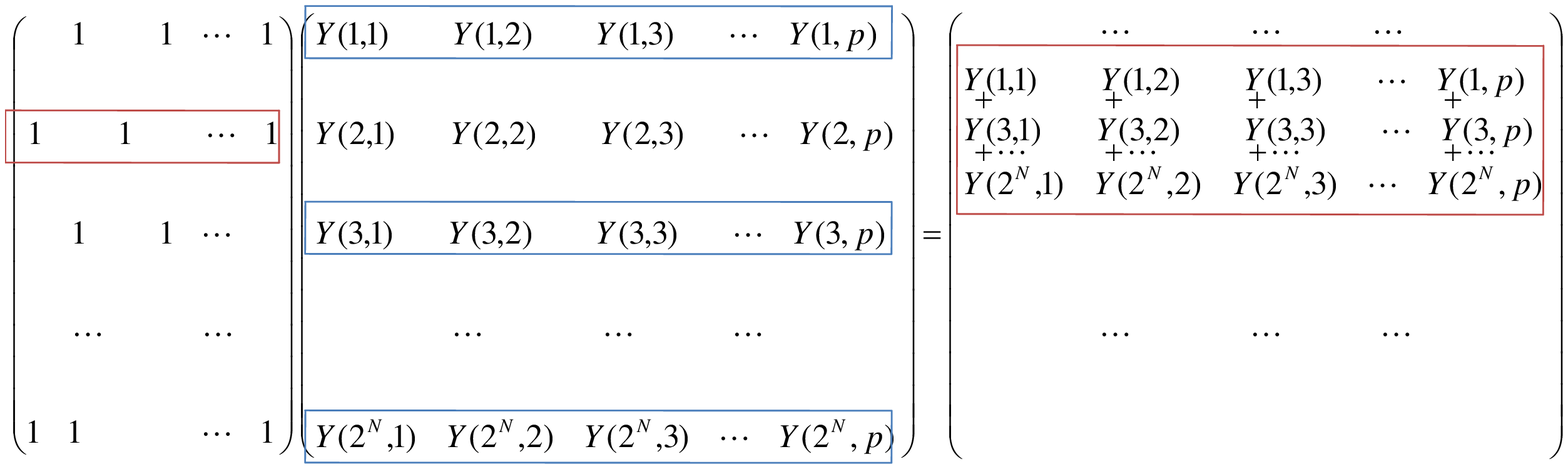}
\caption{The illustration of H*Y.}
\label{fig12}
\end{figure}

When Y is a matrix, we can treat Y ($2^N$ by p) column by column
for $\sum_{<i,j>}I\otimes\cdots \sigma_{i}^{x}\otimes\cdots
\sigma_{j}^{x}\otimes\cdots I$. Also, we can accelerate the
computation by treating every row of Y as a vector and adding these vectors at once. Fig. \ref{fig12} visualized the process.

Notice that the result of the multiplication of the xth row of $\sum_{<i,j>}\sigma_i^x \sigma_j^x$ (delineated by the red
box above) and Y, is equivalent
to the sum of rows of Y, whose row numbers are the column indecis of non-zero
elements' of the xth row. Such that we transform a matrix operation to straight forward summation \& multiplication of numbers.
\subsubsection{Exact entanglement}
\begin{figure}
\centering{}
\includegraphics[scale=0.7]{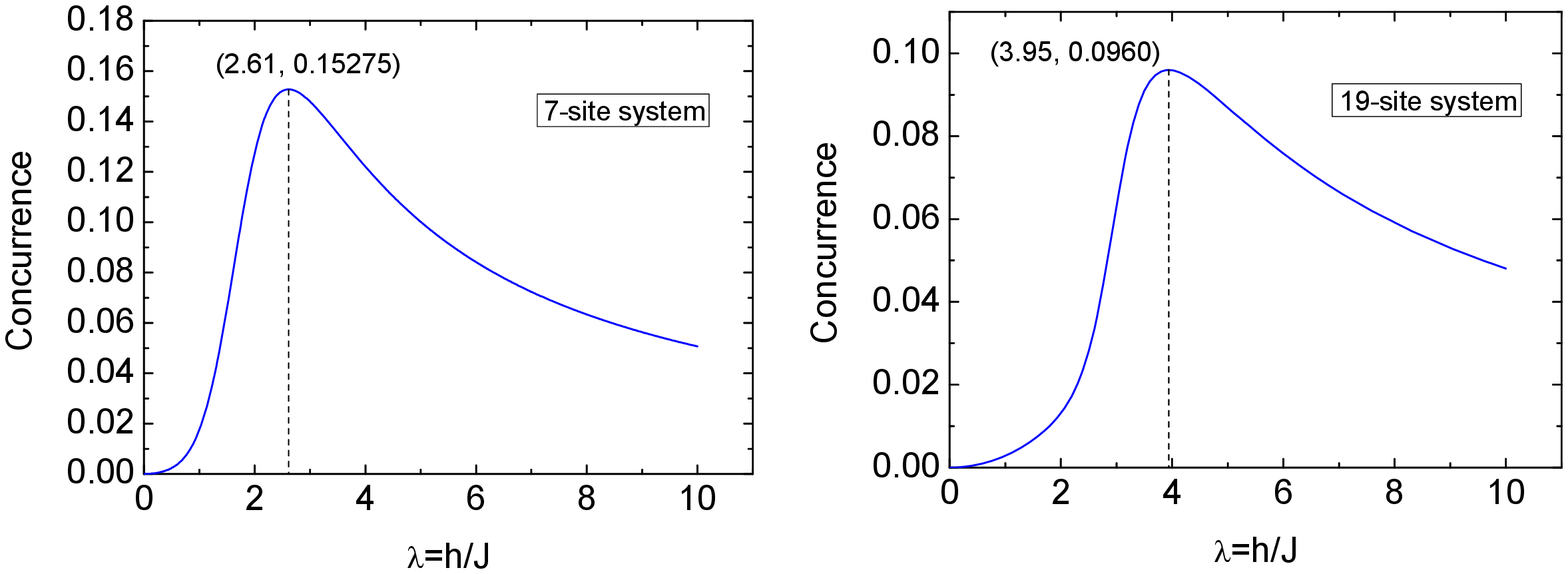}
\caption{Concurrence  of center spin and its nearest neighbor as a
function of $\lambda$ for both 7-site and 19-site system.
In the 7-site system, concurrence reaches maximum  0.15275 when $\lambda=2.61$.
In the 19-site system, concurrence reaches the maximum  0.0960 when $\lambda=3.95$. }
\label{fig1}
\end{figure}
We examine the change of concurrence between the center spin and its nearest neighbor as a function of $\lambda=h/J$ for both the 7-site and 19-site systems.
In Fig.\ref{fig1}, the concurrence of the 7-site system reaches its maximum  0.15275
when $\lambda=2.61$. In the 19-site system, the concurrence reaches
0.0960 when $\lambda=3.95$ \cite{XuQ2010}. The maximum value of concurrence in the 19-site model, where each site interacts with six neighbors, is roughly 1/3 of the maximum concurrence in the one-dimensional transverse Ising model with size N=201 \cite{Osenda2003}, where it has only two neighbors for each site. It is the monogamy \cite{Coffman2000,Osborne2006} that limits the entanglement shared among the number of neighboring sites. This property also shown in the fact that fewer the number of neighbors of a pair the larger the entanglement among other nearest neighbors. Our numerical calculation shows that the maximum concurrence of next-nearest neighbor is less than $10^{-8}$. It shows that the entanglement is short ranged, though global.

\subsection{Time evolution of the spin system}

Decoherence is considered as one of
the main obstacles toward realizing an effective quantum
computing system \cite{Zurek1991}. The main effect of decoherence is to
randomize the relative phases of the possible states of the considered system.
Quantum error correction \cite{Shor1995} and decoherence free subspace \cite{Bacon2000,Divincenzo2000} have been proposed to protect the quantum property during the computation process. Still offering a potentially ideal protection against environmentally induced decoherence is difficult. In NMR quantum computers, a series of magnetic pulses were applied to a selected nucleus of a molecule to implement quantum gates \cite{Doronin2002}. Moreover, a spin-pair entanglement is a reasonable measure for decoherence between the considered two-spin system and the environmental spins. The coupling between the system and its environment leads to decoherence in the system and vanishing of entanglement between the two spins. Evaluating the entanglement remaining in the considered system helps us to understand the behavior of the decoherence between the considered two spins and their environment \cite{Lages2005}.

The study of quantum entanglement in two-dimensional systems possesses a number of extra problems compared with systems of one dimension. The particular one is the lack of exact solutions. The existence of exact solutions has contributed enormously to the understanding of the entanglement for 1D systems \cite{Lieb1961,Sachdev2001,HuangZ2005,Sadiek2010}. Studies can be carried out on interesting but complicated properties, applied to infinitely large system, and so forth use finite scaling method to eliminate the size effects, etc. Some approximation methods, like Density matrix renormalization group (DMRG), are also only workable in one dimension \cite{Doronin2002,Xavier2010,Silva-Valencia2005,Capraro2002}. So when we carry out this two-dimensional study, no methods can be inherited from previous researches.  They heavily rely on numerical calculations, resulting in severe limitations on the system size and properties. For example, dynamics of the system is a computational-costing property. We have to think of a way to improve the effectiveness of computation in order to increase the size of research objects; mean while dig the physics in the observable systems. It may show the general physics, or tell us the direction of less resource-costing large scale calculations.

To tackle down the problem, we introduce two calculation methods: step by step time-evolution matrix transformation and step by step projection. We compare them side by side, in short, besides the exactly same results, step by step projection method turned out to be twenty times faster than the matrix transformation.

\subsubsection{The evolution operator}

According to quantum mechanics the transformation of $|\psi_i(t_{0})\rangle$, the state vector at the initial instant $t_0$, into $|\psi_i(t)\rangle$, the state vector at an arbitrary instant, is linear \cite{Cohen-Tannoudji2005}. Therefore there exists a linear operator $U(t,\, t_{0})$ such that:
\be
\label{U_vector}
|\psi_i(t)\rangle=U(t,\, t_0)\,|\psi_i(t_0)\rangle.
\ee
This is, by definition, the evolution operator of the system. Substituting Eq. (\ref{U_vector}) into the Schor\"{o}dinger equation, we obtain:
\be
i\hbar\frac{\partial}{\partial t}U(t,t_0)|\psi(t_0)\rangle=H(t)U(t,t_0)|\psi(t_0)\rangle,
\ee
which means
\be
i\hbar\frac{\partial}{\partial t}U(t,t_0)=H(t)U(t,t_0).
\ee
Further taking the initial condition
\be\label{U_initial}
U(t_0,t_0)=\mathbb{I},
\ee
the evolution operator can be condensed into a single integral equation:
\be\label{U_integral}
U(t,t_0)=\mathbb{I}-\frac{i}{\hbar}\int_{t_0}^t H(t')U(t',t_0)dt'.
\ee

When the operator $H$ does not depend on time, Eq. (\ref{U_integral}) can easily be integrated and at finally gives out
\be\label{U_time_independent}
U(t,t_0)=e^{-iH(t-t_0)/\hbar}.
\ee

\subsubsection{Step by step time-evolution matrix transformation}

To unveil the behavior of concurrence at time $t$, we need to find the density matrix of the system at that moment, which can be obtained from
\be
\rho(t)=U(t)\rho(0)U^{\dagger}(t).
\ee
\begin{figure}[htbp]
 \centering
   \includegraphics[width=7 cm]{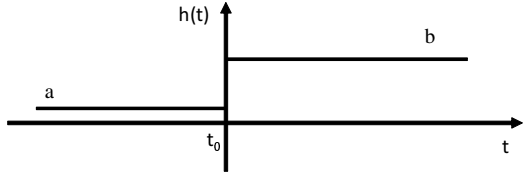}
   \caption{{\protect\footnotesize The external magnetic field in a step function form $h(t)=a\;@\;t\leq t_0$, $h(t)=b\;@\;t>t_0$.}}
 \label{Magnetic_fields_Step}
\end{figure}
Although Eq. (\ref{U_integral}) gives a beautiful expression for the evolution operator, in reality $U$ is hard to be obtained because of the integration involved. In order to overcome this obstacle, let us first consider the simplest time-dependent magnetic field: a step function of the form (Fig.~\ref{Magnetic_fields_Step})
\be
h(t)= a+(b-a)\theta(t-t_0) \; ,
\ee
where $\theta(t-t_0)$ is the usual mathematical step function defined by
\be
\theta(t-t_0)=\left\{
\begin{array}{lr}
0 & \qquad t\leq t_0 \\
1 & \qquad t>t_0
\end{array}.
\right.
\ee
At $t_0$ and before, the system is time-independent since $H_a \equiv H(t\leq t_0)=-\sum_{<i,j>}\sigma_{i}^{x}\sigma_{j}^{x}-a\Sigma_{i}\sigma_{i}^{z}$. Therefore we are capable of evaluating its ground state and density matrix at $t_0$ straightforwardly. For the interval $t_0$ to $t$, the Hamiltonian $H_b \equiv H(t>t_0)=-\sum_{<i,j>}\sigma_{i}^{x}\sigma_{j}^{x}-b\Sigma_{i}\sigma_{i}^{z}$ does not depend on time either, so Eq. (\ref{U_time_independent}) enables us to write out
\be
U(t,t_0)=e^{-iH(t>t_0)(t-t_0)/\hbar},
\ee
and therefore
\be
\rho(t)=U(t,t_0)\rho(t_0)U^{\dagger}(t,t_0).
\ee
Starting from here, it is not hard to think of breaking an arbitrary magnetic function into small time intervals, and treating every neighboring intervals as a step function. Comparing the two graphs in Fig.~\ref{Step_by_step}, the method has just turned a ski sliding into a mountain climbing.
\begin{figure}[htbp]
\begin{minipage}[c]{\textwidth}
 \centering
   \subfigure[]{\label{Step_by_step1}\includegraphics[width=7 cm]{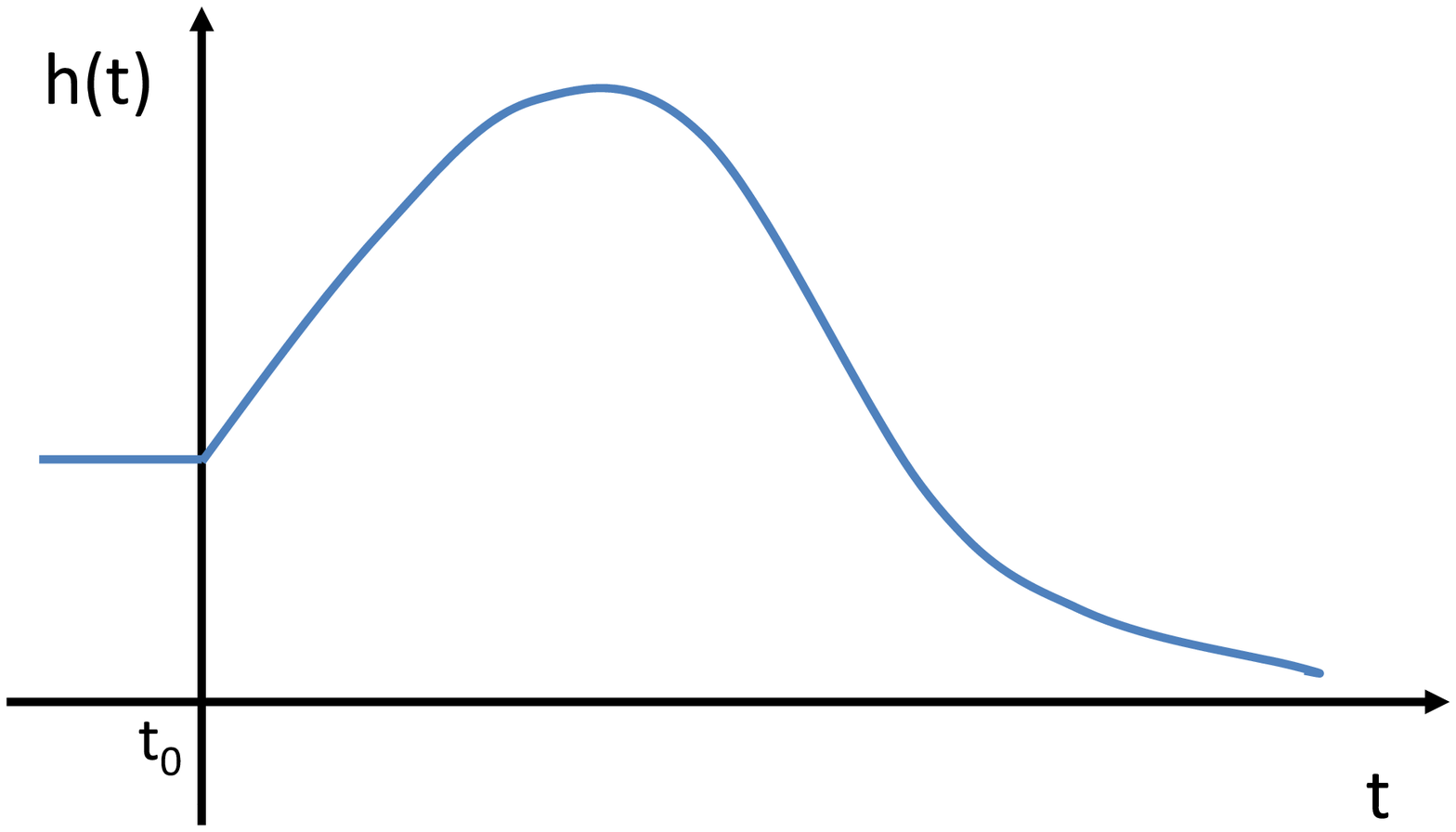}}\quad
   \subfigure[]{\label{Step_by_step2}\includegraphics[width=7 cm]{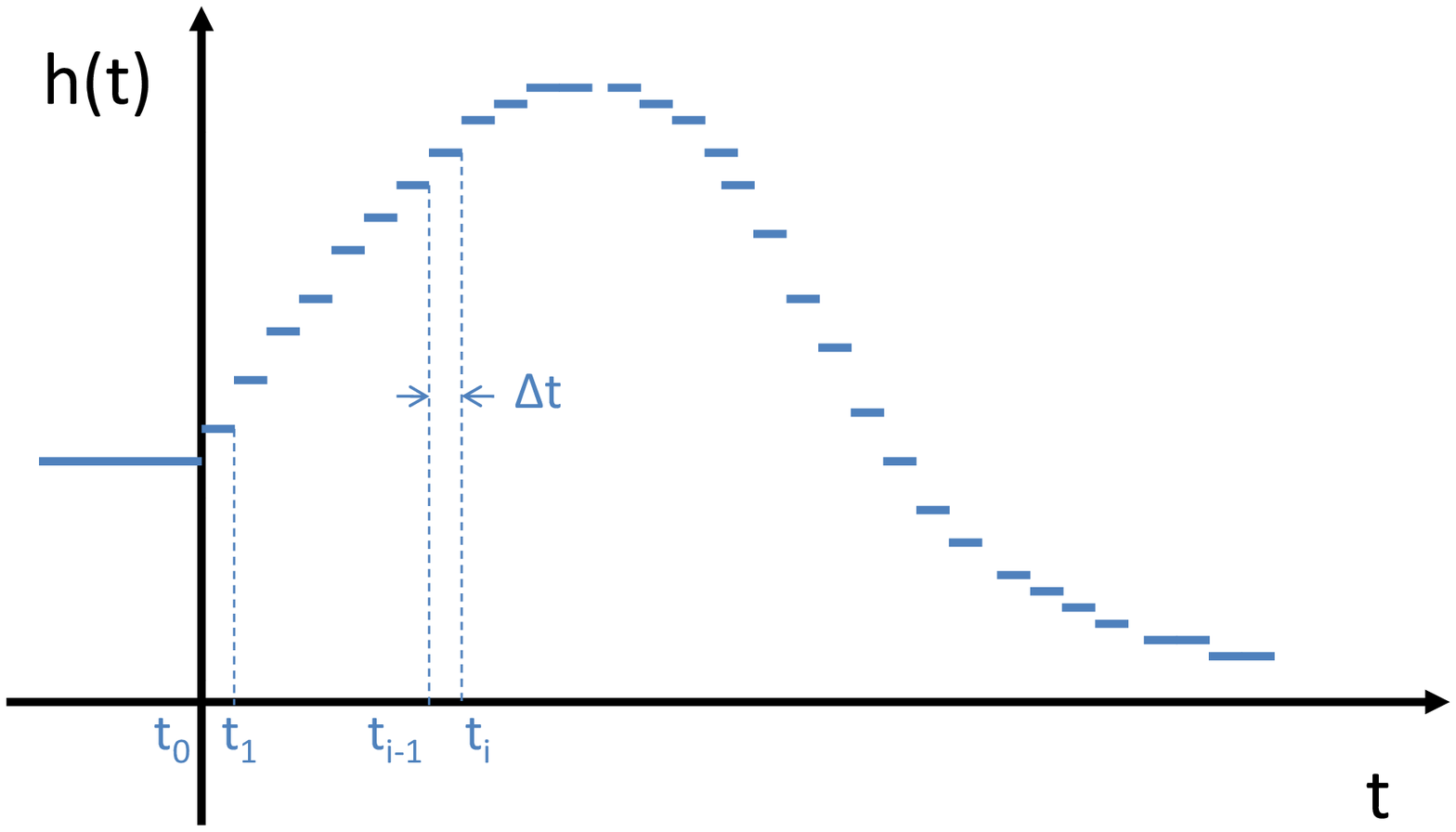}}
   \caption{{\protect\footnotesize Divide an arbitrary magnetic field function into small time intervals. Every time step is $\Delta t$. Treat the field within the interval as a constant. In all, we turn a smooth function into a collection of step functions, which makes the calculation of dynamics possible.}}
 \label{Step_by_step}
 \end{minipage}
\end{figure}
Assuming each time interval is $\Delta t$, setting $\hbar=1$ then
\bea
U(t_i,t_0)|\psi_0\rangle&=&U(t_{i},t_{i-1})U(t_{i-1},t_{i-2})...U(t_1,t_0)|\psi_0\rangle,\\
U(t_i,t_0)&=&\prod_{k=1}^{i}exp[-iH(t_k)\Delta t],\\
U(t_i,t_0)&=&exp[-iH(t_i)\Delta t]U(t_{i-1}-t_0).
\eea
Here we avoided integration, instead we have chain multiplications which can be easily realized as loops in computational calculations. This is a common numerical technique; desired precisions can be achieved via proper time step length adjustment.

\subsubsection{Step by step projection}

Step by step matrix transformation method successfully breaks down the integration, but still involves matrix exponential, which is numerically resource costing. We propose a projection method to accelerate the calculations. Let us look at the step magnetic field again Fig. \ref{Magnetic_fields_Step}. For $H_a$, after enough long time, the system at zero temperature is in the ground state $|\phi\rangle$ with energy, say, $\varepsilon$. We want to ask how will this state evolves after the magnetic field is turned to the value b? Assuming the new Hamiltonian $H_b$ has $N$ eigenpairs $E_i$ and $|\psi_{i}\rangle$. The original state $|\phi\rangle$ can be expanded in the basis $\{|\psi_{i}\rangle\}$:
\be
|\phi\rangle=c_{1}|\psi_{1}\rangle+c_{2}|\psi_{2}\rangle+...+c_{N}|\psi_{N}\rangle,
\ee
where
\be
c_i=\langle\psi_i|\phi\rangle.
\ee
When $H$ is independent of time between $t$ and $t_0$ then we can write
\be
U(t,\, t_{0})\,|\psi_{i,t_0}\rangle=e^{-iH(t>t_0)(t-t_{0})/\hbar}|\psi_{i,t_0}\rangle=e^{-iE_{i}(t-t_{0})/\hbar}|\psi_{i,t_0}\rangle,
\ee
Now the exponent in the evolution operator is a number no longer a matrix. The ground state will evolve with time as
\bea\label{projection_sum}
|\phi(t)\rangle &=& c_{1}|\psi_{1}\rangle e^{-iE_{1}(t-t_0)}+c_{2}|\psi_{2}\rangle e^{-iE_{2}(t-t_0)}+...+c_{N}|\psi_{N}\rangle e^{-iE_{N}(t-t_0)}\nonumber\\
&=& \sum_{i=1}^{N}c_{i}|\psi_{i}\rangle e^{-iE_{i}(t-t_0)}.
\eea
and the pure state density matrix becomes
\be
\rho(t)=|\phi(t)\rangle\langle\phi(t)|.
\ee

Again any complicated function can be treated as a collection of step functions. When the state evolves to the next step just repeat the procedure to get the following results. Our test shows, for the same magnetic field both methods give the same results, but projection is much (about 20 times faster) than matrix transformation. This is a great advantage when the system size increases. But this is not the end of the problem. The summation is over all the eigenstates. Extending one layer out to $19$ sites, fully diagonalizing the $2^{19}$ by $2^{19}$ Hamiltonian and summing over all of them in every time step is breath taking.

\subsubsection{Dynamics of the spin system in a time-dependent magnetic field}

We consider the dynamics of entanglement in a two-dimensional spin system, where spins are coupled through an exchange interaction and subject to an external time-dependent magnetic field. Four forms of time-dependent magnetic field are considered: step, exponential, hyperbolic and periodic.
\be
h(t)= a+(b-a)\theta(t-t_0) \; ,
\ee
\be\label{eq_magnetic fields_exp}
h(t)=\left\{
\begin{array}{lr}
a & \qquad t\leq t_0 \\
b+(a-b)e^{-\omega t} & \qquad t>t_0
\end{array}
\right.
\ee
\be\label{eq_magnetic fields_hyper}
h(t)=\left\{
\begin{array}{lr}
a & \qquad t\leq t_0 \\
\frac{(b-a)}{2}[\tanh(\omega t)+1]+a & \qquad t>t_0
\end{array}
\right.
\ee
\be
h(t)=\left\{
\begin{array}{lr}
a & \qquad t\leq t_0 \\
a-a\sin(\omega t+\phi) & \qquad t>t_0
\end{array}
\right.
\ee

We show in the following figures that the system entanglement behaves in an ergodic way in contrary to the one-dimensional Ising system. The system shows great controllability under all forms of external magnetic field except the step function one which creates rapidly oscillating entanglement. This controllability is shown to be breakable as the different magnetic field parameters increase. Also it will be shown that the mixing of even a few excited states by small thermal fluctuation is devastating to the entanglement of the ground state of the system. These can be explained by the Fermi's golden rule and adiabatic approximation.

\begin{figure}[htbp]
\begin{minipage}[c]{\textwidth}
 \centering
   \includegraphics[width=15 cm]{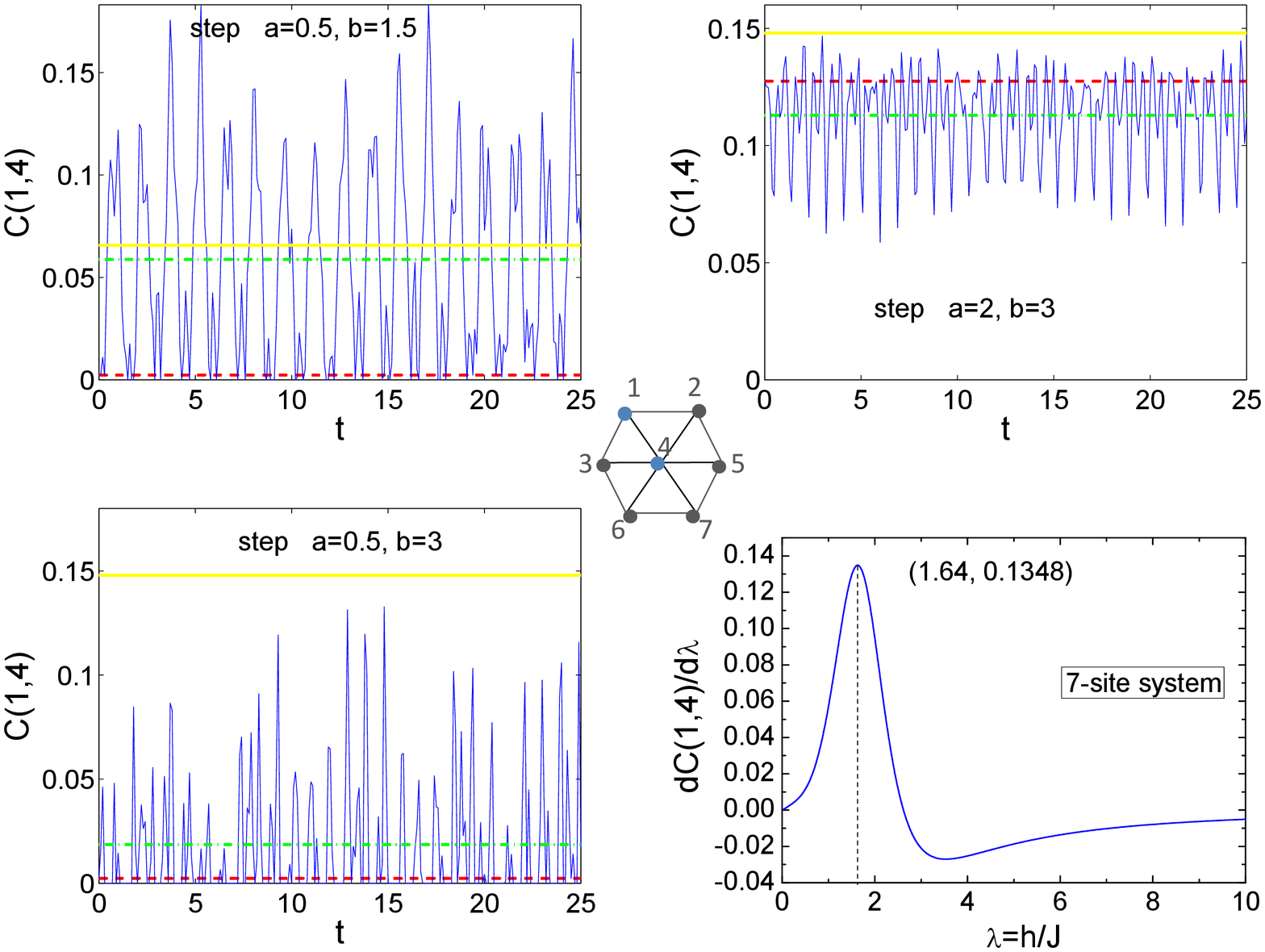}
   \caption{{\protect\footnotesize Dynamics of C(1,4) (solid blue line) in the 7-site system when the step magnetic field is changed from $a=0.5$ to $b=1.5$ (before the ``critical point'' $h=1.64$), from $a=2$ to $b=3$ (after) and from $a=0.5$ to $b=3$ (big step cross the ``critical point''), where time $t$ is in the unit of $J^{-1}$, the dashed red line stand for the concurrence corresponding to a constant magnetic field $h=a$, the straight solid yellow line for $h=b$ and the dot-dashed green line for the average value of the oscillating concurrence.}}
 \label{before_after_cross_cp_14}
 \end{minipage}
\end{figure}
\begin{figure}[htbp]
\begin{minipage}[c]{\textwidth}
 \centering
   \subfigure[]{\label{fig:he_a}\includegraphics[width=6 cm]{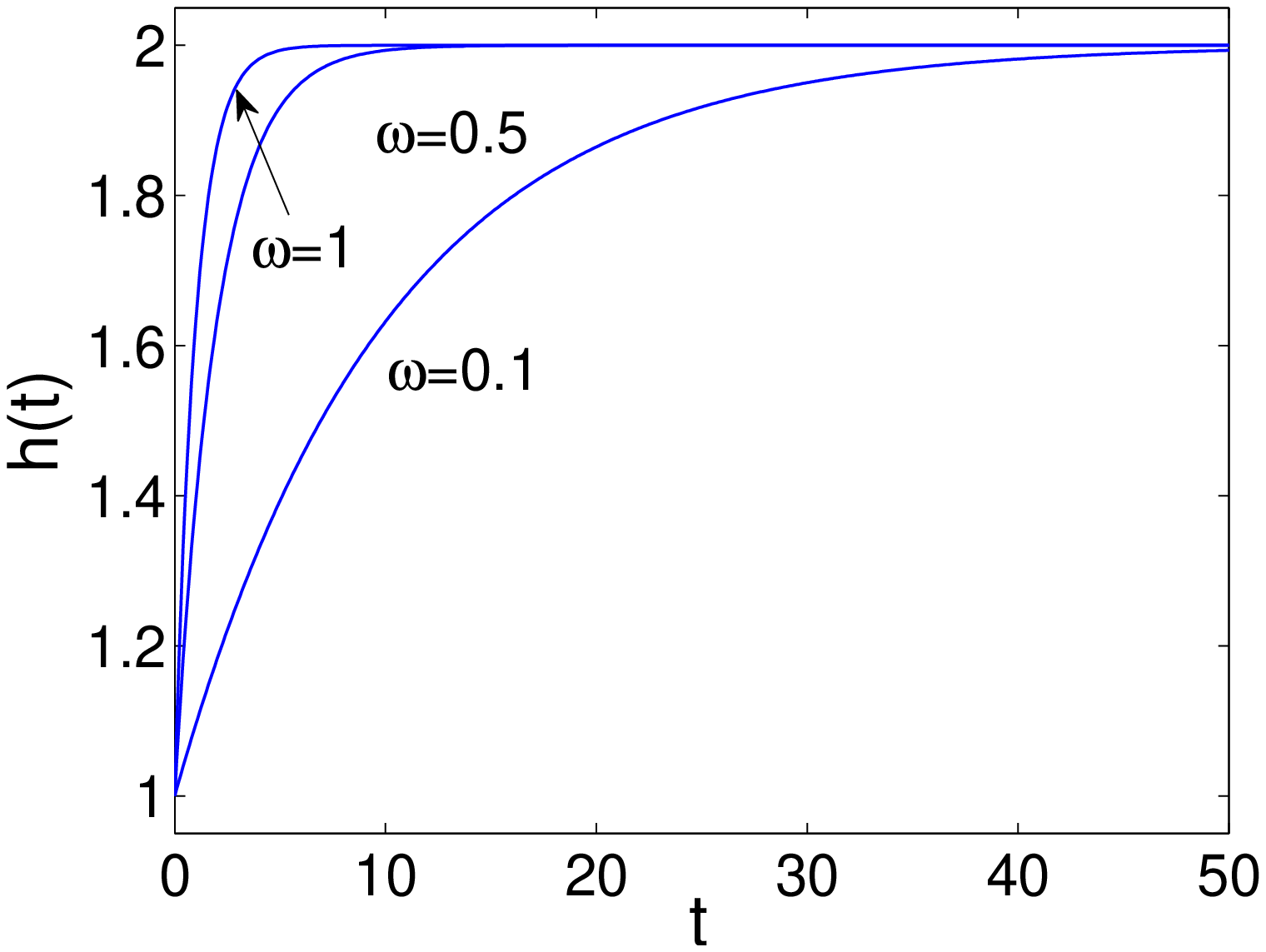}}\quad
   \subfigure[]{\label{fig:he_b}\includegraphics[width=6 cm]{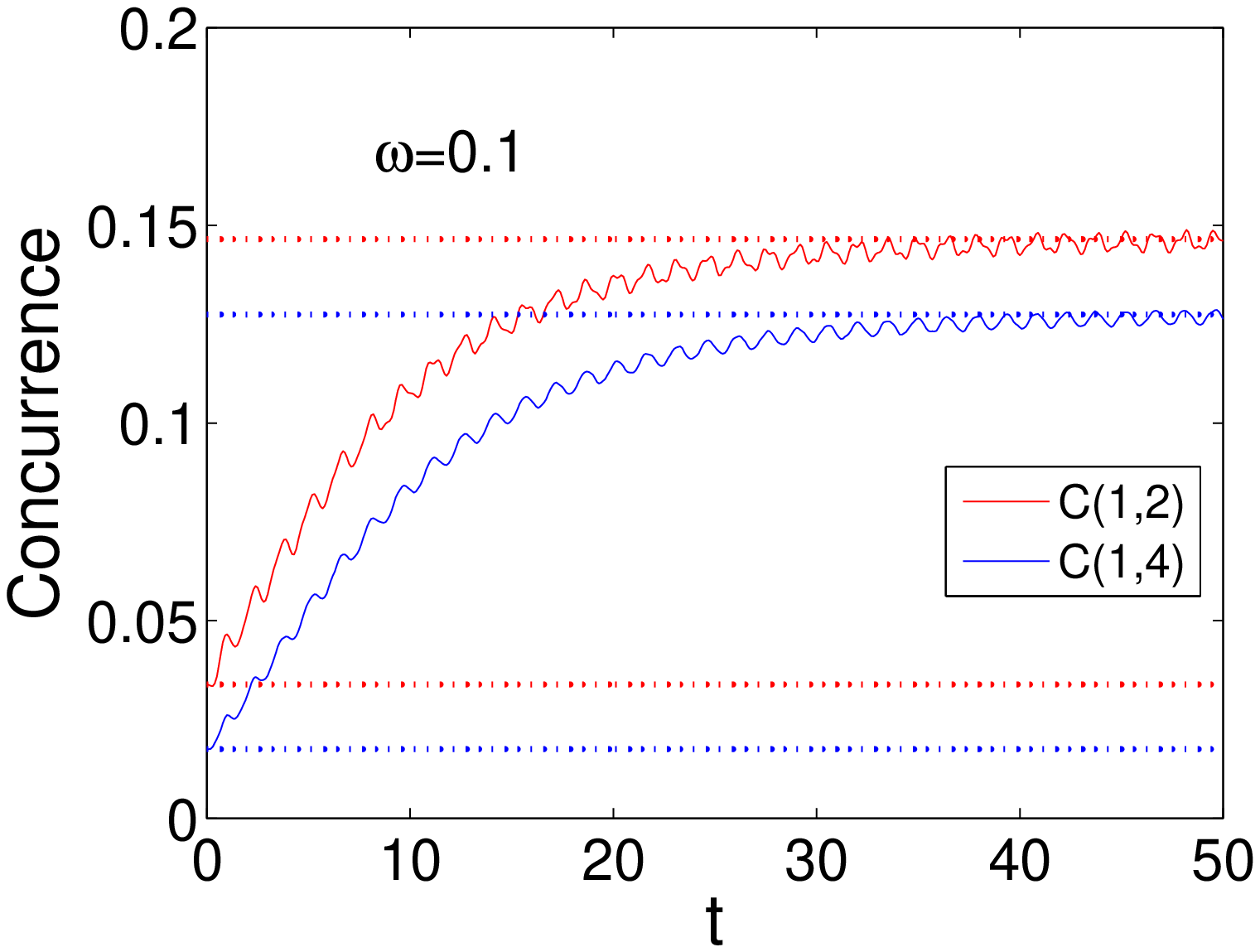}}\\
   \subfigure[]{\label{fig:he_c}\includegraphics[width=6 cm]{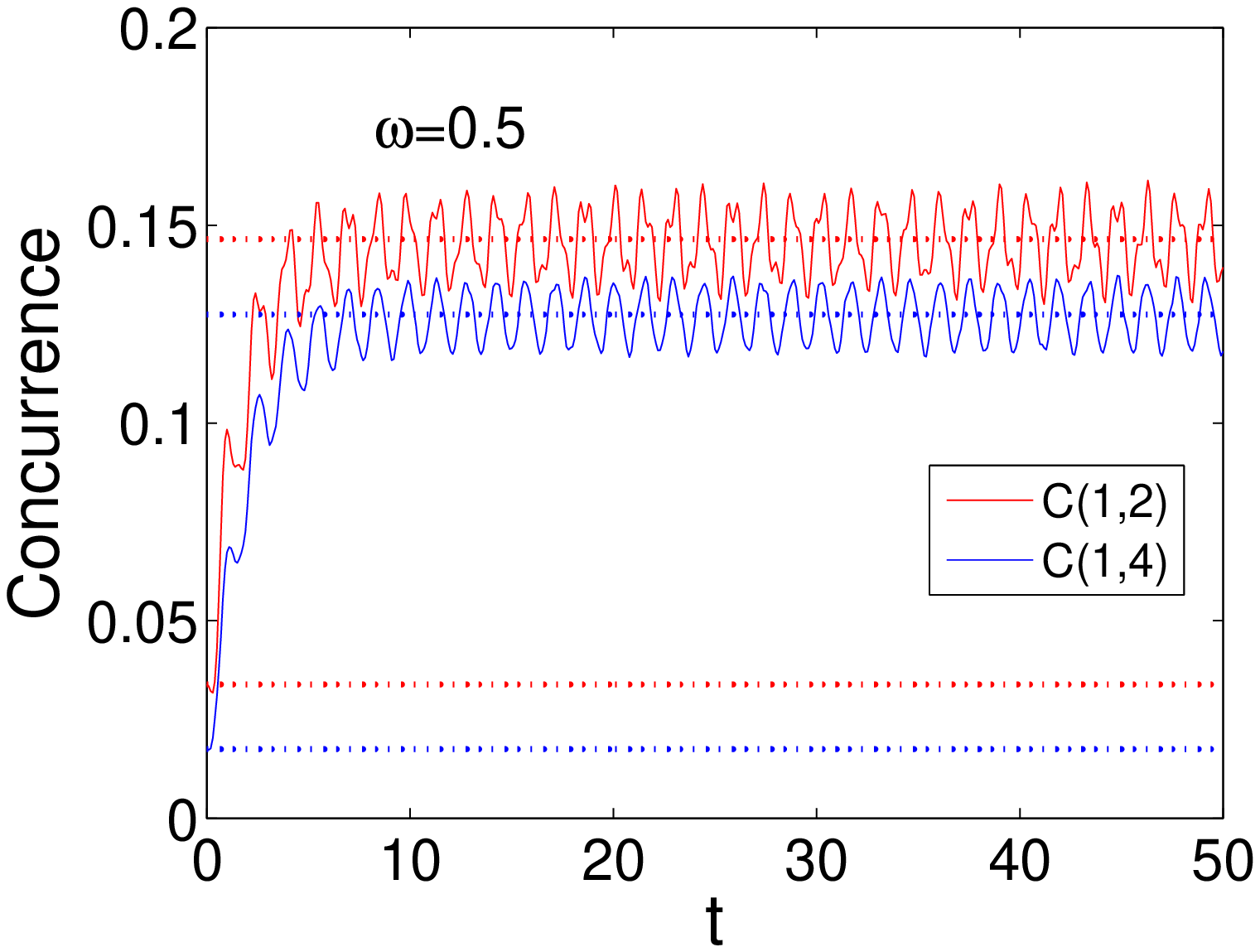}}\quad
   \subfigure[]{\label{fig:he_d}\includegraphics[width=6 cm]{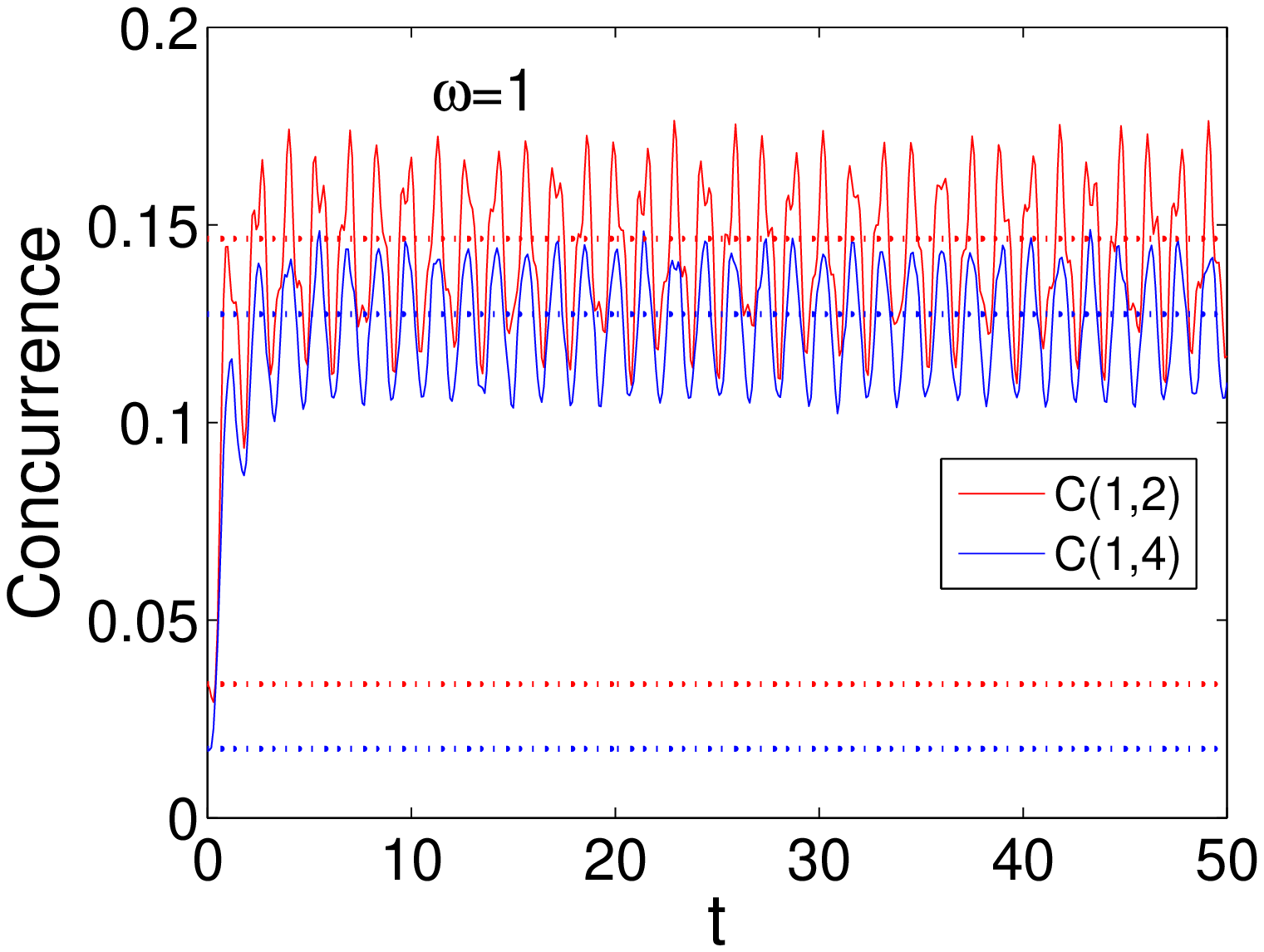}}
   \caption{{\protect\footnotesize Dynamics of the concurrences C(1,2) (solid red (upper) line) and C(1,4) (solid blue (lower) line) in applied exponential magnetic fields of various frequencies $\omega=0.1$, $0.5$ \& $1$, with strength $a=1$, $b=2$, where time $t$ is in the unit of $J^{-1}$. The straight dotted red (upper) lines are concurrences C(1,2) under constant magnetic field $a=1$ and $b=2$, so are the blue (lower) lines for C(1,4). }}
 \label{exp_h}
 \end{minipage}
\end{figure}
\begin{figure}[htbp]
\begin{minipage}[c]{\textwidth}
 \centering
   \subfigure[]{\label{fig:ht_a}\includegraphics[width=6 cm]{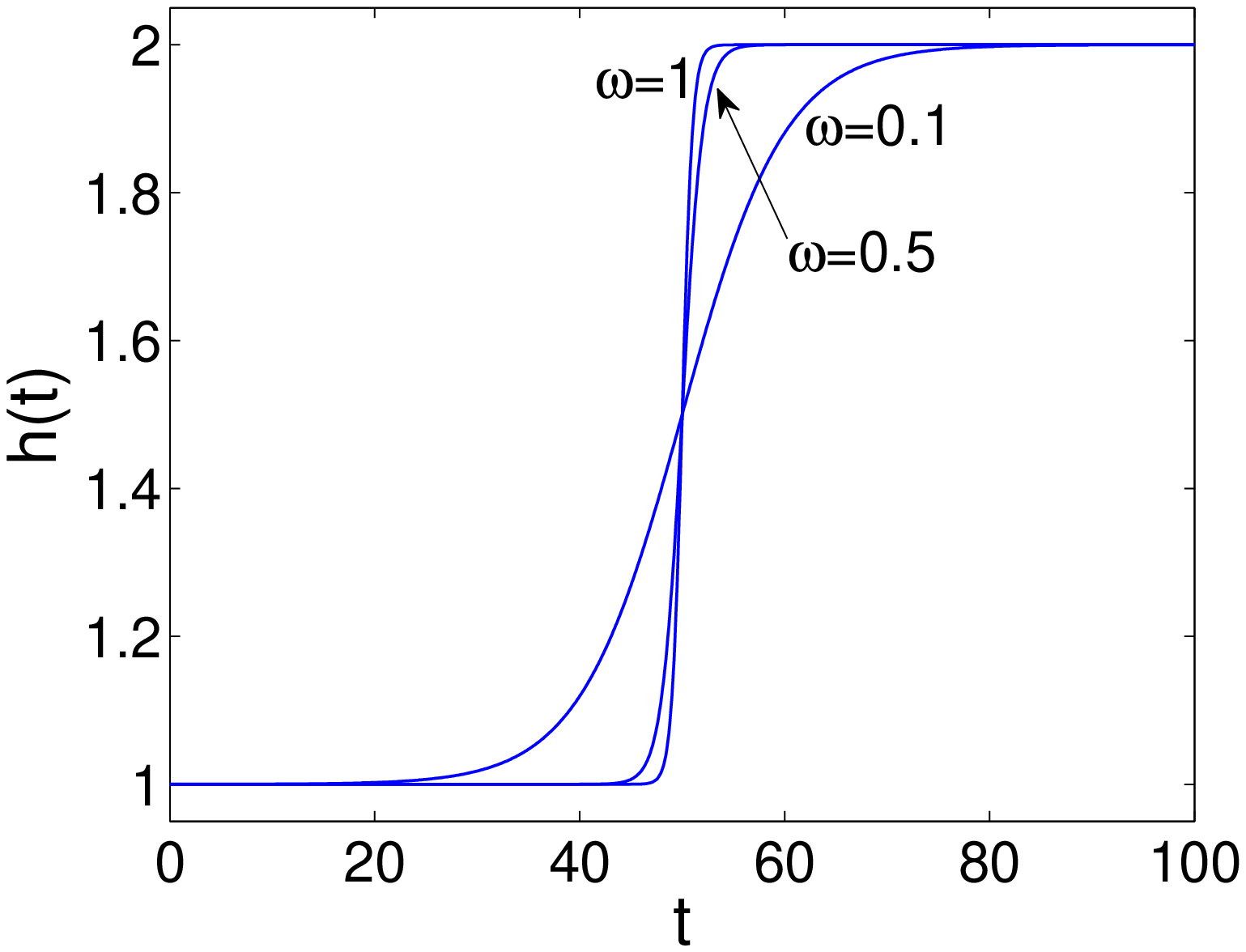}}\quad
   \subfigure[]{\label{fig:ht_b}\includegraphics[width=6 cm]{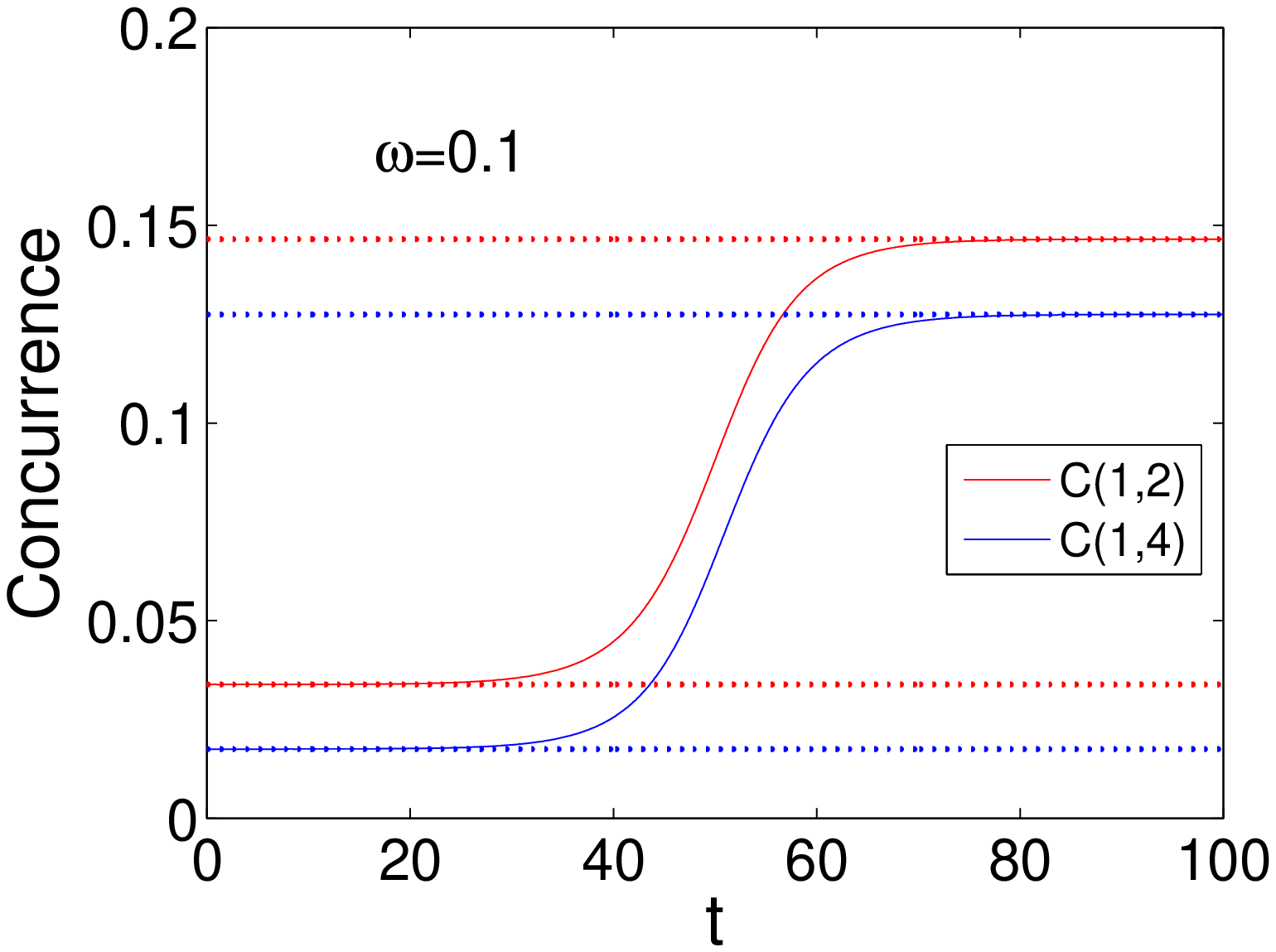}}\\
   \subfigure[]{\label{fig:ht_c}\includegraphics[width=6 cm]{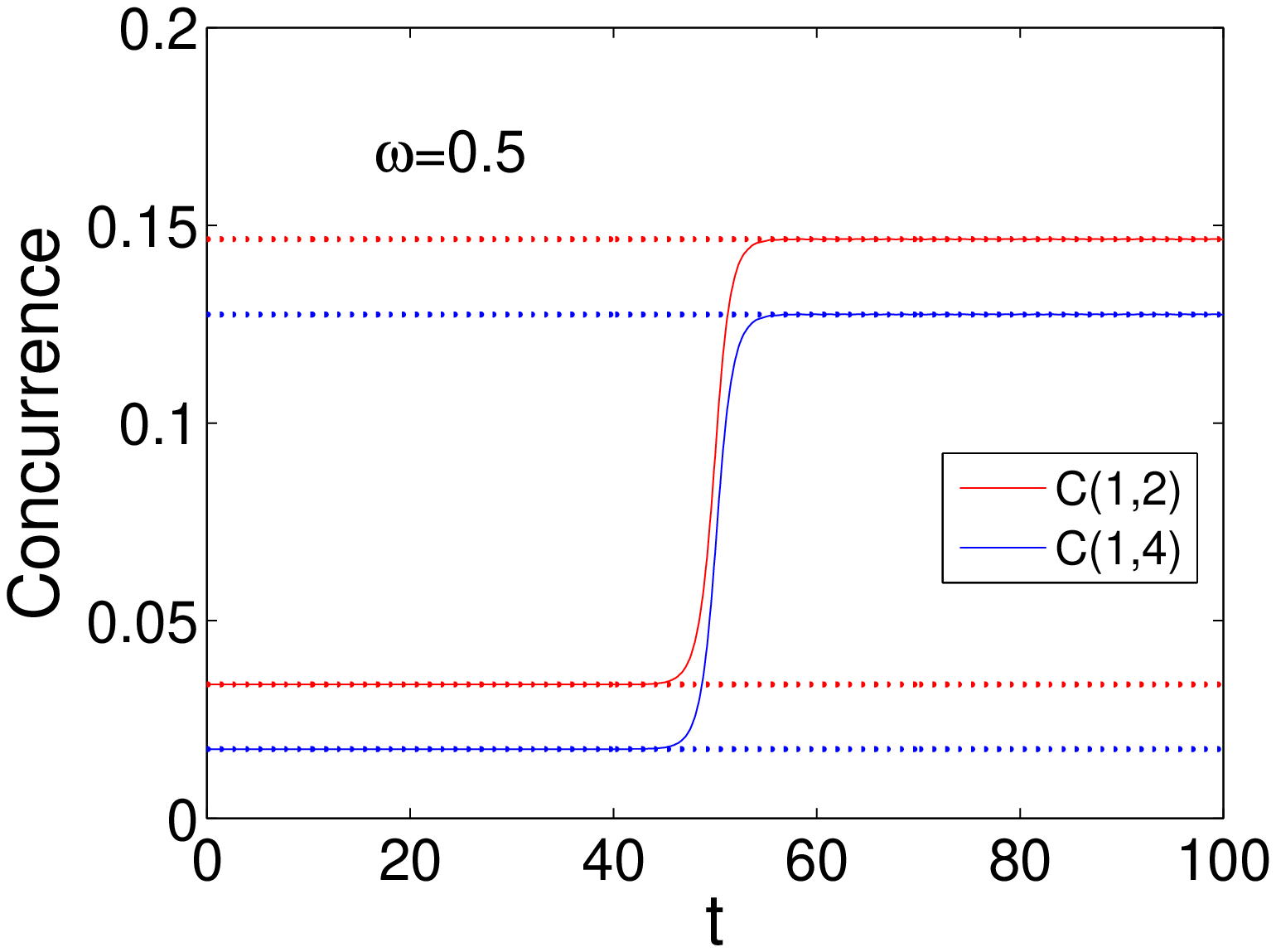}}\quad
   \subfigure[]{\label{fig:ht_d}\includegraphics[width=6 cm]{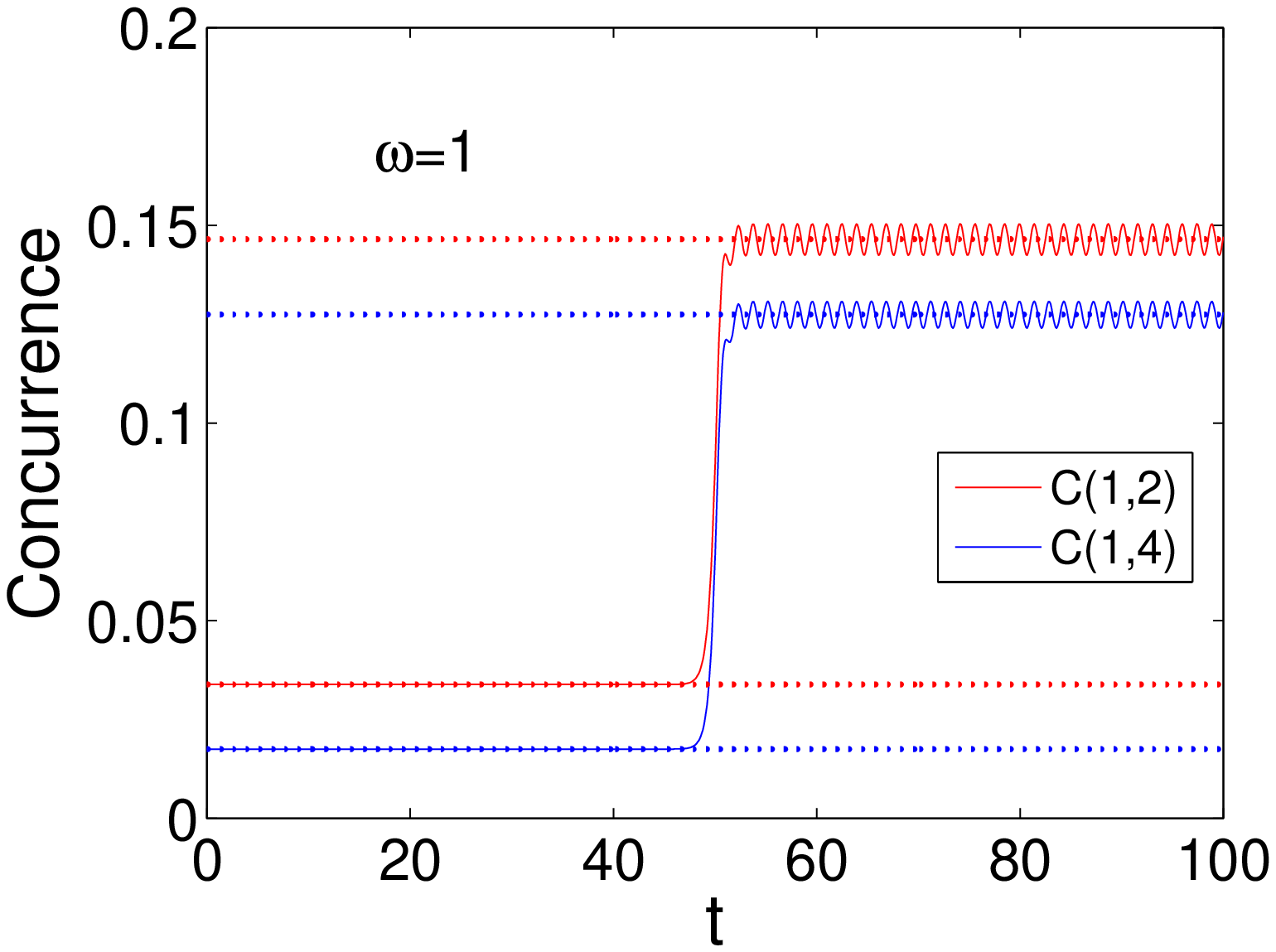}}
   \caption{{\protect\footnotesize Dynamics of the concurrences C(1,2) (solid red (upper) line) and C(1,4) (solid blue (lower) line) in applied hyperbolic magnetic fields of various frequencies $\omega=0.1$, $0.5$ \& $1$, with strength $a=1$, $b=2$, where time $t$ is in the unit of $J^{-1}$. The straight dotted red (upper) lines are concurrences C(1,2) under constant magnetic field $a=1$ and $b=2$, so are the blue (lower) lines for C(1,4).}}
 \label{tanh_h}
 \end{minipage}
\end{figure}
\begin{figure}[htbp]
\begin{minipage}[c]{\textwidth}
 \centering
   \subfigure[]{\label{fig:hs_a}\includegraphics[width=6 cm]{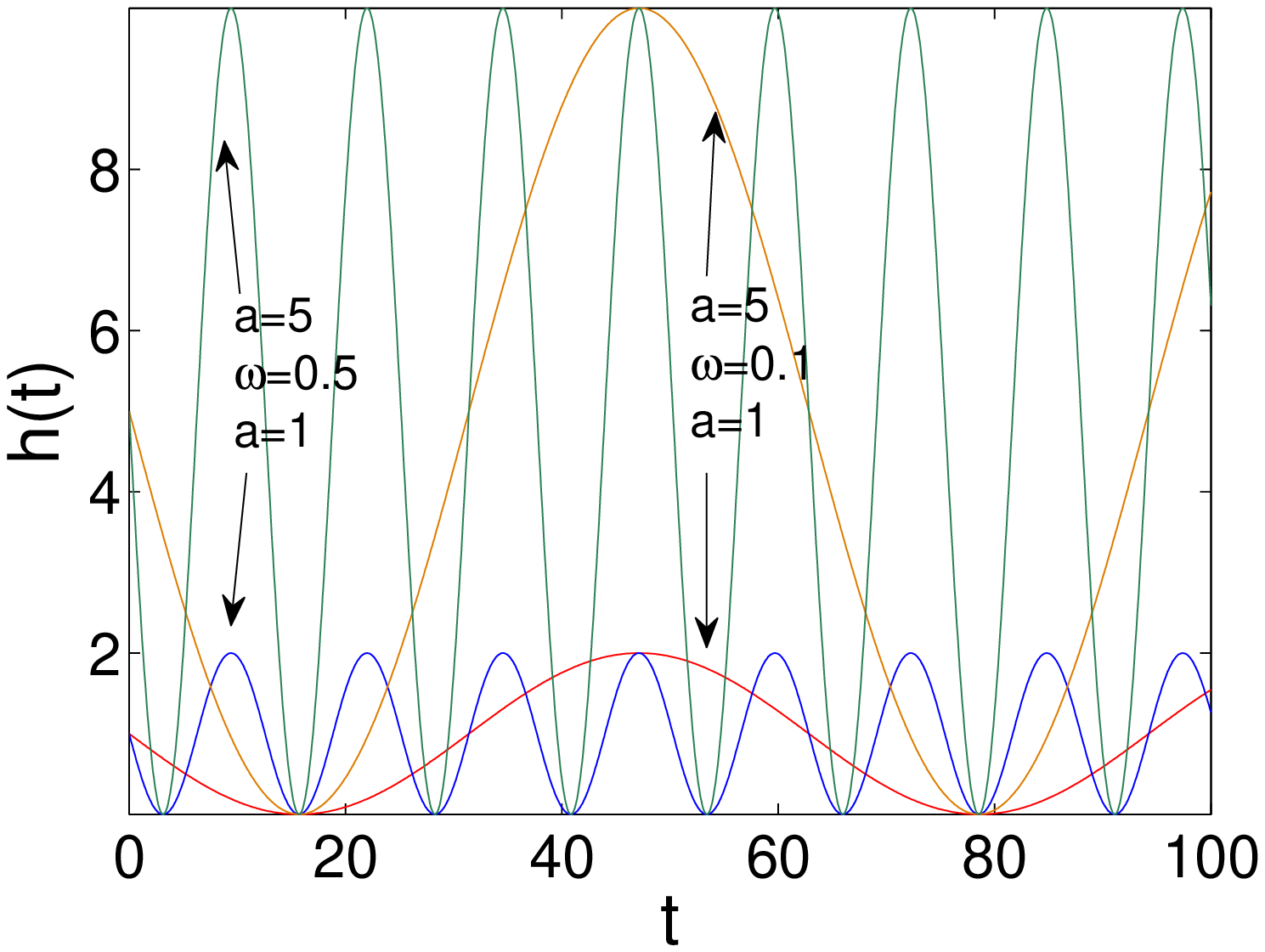}}\quad
   \subfigure[]{\label{fig:hs_b}\includegraphics[width=6 cm]{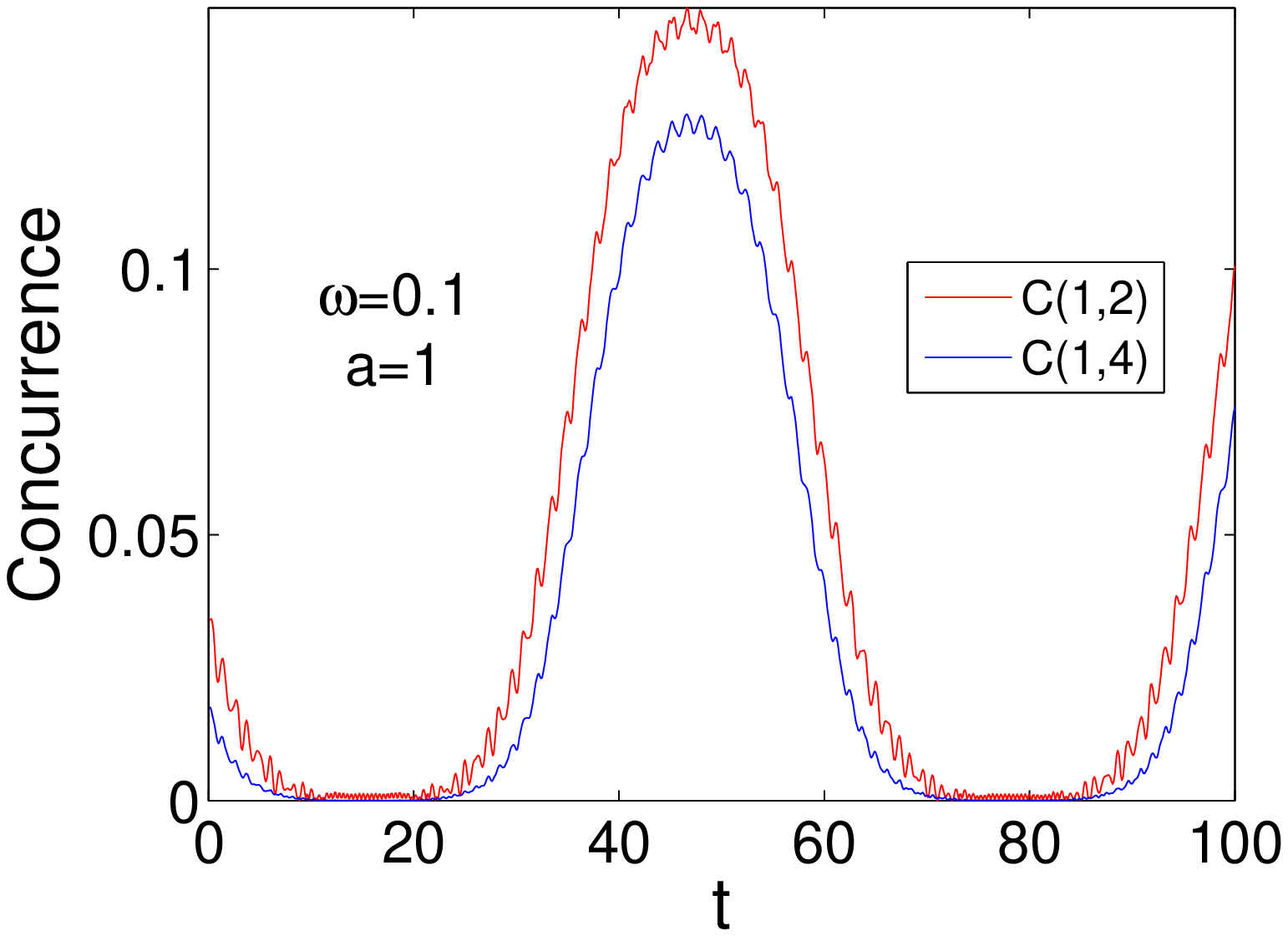}}\\
   \subfigure[]{\label{fig:hs_c}\includegraphics[width=6 cm]{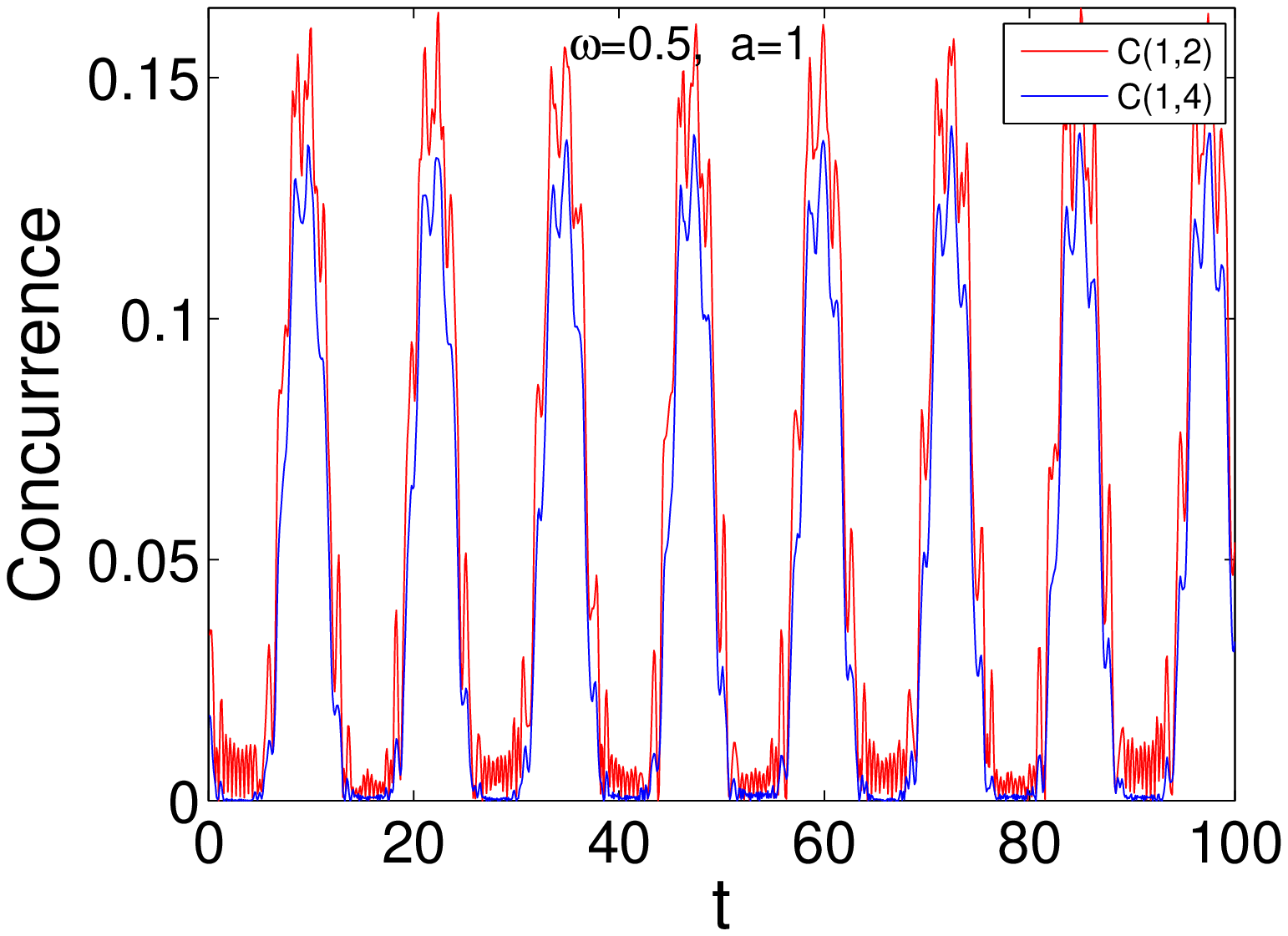}}\quad
   \subfigure[]{\label{fig:hs_d}\includegraphics[width=6 cm]{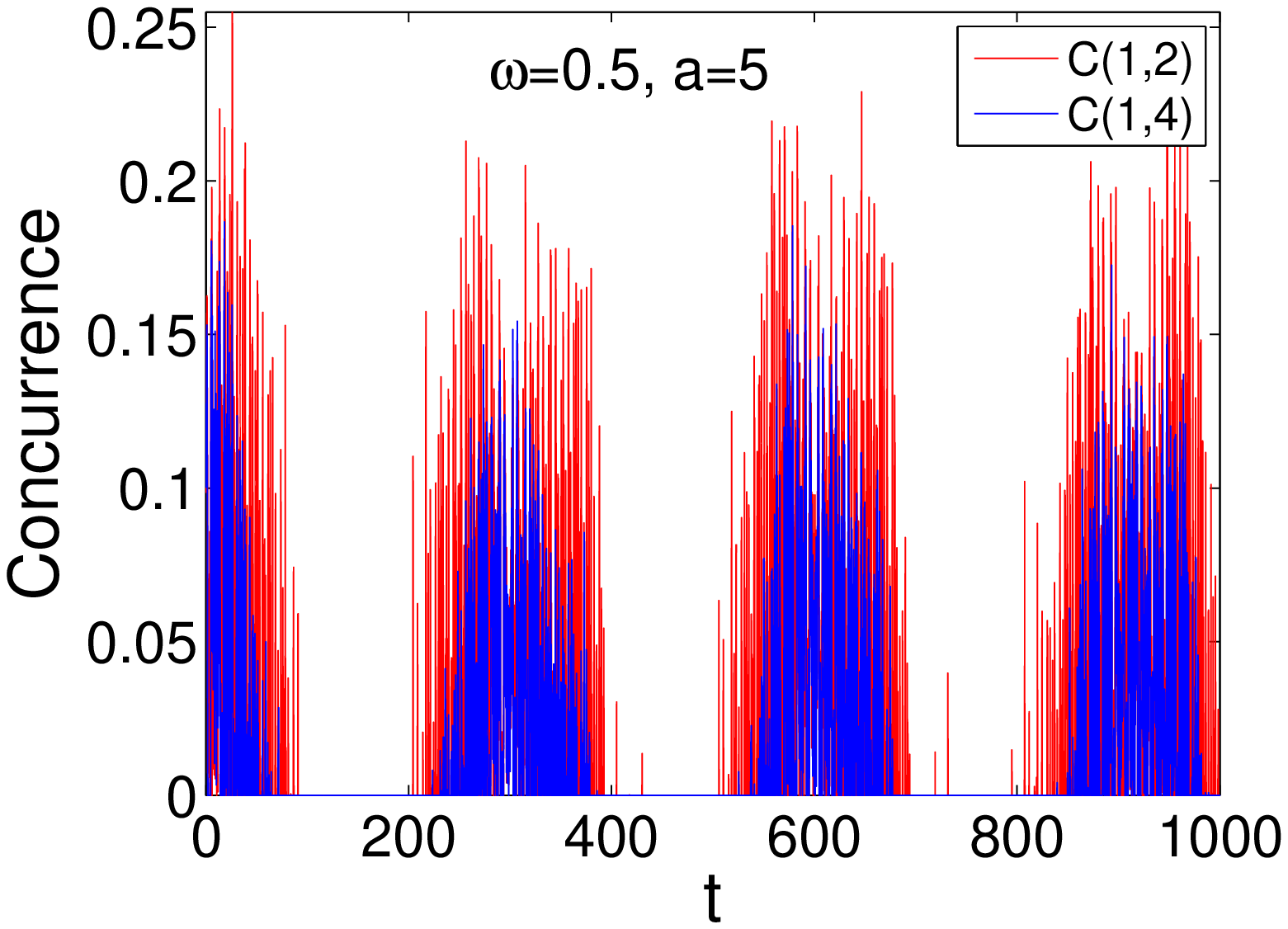}}
   \caption{{\protect\footnotesize  Dynamics of the concurrences C(1,2) (red (upper) line) and C(1,4) (blue (lower) line) in applied sine magnetic fields of various frequencies and field strength $\omega=0.1$ \& $a=1$, $\omega=0.5$ \& $a=1$ and $\omega=0.5$ \& $a=5$, where time $t$ is in the unit of $J^{-1}$.}}
 \label{sin_h}
 \end{minipage}
\end{figure}

\subsection{Tuning entanglement and ergodicity of spin systems using impurities and anisotropy}

There has been great interest in studying the different sources of errors in quantum computing and their effect on quantum gate operations \cite{Jones2003, Cummins2003}. Different approaches have been proposed for protecting quantum systems during the computational implementation of algorithms such as quantum error correction \cite{Shor1995} and  decoherence-free subspace \cite{Bacon2000}. Nevertheless, realizing a practical protection against the different types of induced decoherence is still a hard task. Therefore, studying the effect of naturally existing sources of errors such as impurities and lack of isotropy in coupling between the quantum systems implementing the quantum computing algorithms is a must. Furthermore, considerable efforts should be devoted to utilizing such sources to tune the entanglement rather than eliminating them. The effect of impurities and anisotropy of coupling between neighbor spins in a one dimensional spin system has been investigated \cite{Osenda2003}. It was demonstrated that the entanglement can be tuned in a class of one-dimensional systems by varying the anisotropy of the coupling parameter as well as by introducing impurities into the spin system. For a physical quantity to be eligible for an equilibrium statistical mechanical description it has to be ergodic, which means that its time average coincides with its ensemble average. To test ergodicity for a physical quantity one has to compare the time evolution of its physical state to the corresponding equilibrium state. There has been an intensive efforts to investigate ergodicity in one-dimensional spin chains where it was demonstrated that the entanglement, magnetization, spin-spin correlation functions are non-ergodic in Ising and XY spin chains for finite number of spins as well as at the thermodynamic limit \cite{Barouch1970,Sen(De)2004,HuangZ2006,Sadiek2010}.

In this part, we consider the entanglement in a two-dimensional $XY $triangular spin system, where the nearest neighbor spins are coupled through an exchange interaction $J$ and subject to an external magnetic field $h$. We consider the system at different degrees of anisotropy to test its effect on the system entanglement and dynamics. The number of spins in the system is 7, where all of them are identical
except one (or two) which are considered impurities. The Hamiltonian for such a system is given by
\begin{equation}
\label{Sch_equ2}
H=-\frac{(1+\gamma)}{2}\sum_{<i,j>}J_{i,j}\sigma_{i}^x\sigma_{j}^x -\frac{(1-\gamma)}{2}\sum_{<i,j>}J_{i,j}\sigma_{i}^y\sigma_{j}^y  - h(t) \sum_{i} \sigma_{i}^z,
\end{equation}
where $<i,j>$ is a pair of nearest-neighbors sites on the lattice, $J_{i,j}=J$ for all sites except the sites nearest to an impurity site. For a single impurity, the coupling between the impurity and its neighbors $J_{i,j}=J'=(\alpha+1)J$, where $\alpha$ measures the strength of the impurity. For double impurities $J_{i,j}=J'=(\alpha_1+1)J$ is the coupling between the two impurities and $J_{i,j}=J''=(\alpha_2+1)J$ is the coupling between any one of the two impurities and its neighbors while the coupling is just $J$ between the rest of the spins. For this model it is convenient to set $J=1$. Exactly solving Schrodinger equation of the Hamiltonian (\ref{Sch_equ2}), yielding the system energy eigenvalues ${E_i}$ and eigenfunctions ${\psi_i}$. The density matrix of the system is defined by
\begin{equation}
\label{dens_matrix}
\rho = |\psi_0 \rangle \langle \psi_0 | \; ,
\end{equation}
where $|\psi_0\rangle$ is the ground state energy of the entire spin system.
We confine our interest to the entanglement between two spins, at any sites $i$ and $j$ \cite{Osterloh2002}. All the information needed in this case, at any moment $t$, is contained in the reduced density matrix $\rho_{i, j}(t)$ which can be obtained from the entire system density matrix by integrating out all the spins states except $i$ and $j$. We adopt the entanglement of formation (or equivalently the concurrence C), as a well known  measure of entanglement \cite{Wootters1998}. The dynamics of entanglement is evaluated using the the step-by-step time-evolution projection technique introduced previously \cite{XuQ2011}.
\subsubsection{Single impurity}
\begin{figure}[htbp]
\begin{minipage}[c]{\textwidth}
 \centering
   \includegraphics[width=18 cm]{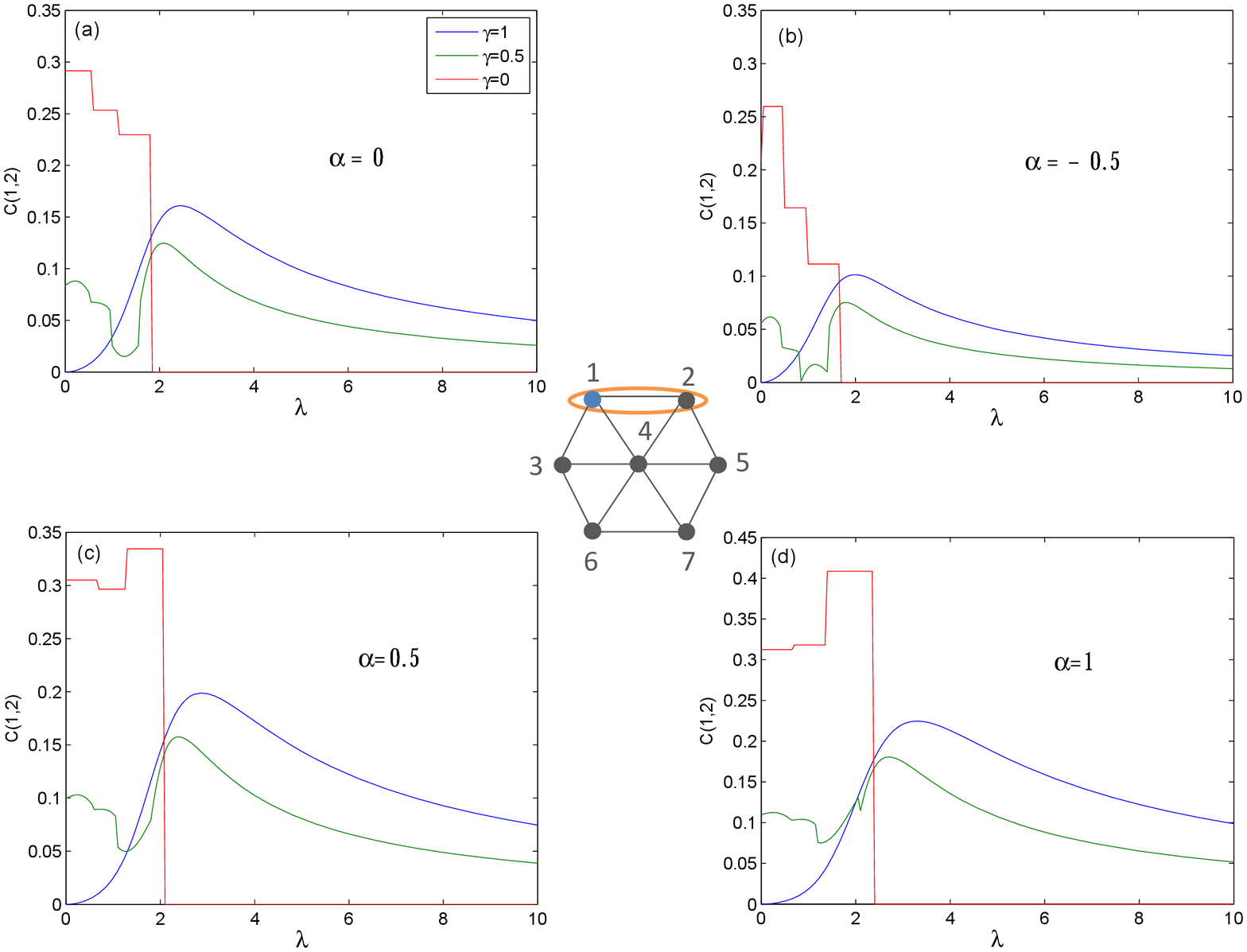}
   \caption{{\protect\footnotesize The concurrence $C(1,2)$ versus the parameter $\lambda$ with a single impurity at the border site 1 with different impurity coupling strengths $\alpha = -0.5, 0, 0.5, 1$ for different degrees of anisotropy $\gamma = 1, 0.5, 0$ as shown in the subfigures. The legend for all subfigures is as shown in subfigure (a).}}
 \label{B_C12}
 \end{minipage}
\end{figure}

We start by considering the effect of a single impurity located at the border site 1. The concurrence between the impurity site 1 and site 2, $C(1,2)$, versus the parameter $\lambda$ for the three different models, Ising ($\gamma=1$), partially anisotropic ($\gamma=0.5$) and isotropic XY ($\gamma=0$) at different impurity strengths ($\alpha= -0.5, 0, 0.5, 1$) is in fig.~\ref{B_C12}.
Firstly, the impurity parameter $\alpha$ is set to zero. For the corresponding Ising model, the concurrence $C(1,2)$, in fig.~\ref{B_C12}(a), demonstrates the usual phase transition behavior where it starts at zero value and increases gradually as $\lambda$ increases reaching a maximum at $\lambda \approx 2$ then decays as $\lambda$ increases further. As the degree of anisotropy decreases the behavior of the entanglement changes, where it starts with a finite value at $\lambda=0$ and then shows a step profile for the small values of $\lambda$. For the partially anisotropic case, the step profile is smooth and the entanglement mimics the Ising case as $\lambda$ increases but with smaller magnitude. The entanglement of isotropic XY system shows a sharp step behavior then suddenly vanishes before reaching $\lambda=2$. Interestingly, the entanglement behavior of the two-dimensional spin system at the different degrees of anisotropy mimics the behavior of the one-dimensional spin system at the same degrees of anisotropy at the extreme values of the parameter $\lambda$. Comparing the entanglement behavior in the two-dimensional Ising spin system with the one-dimensional system, one can see a great resemblance except that the critical value becomes $h/J \approx 2$ in the two dimensional case as shown in fig.~\ref{B_C12}. On the other hand, for the partially anisotropic and isotropic $XY$ systems, the entanglements of the two-dimensional and one-dimensional system agrees at the extreme values of $\lambda$ where it vanishes for $h >> J$ and reaches a finite value for $h << J$. The former case corresponds to an alignment of the spins in the $z$-direction, paramagnetic state, while the latter case corresponds to alignment in the $x$ and $y$-directions which a ferromagnetic state. The effect of a weak impurity ($J'<J$), $\alpha=-0.5$, is shown in fig.~\ref{B_C12}(b) where the entanglement behavior is the same as before except that the entanglement magnitude is reduced compared with the pure case. On the other hand,  considering the effect of a strong impurity ($J'> J$), where $\alpha=0.5$ and 1, as shown in fig.~\ref{B_C12}(c) and fig.~\ref{B_C12}(d) respectively, one can see that the entanglement profile for $\gamma =1$ and 0.5 have the same overall behavior as in the pure and weak impurity cases except that the entanglement magnitude becomes higher as the impurity gets stronger and the peaks shift toward higher $\lambda$ values. Nevertheless, the isotropic $XY$ system behaves differently from the previous cases where it starts to increase first in a step profile before suddenly dropping to zero again, which will be explained latter.

\begin{figure}[htbp]
\begin{minipage}[c]{\textwidth}
 \centering
   \includegraphics[width=18 cm]{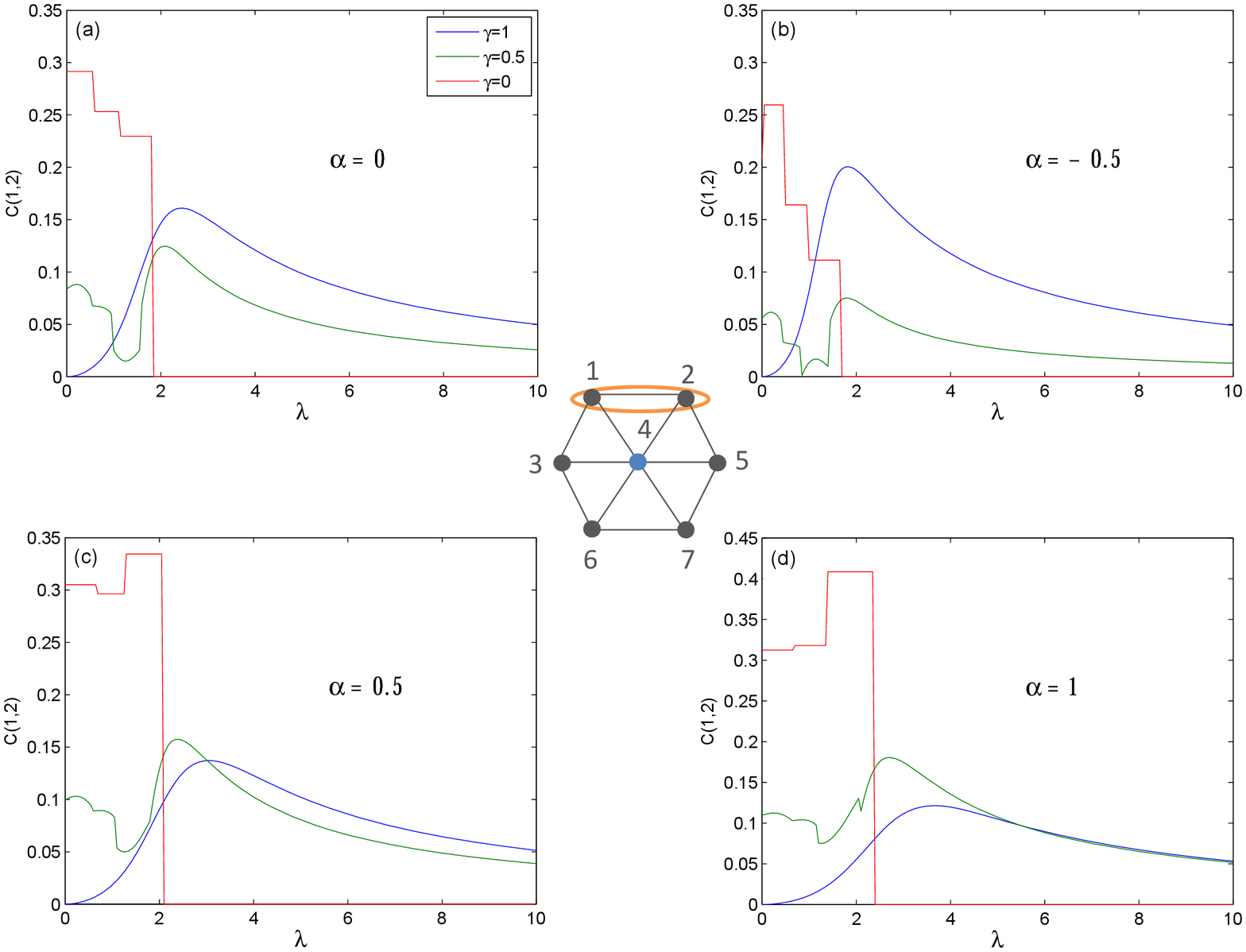}
    \caption{{\protect\footnotesize The concurrence $C(1,2)$ versus the parameter $\lambda$ with a single impurity at the central site 4 with different impurity coupling strengths $\alpha = -0.5, 0, 0.5, 1$ for different degrees of anisotropy $\gamma = 1, 0.5, 0$ as shown in the subfigures. The legend for all subfigures is as shown in subfigure (a).}}
 \label{C_C12}
 \end{minipage}
\end{figure}

To explore the effect of the impurity location we investigate the case of a single impurity spin located at site 4, instead of site 2, where we plot the concurrences $C(1,2)$ in figs.~\ref{C_C12}.
Interestingly, while changing the impurity location has almost no effect on the behavior of the entanglement $C(1,2)$ of the partially anisotropic and isotropic XY systems, it has a great impact on that of the Ising system where the peak value of the entanglement increases significantly in the weak impurity case and decreases as the impurity gets stronger as shown in fig.~\ref{C_C12}.

\begin{figure}[htbp]
\begin{minipage}[c]{\textwidth}
 \centering
   \includegraphics[width=18 cm]{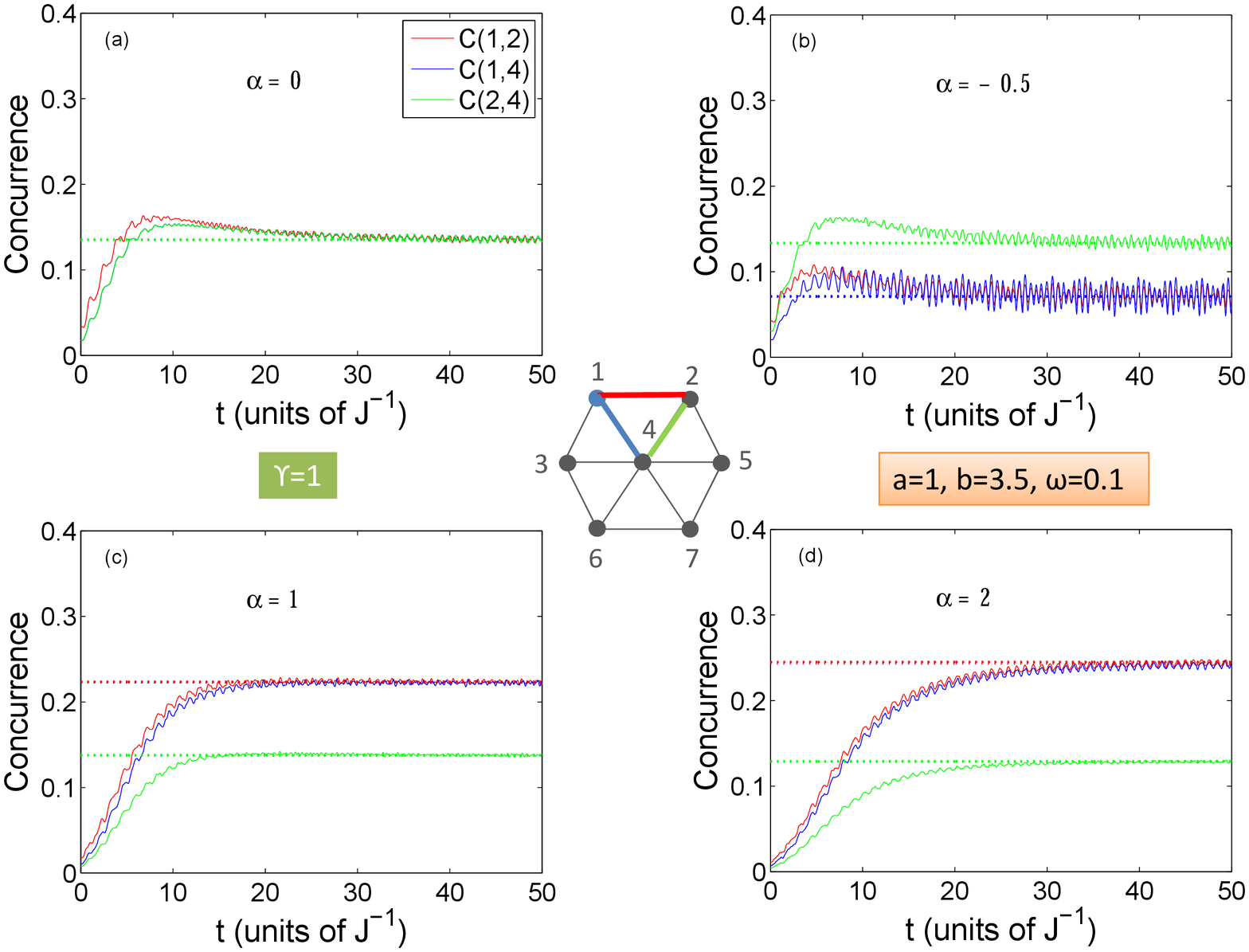}
   \caption{{\protect\footnotesize Dynamics of the concurrences $C(1,2), C(1,4), C(2,4)$ with a single impurity at the border site 1 with different impurity coupling strengths $\alpha = -0.5, 0, 1, 2$ for the two dimensional Ising lattice ($\gamma = 1$) under the effect of an exponential magnetic field with parameters values a=1, b=3.5 and $\omega=0.1$. The straight dashed lines represent the equilibrium concurrences corresponding to constant magnetic field $h=3.5$. The legend for all subfigures is as shown in subfigure (a).}}
 \label{B_Dyn_G_1}
 \end{minipage}
\end{figure}

Now we turn to the dynamics of the two dimensional spin system under the effect of a single impurity and different degrees of anisotropy. We investigate the dynamical reaction of the system to an applied time-dependent magnetic field with exponential form $h(t)= b + (a-b) e^{-w \; t}$ for $t > 0$ and $h(t)=a$ for $t \leq 0$. We start by considering the Ising system, $\gamma=1$ with a single impurity at the border site 1, which is explored in fig.~\ref{B_Dyn_G_1}. For the pure case, $\alpha=0$ shown in fig.~\ref{B_Dyn_G_1}(a), the results confirms the ergodic behavior of the system that was demonstrated in our previous work \cite{XuQ2011}, where the asymptotic value of the entanglement coincide with the equilibrium state value at $h(t)=b$. As can be noticed from figs.~\ref{B_Dyn_G_1}(b), \ref{B_Dyn_G_1}(c) and \ref{B_Dyn_G_1}(d) neither the weak nor strong impurities have effect on the ergodicity of the Ising system. Nevertheless, there is a clear effect on the asymptotic value of entanglements $C(1,2)$ and $C(1,4)$ but not on $C(2,4)$ which relates two regular sites. The weak impurity, $\alpha=-0.5$ reduces the asymptotic value of $C(1,2)$ and $C(1,4)$ while the strong impurities, $\alpha= 1, 2$ raise it compared to the pure case.

\begin{figure}[htbp]
\begin{minipage}[c]{\textwidth}
 \centering
   \includegraphics[width=18 cm]{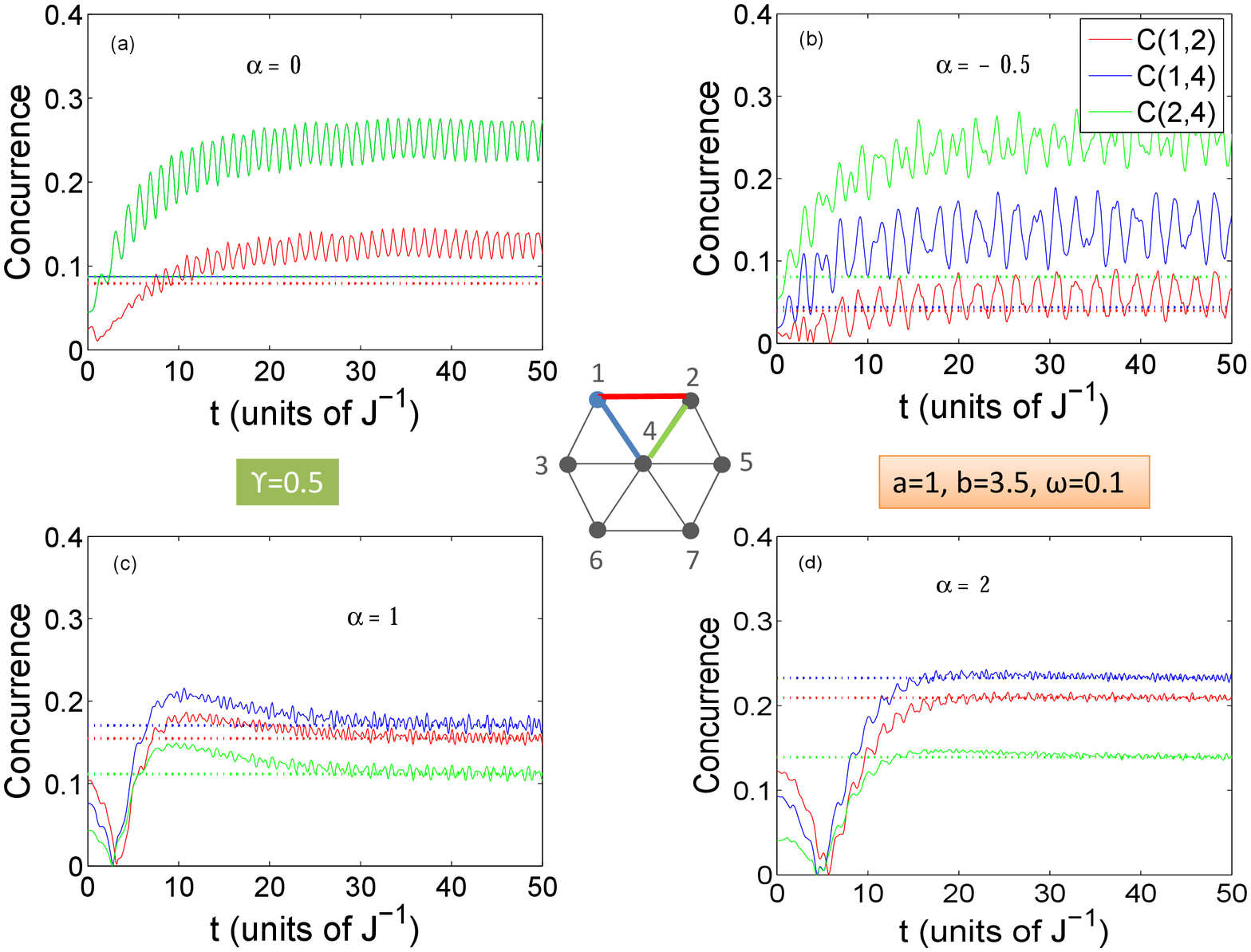}
   \caption{{\protect\footnotesize Dynamics of the concurrences $C(1,2), C(1,4), C(2,4)$ with a single impurity at the border site 1 with different impurity coupling strengths $\alpha = -0.5, 0, 1, 2$ for the two dimensional partially anisotropic lattice ($\gamma = 0.5$) under the effect of an exponential magnetic field with parameters values a=1, b=3.5 and $\omega=0.1$. The straight dashed lines represent the equilibrium concurrences corresponding to constant magnetic field $h=3.5$. The legend for all subfigures is as shown in subfigure (b).}}
 \label{B_Dyn_G_05}
 \end{minipage}
\end{figure}
The dynamics of the partially anisotropic XY system under the effect exponential magnetic field with parameters $a =1$, $b=3.5$ and $\omega = 0.1$, is explored in fig.~\ref{B_Dyn_G_05}. It is remarkable to see that while for both the pure and weak impurity cases, $\alpha=0$ and $-0.5$ , the system is nonergodic as shown in figs.~\ref{B_Dyn_G_05}(a) and ~\ref{B_Dyn_G_05}(b), and it is ergodic in the strong impurity cases $\alpha= 1$ and 2 as illustrated in figs.~\ref{B_Dyn_G_05}(c) and ~\ref{B_Dyn_G_05}(d).
\subsubsection{Double impurities}
\begin{figure}[htbp]
\begin{minipage}[c]{\textwidth}
\centering
   \subfigure{\includegraphics[width=8 cm]{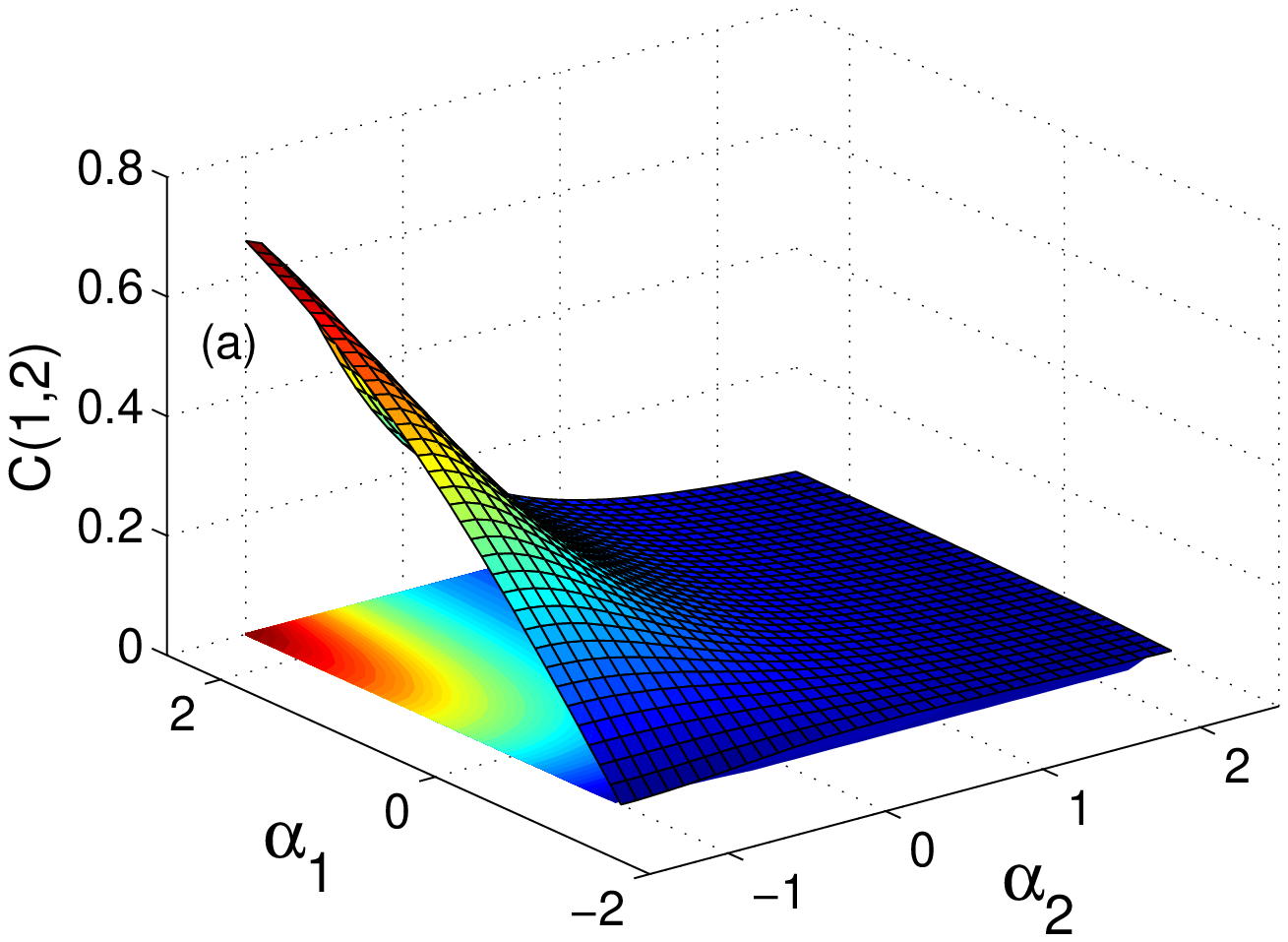}}\quad
   \subfigure{\includegraphics[width=8 cm]{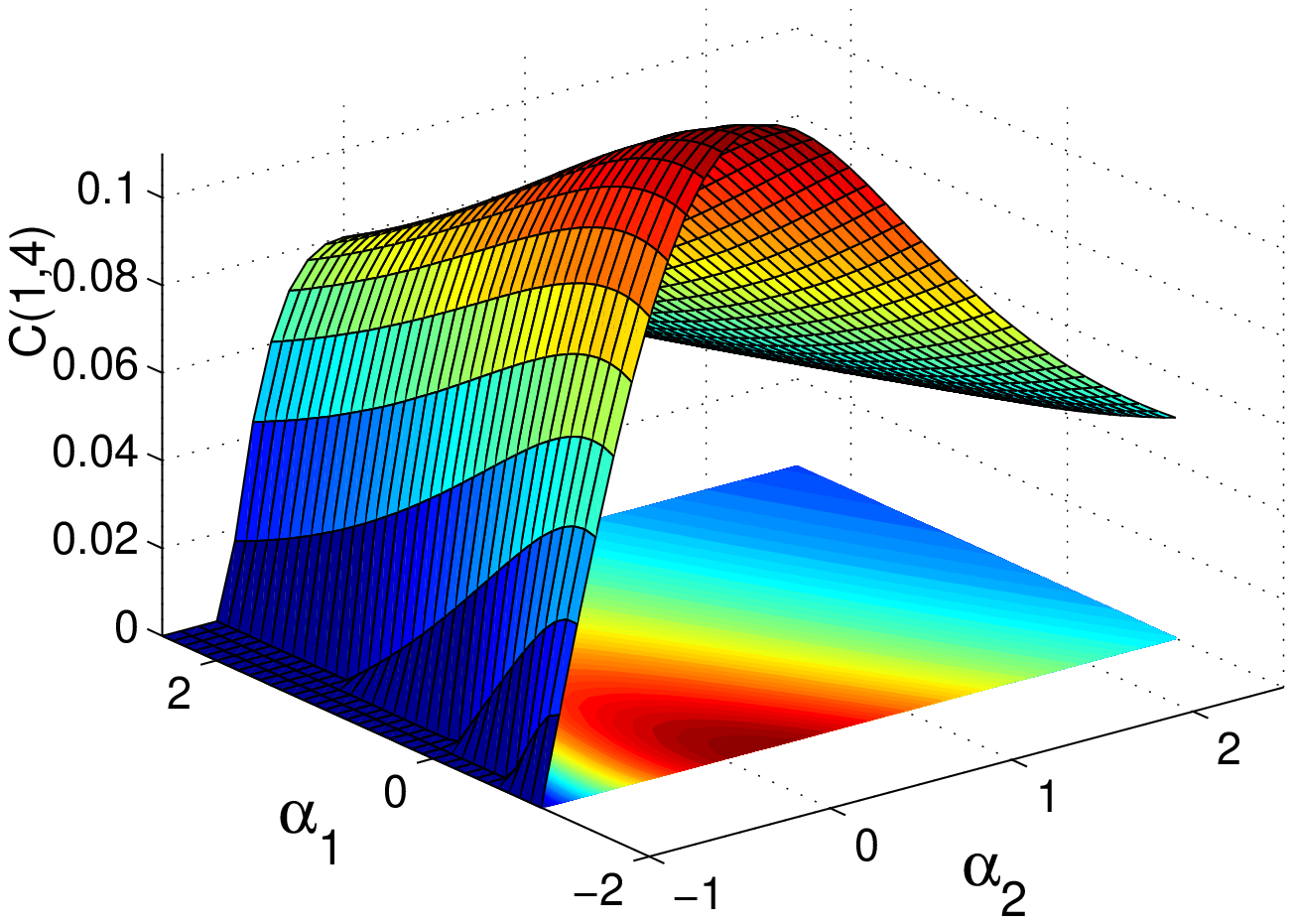}}\\
   \subfigure{\includegraphics[width=8 cm]{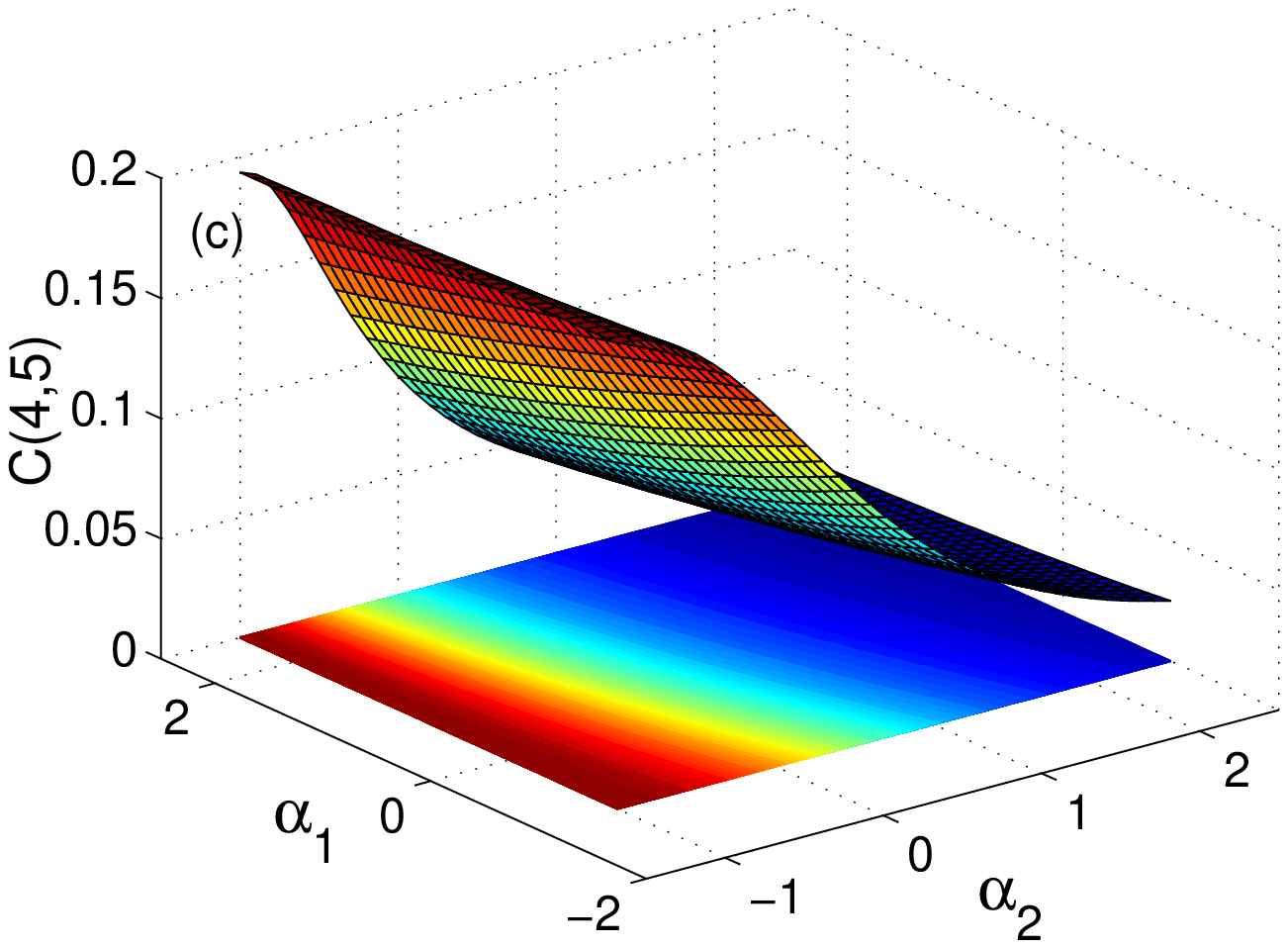}}\quad
   \subfigure{\includegraphics[width=8 cm]{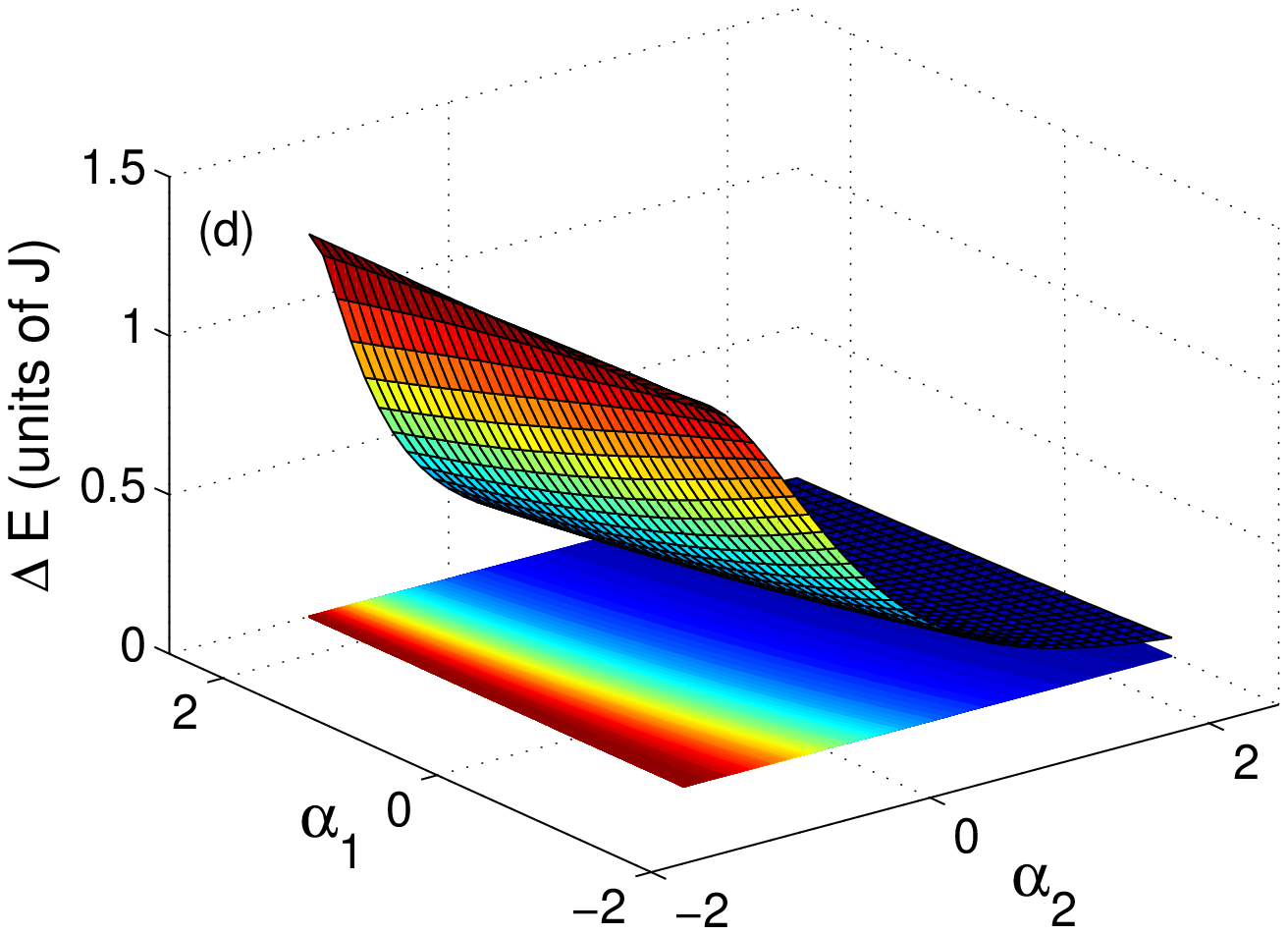}}
   \caption{{\protect\footnotesize The concurrence $C(1,2)$, $C(1,4)$, $C(4,5)$ versus the impurity coupling strengths $\alpha_1$ and $\alpha_2$ with double impurities at sites 1 and 2 for the two dimensional Ising lattice ($\gamma = 1$) in an external magnetic field h=2.}}
 \label{Imp12_G1_h2}
 \end{minipage}
\end{figure}
In this subsection we study the effect of double impurity, where we start with two located at the border sites 1 and 2. We set the coupling strength between the two impurities as $J' = (1+\alpha_1) J$, between any one of the impurities and its regular nearest neighbors as $J'' = (1+\alpha_2) J$ and between the rest of the nearest neighbor sites on the lattice as $J$. The effect of the impurities strength on the concurrence between different pairs of sites for the Ising lattice is shown in fig.~\ref{Imp12_G1_h2}. In fig.~\ref{Imp12_G1_h2}(a) we consider the entanglement between the two impurity sites 1 and 2 under a constant external magnetic field $h=2$. The concurrence $C(1,2)$ takes a large value when the impurity strengths $\alpha_1$, controlling the coupling between the impurity sites, is large and when $\alpha_2$, controlling coupling between impurities and their nearest neighbors, is weak. As $\alpha_1$ decreases and $\alpha_2$ increases, $C(1,2)$ decreases monotonically until it vanishes. As one can conclude, $\alpha_1$ is more effective than $\alpha_2$ in controlling the entanglement in this case. On the other hand, the entanglement between the impurity site 1 and the regular central site 4 is illustrated in fig.~\ref{Imp12_G1_h2}(b) which behaves completely different from C(1,2). The concurrence C(1,4) is mainly controlled by the impurity strength $\alpha_2$ where it starts with a very small value when the impurity is very weak and increases monotonically until it reaches a maximum value at $\alpha_2=0$, i.e. with no impurity, and decays again as the impurity strength increases. The effect of $\alpha_1$ in that case is less significant and makes the concurrence slowly decreases as $\alpha_1$ increases which is expected since as the coupling between the two border sites 1 and 2 increases the entanglement between 1 and 4 decreases. It is important to note that in general $C(1,2)$ is much larger than $C(1,4)$ since the border entanglement is always higher than the central one as the entanglement is shared by many sites. The entanglement between two regular sites is shown in fig.~\ref{Imp12_G1_h2}(c) where the concurrence C(4,5) is depicted against $\alpha_1$ and $\alpha_2$, the entanglement decays gradually as $\alpha_2$ increases while $\alpha_1$ has a very small effect on the entanglement, which slightly decreases as $\alpha_1$ increases as shown. Interestingly, the behavior of the energy gap between the ground state and the first excited state of the Ising system $\Delta E$ versus the impurity strengths $\alpha_1$ and $\alpha_2$, which is explored in fig.~\ref{Imp12_G1_h2}(d) has a strong resemblance to that of the concurrence $C(4,5)$ except that the decay of $\Delta E$ against $\alpha_2$ is more rapid.
\begin{figure}[htbp]
\begin{minipage}[c]{\textwidth}
\centering
   \subfigure{\includegraphics[width=8 cm]{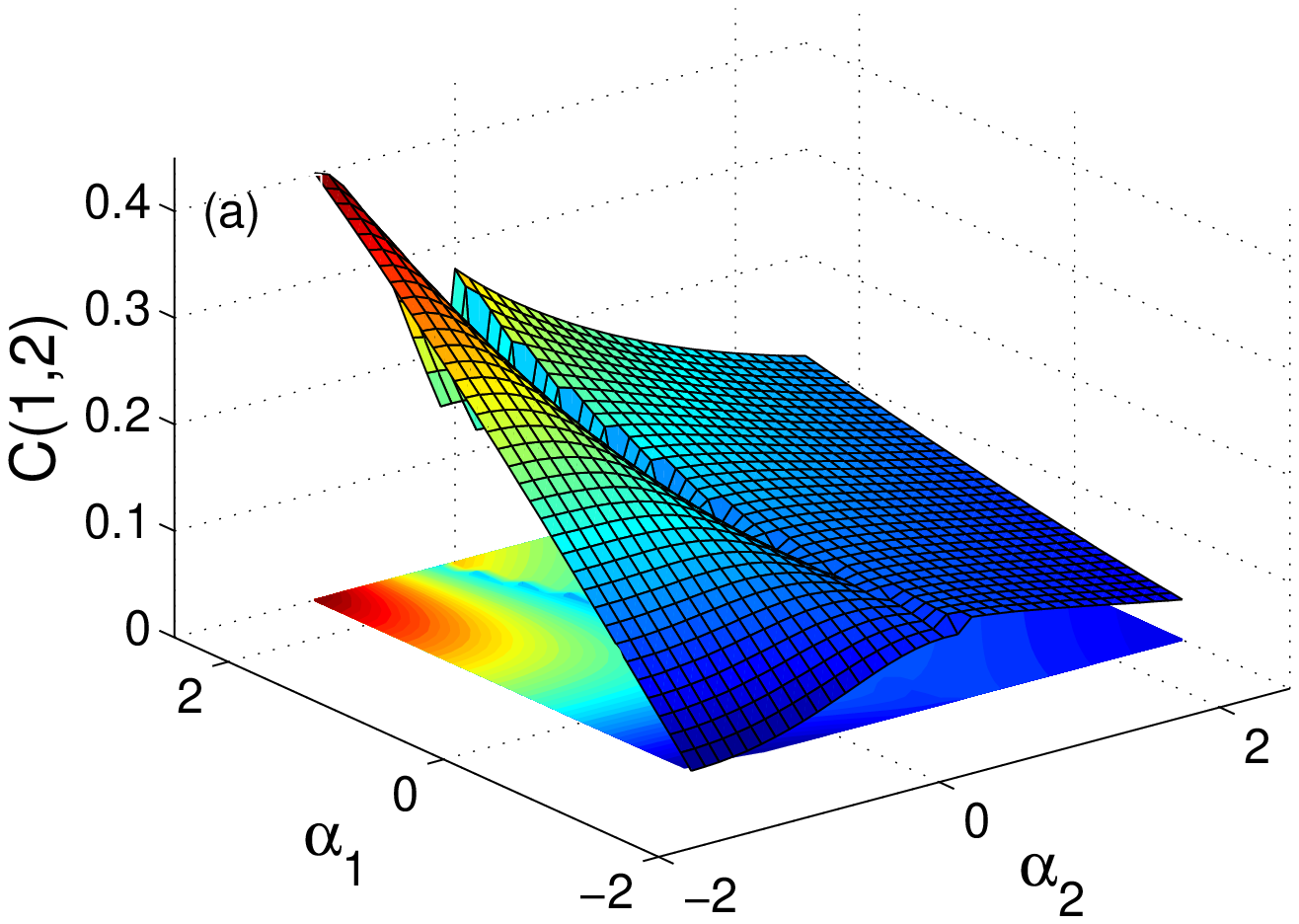}}\quad
   \subfigure{\includegraphics[width=8 cm]{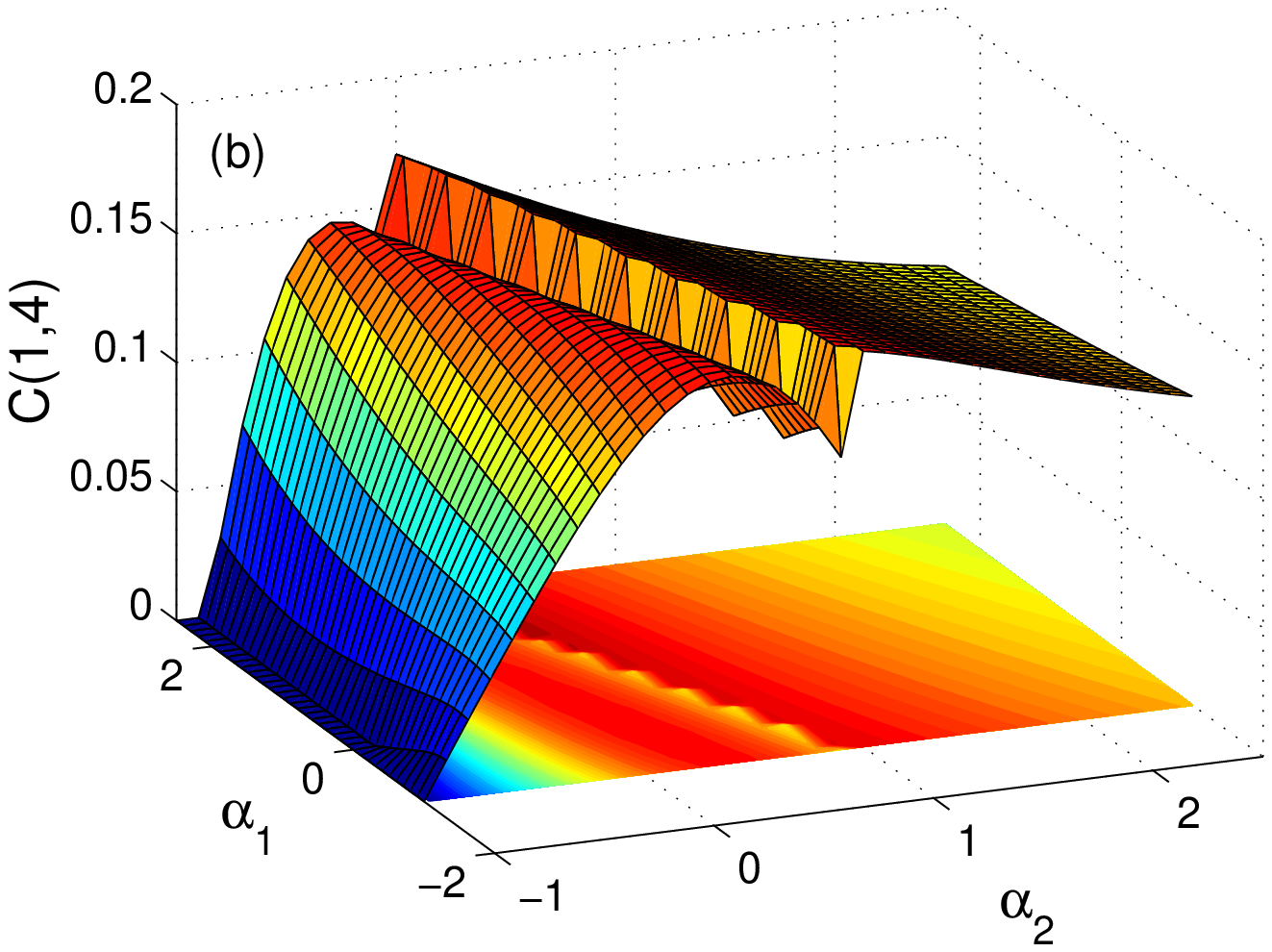}}\\
   \subfigure{\includegraphics[width=8 cm]{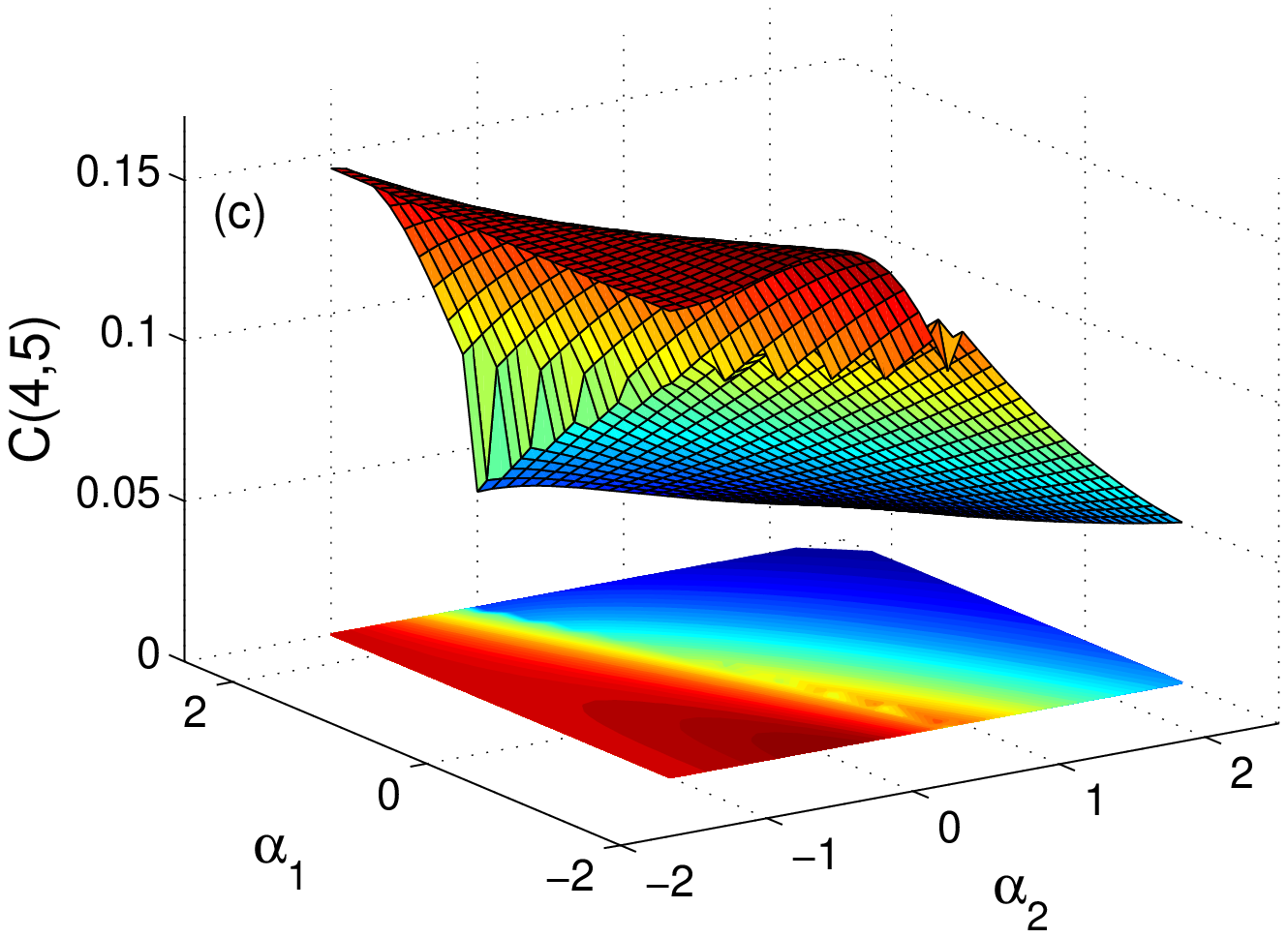}}\quad
   \subfigure{\includegraphics[width=8 cm]{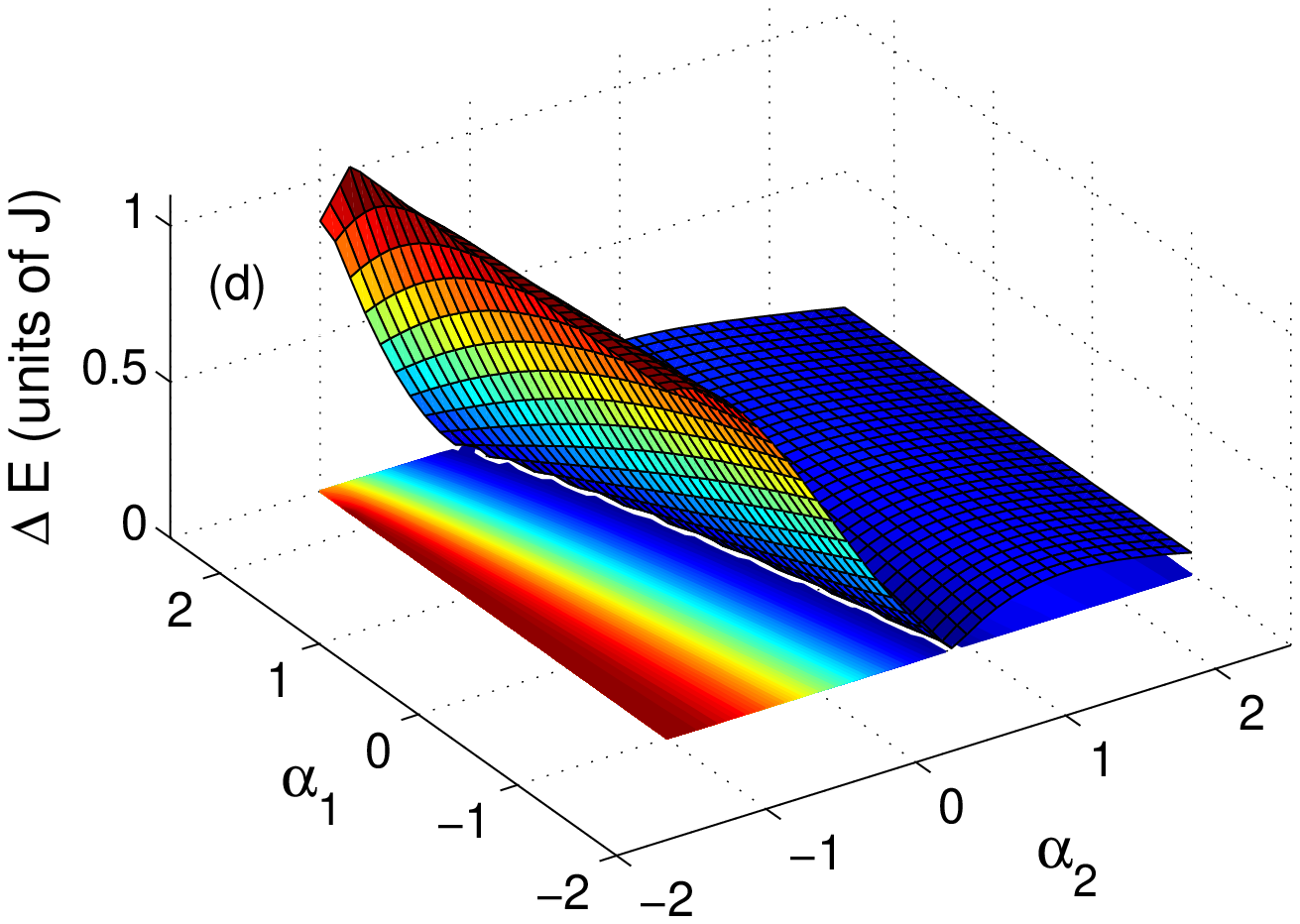}}
   \caption{{\protect\footnotesize The concurrence $C(1,2)$, $C(1,4)$, $C(4,5)$ versus the impurity coupling strengths $\alpha_1$ and $\alpha_2$ with double impurities at sites 1 and 2 for the two dimensional partially anisotropic lattice ($\gamma = 0.5$) in an external magnetic field h=2.}}
 \label{Imp12_G05_h2}
 \end{minipage}
\end{figure}
The partially anisotropic system, $\gamma=0.5$, with double impurity at sites 1 and 2 and under the effect of the external magnetic field $h=2$ is explored in fig.~\ref{Imp12_G05_h2}. As one can see, the overall behavior specially at the border values of the impurity strengths is the same as observed in the Ising case except that the concurrences suffer a local minimum within a small range of the impurity strength $\alpha_2$ between 0 and 1 while corresponding to the whole $\alpha_1$ range. The change of the entanglement around this local minimum takes a step-like profile which is very clear in the case of the concurrence $C(1,4)$ shown in fig.~\ref{Imp12_G05_h2}(b). Remarkably, the local minima in the plotted concurrences coincide with the line of vanishing energy gap as shown in fig.~\ref{Imp12_G05_h2}(d).

\subsubsection{Entanglement and quantum phase transition}
\begin{figure}
\subfigure{\includegraphics[width=0.48\textwidth,height=0.3\textheight]{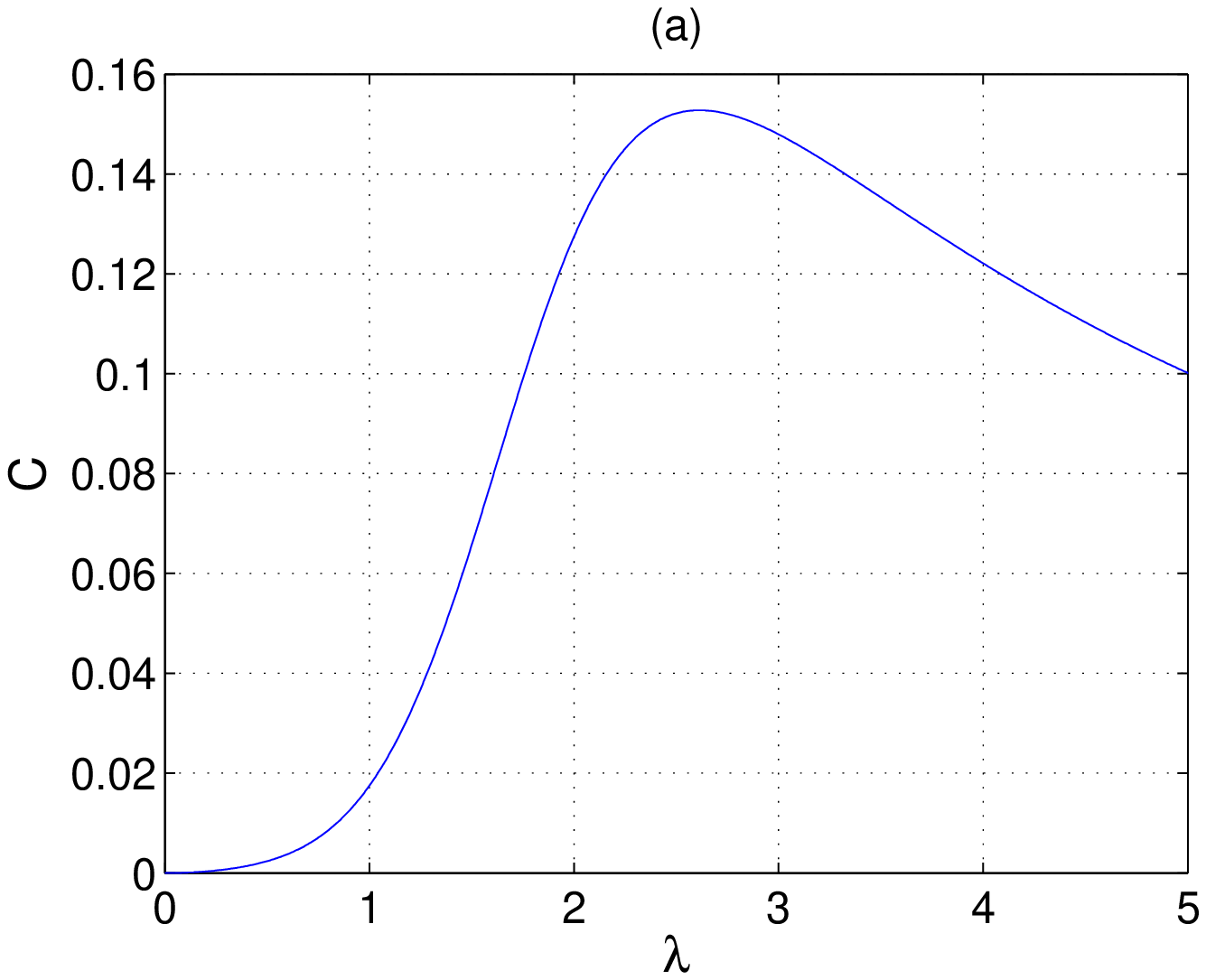}}\quad
\subfigure{\includegraphics[width=0.48\textwidth,height=0.3\textheight]{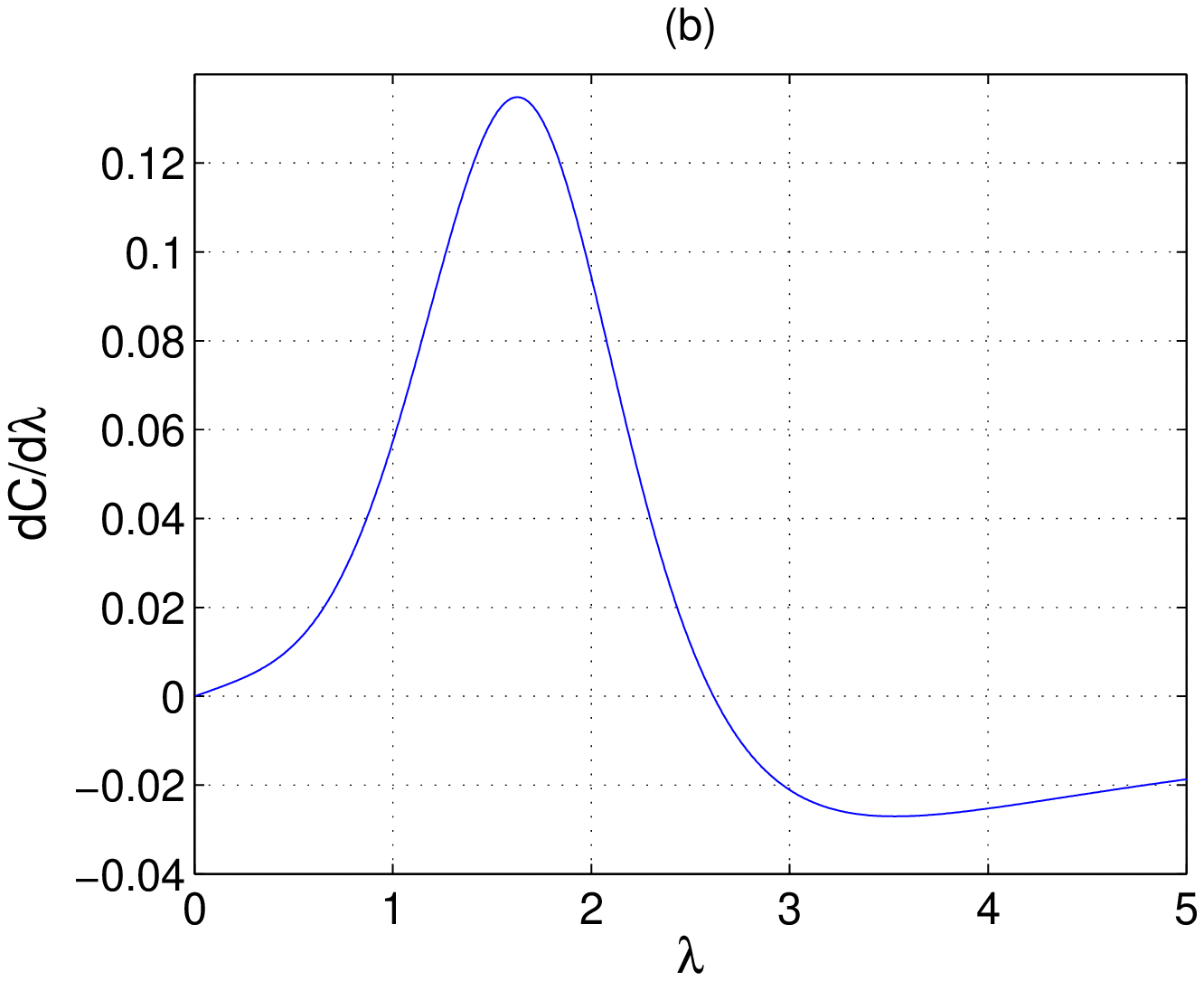}}
\subfigure{\includegraphics[width=0.48\textwidth,height=0.3\textheight]{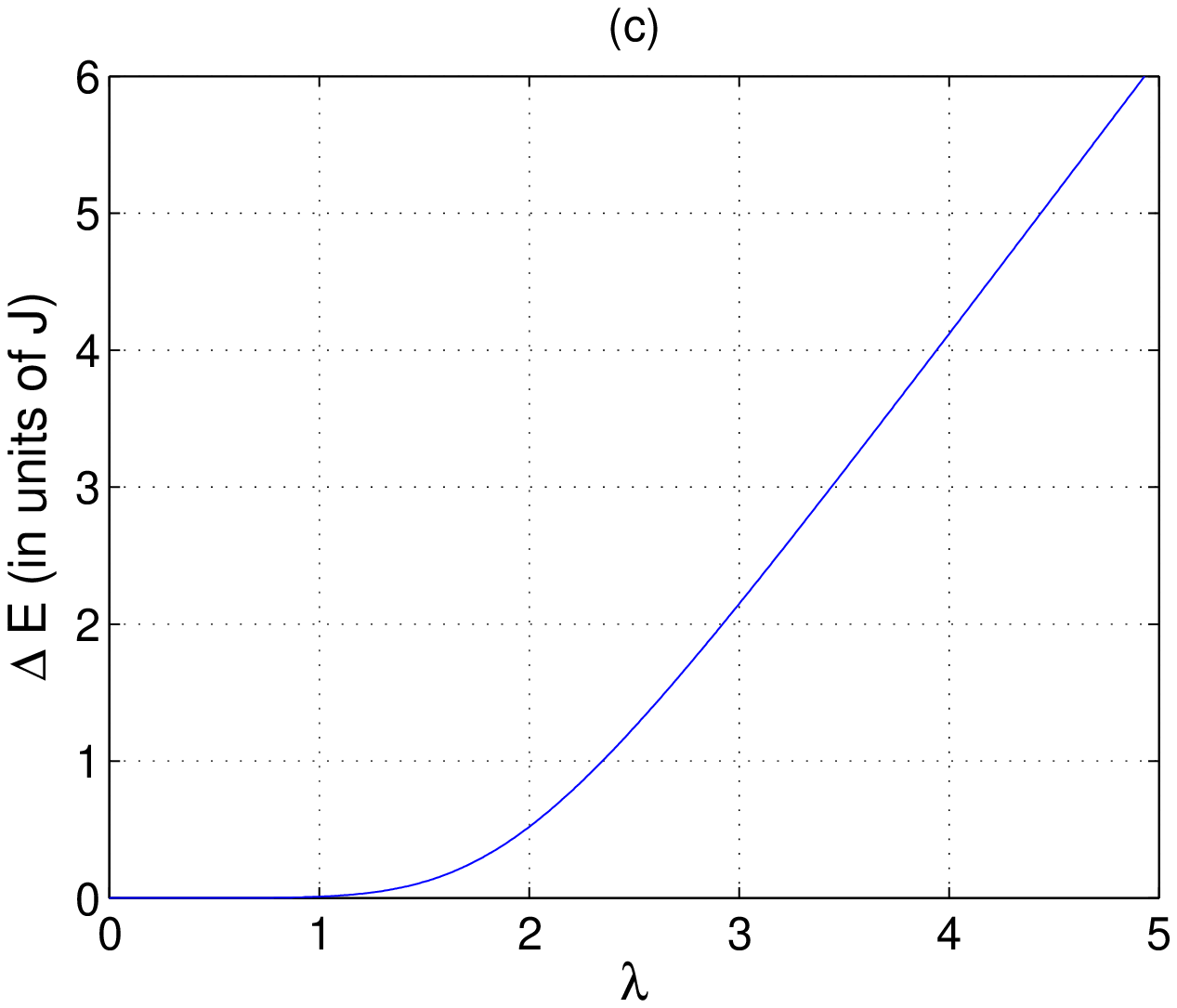}}\quad
\subfigure{\includegraphics[width=0.48\textwidth,height=0.3\textheight]{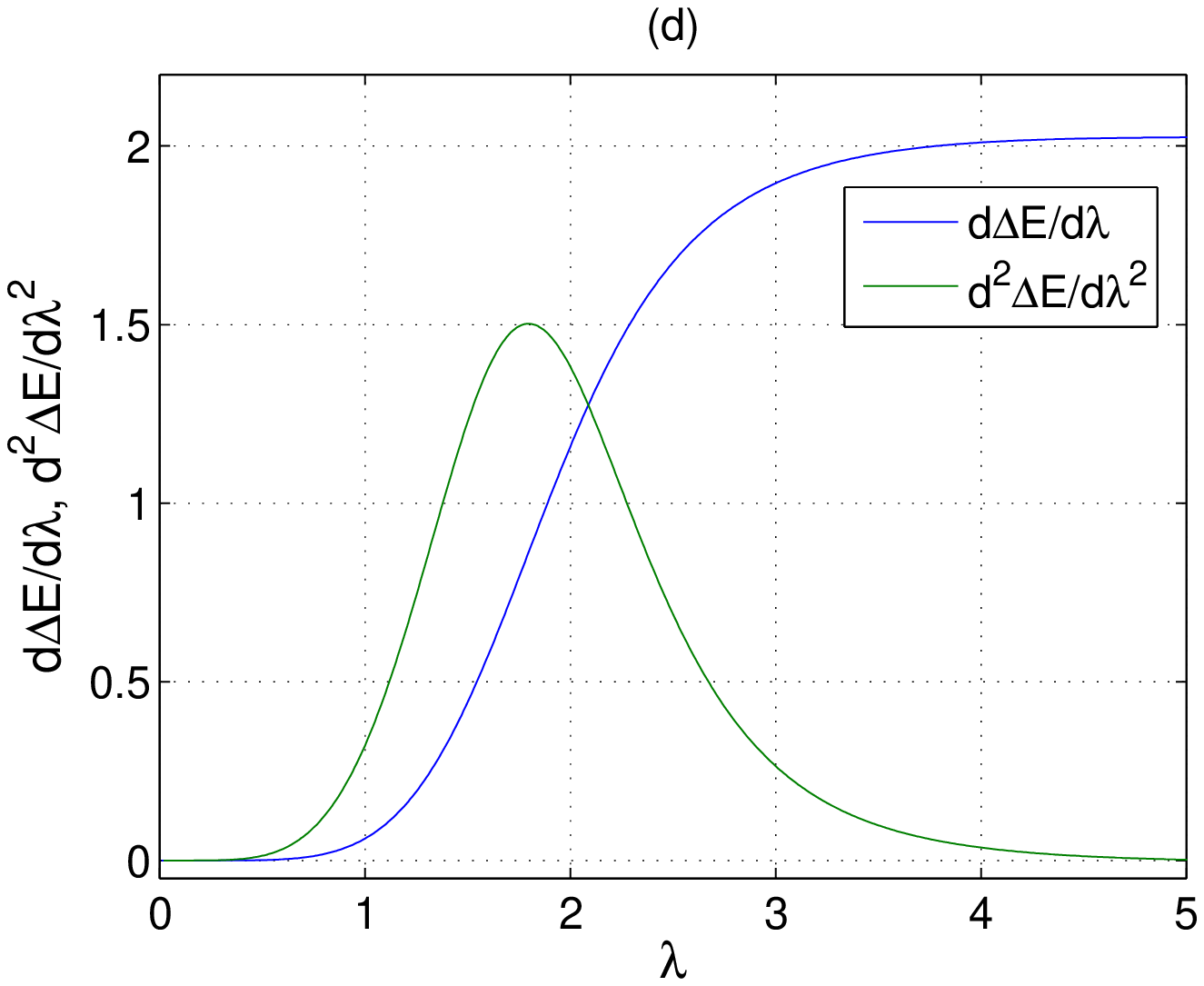}}
\caption{{\protect\footnotesize (a) The concurrence $C_{14}$ versus $\lambda$; (b) the first derivative of the concurrence $C_{14}$ with respect to $\lambda$ versus $\lambda$; (c) the energy gap between the ground state and first excited state versus $\lambda$; (d) the first derivative (in units of $J$) and second derivative (in units of $J^2$) of the energy gap with respect $\lambda$ versus $\lambda$ for the pure Ising system ($\gamma=1$ and $\alpha=0$).}}
\label{QQPT1}
\end{figure}
\begin{figure}
\subfigure{\includegraphics[width=0.48\textwidth,height=0.3\textheight]{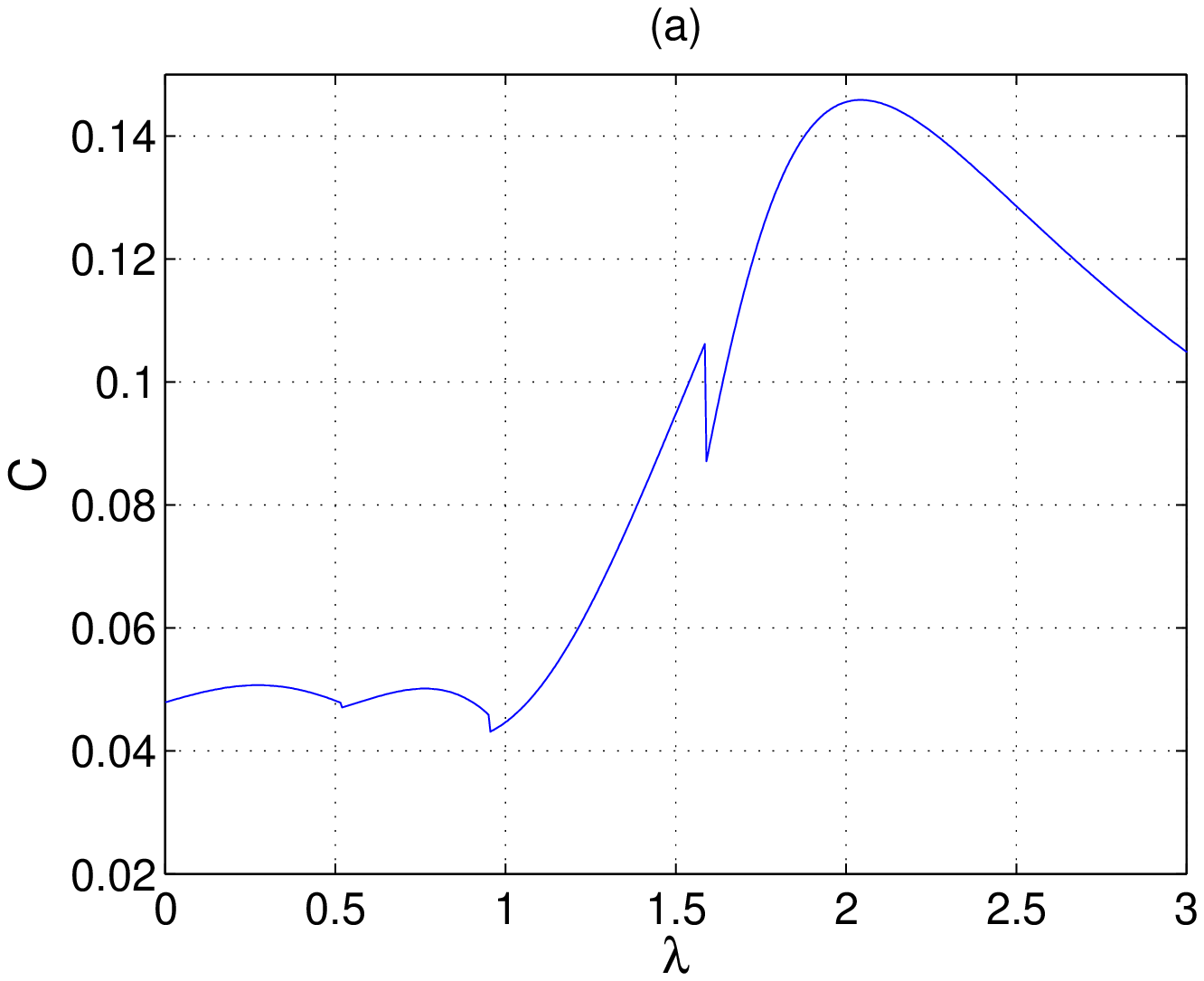}}\quad
\subfigure{\includegraphics[width=0.48\textwidth,height=0.3\textheight]{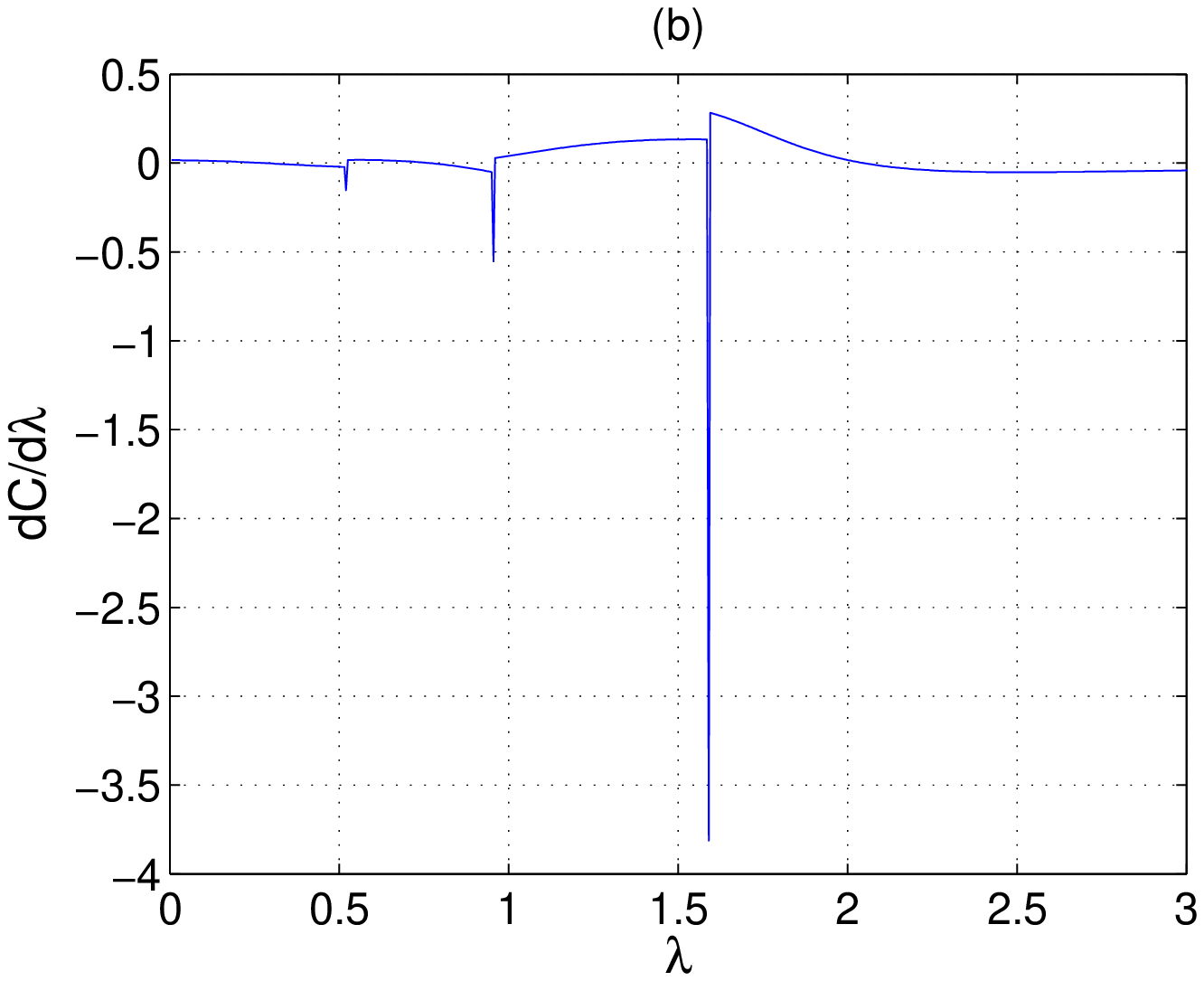}}
\subfigure{\includegraphics[width=0.48\textwidth,height=0.3\textheight]{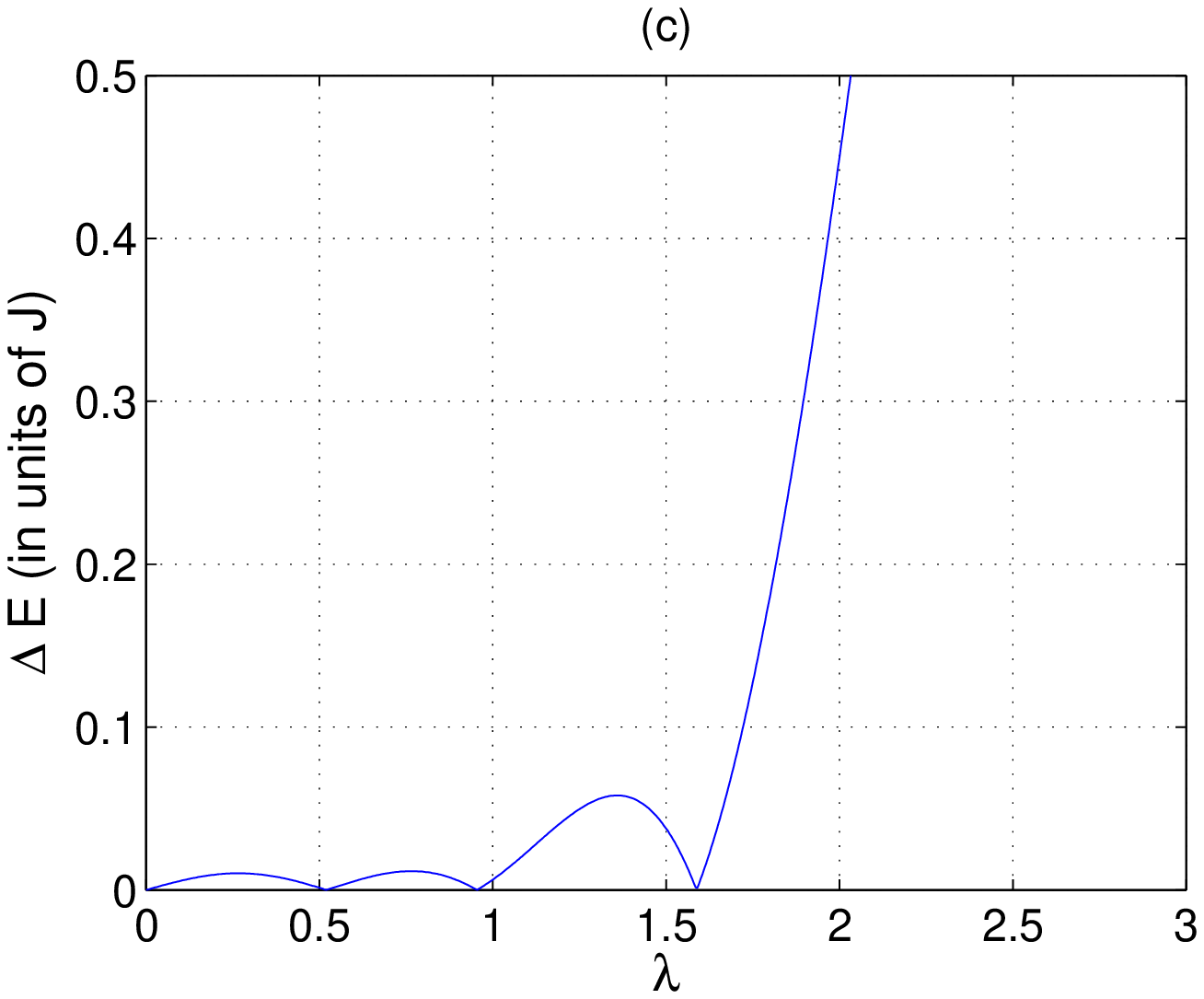}}\quad
\subfigure{\includegraphics[width=0.48\textwidth,height=0.3\textheight]{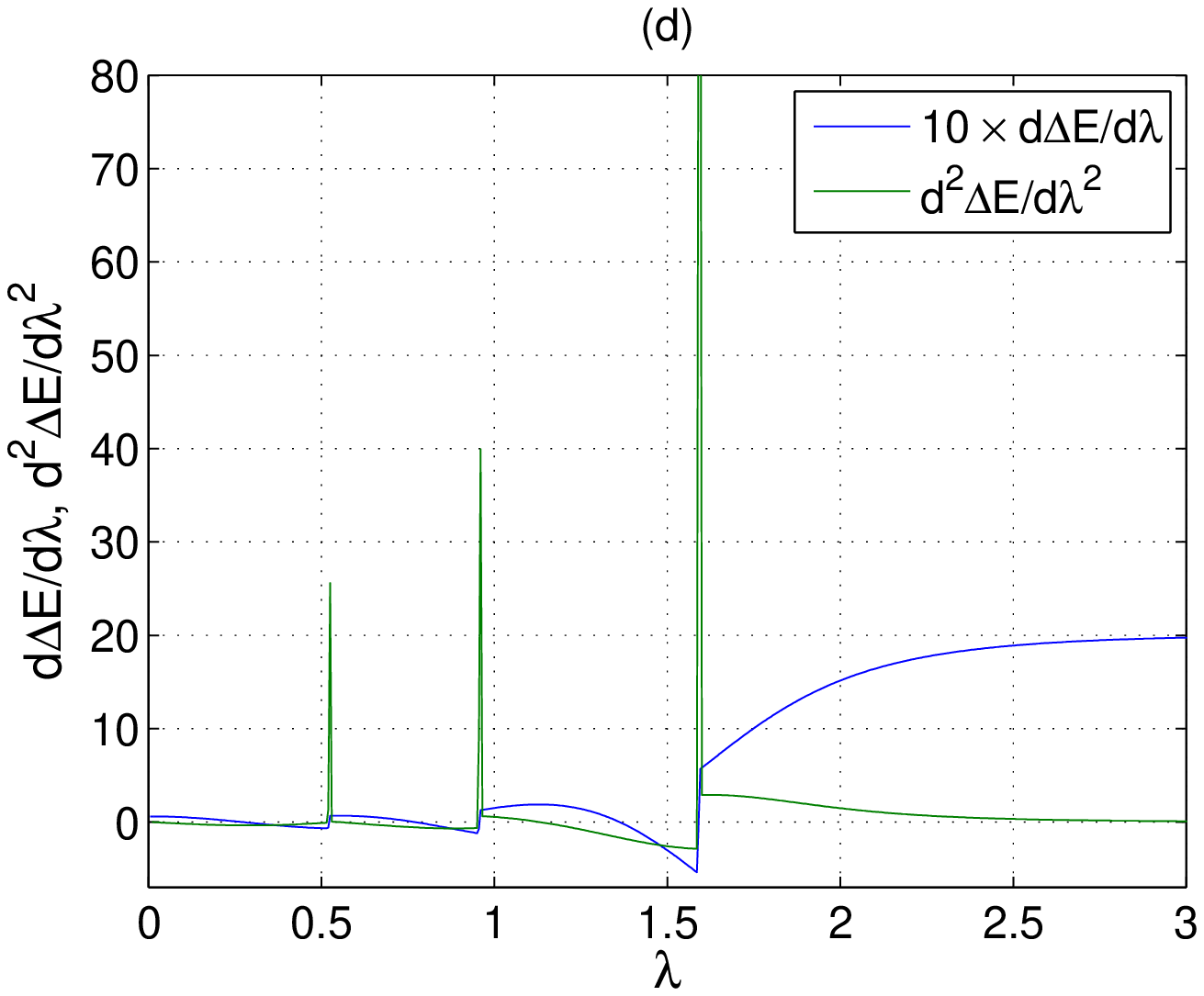}}
\caption{{\protect\footnotesize (a) The concurrence $C_{14}$ versus $\lambda$; (b) the first derivative of the concurrence $C_{14}$ with respect to $\lambda$ versus $\lambda$; (c) the energy gap between the ground state and first excited state versus $\lambda$; (d) the first derivative (in units of $J$) and second derivative (in units of $J^2$) of the energy gap with respect $\lambda$ versus $\lambda$ for the pure partially anisotropic system ($\gamma=0.5$ and $\alpha=0$). Notice that the first derivative of energy gap is enlarged 10 times its actual scale for clearness.}}
\label{QQPT2}
\end{figure}
\begin{figure}
\subfigure{\includegraphics[width=0.48\textwidth,height=0.3\textheight]{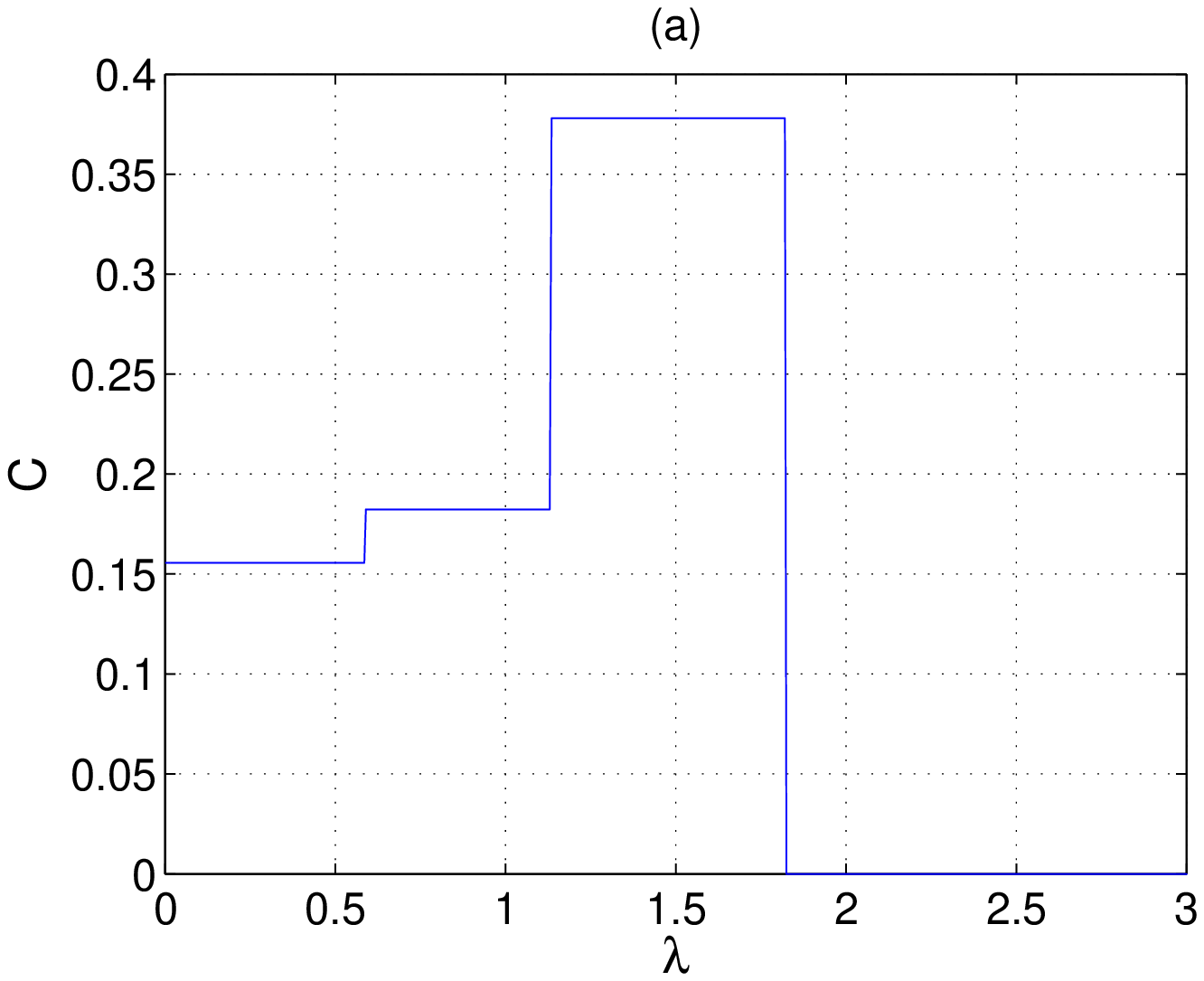}}\quad
\subfigure{\includegraphics[width=0.48\textwidth,height=0.3\textheight]{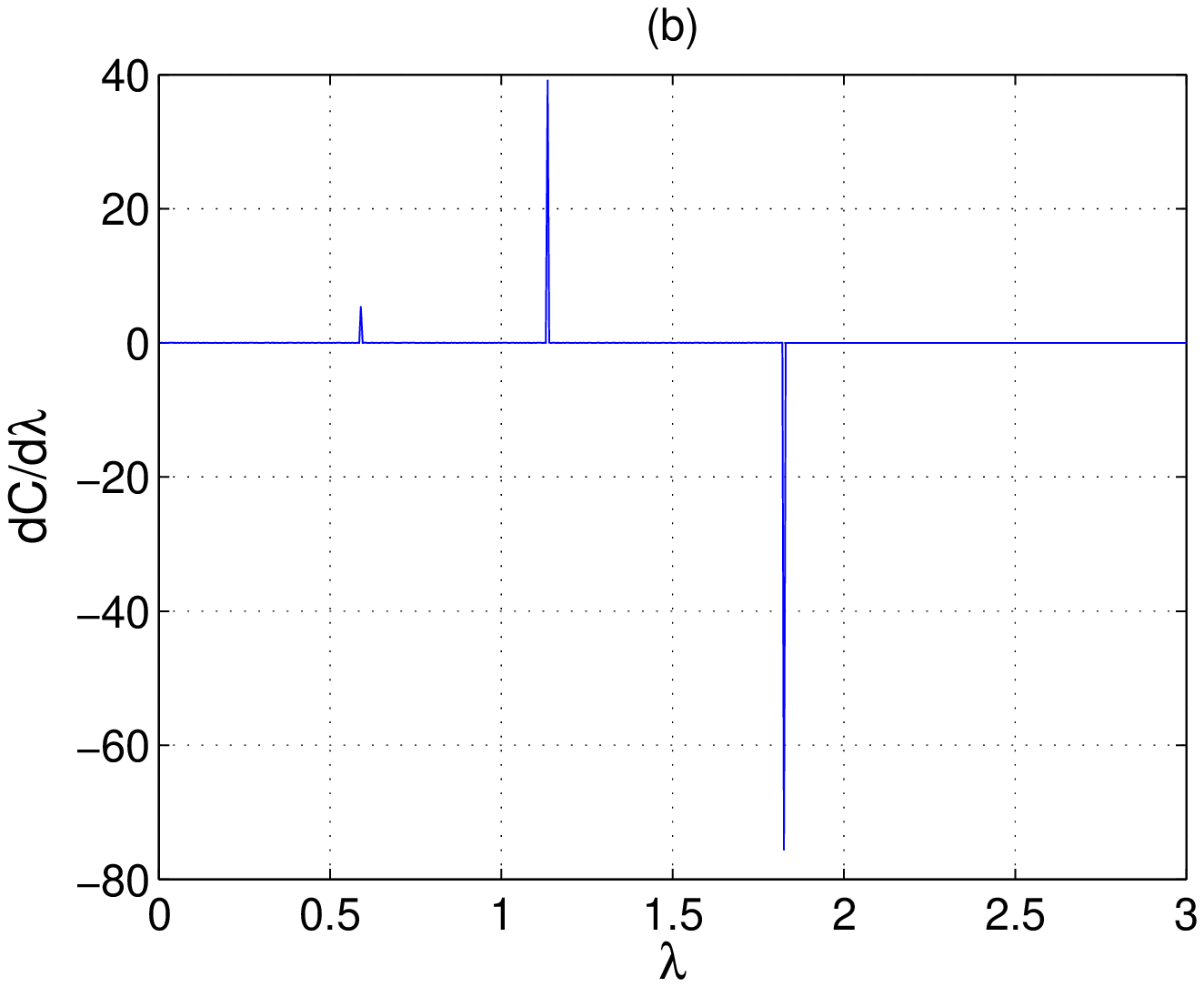}}
\subfigure{\includegraphics[width=0.48\textwidth,height=0.3\textheight]{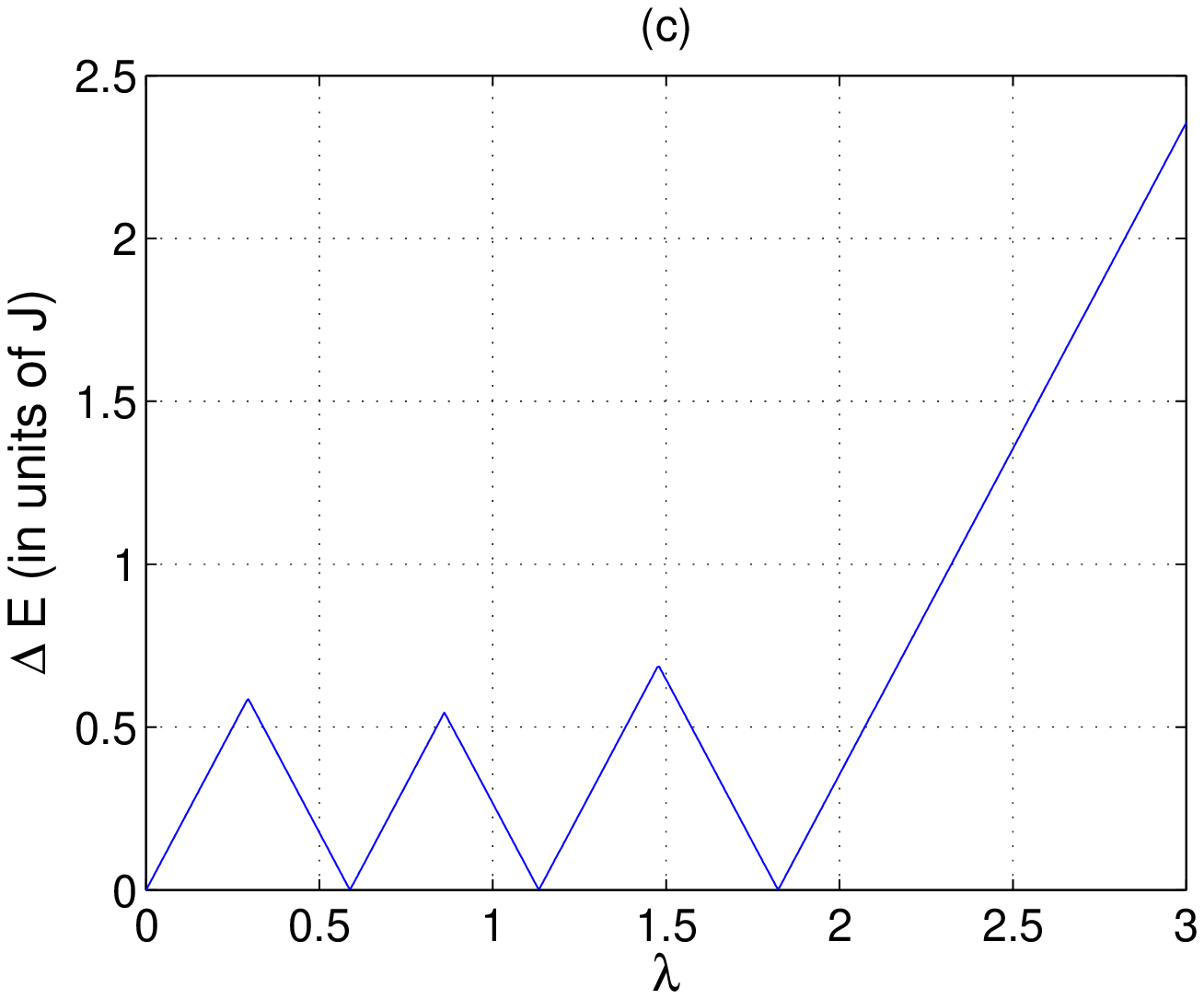}}\quad
\subfigure{\includegraphics[width=0.48\textwidth,height=0.3\textheight]{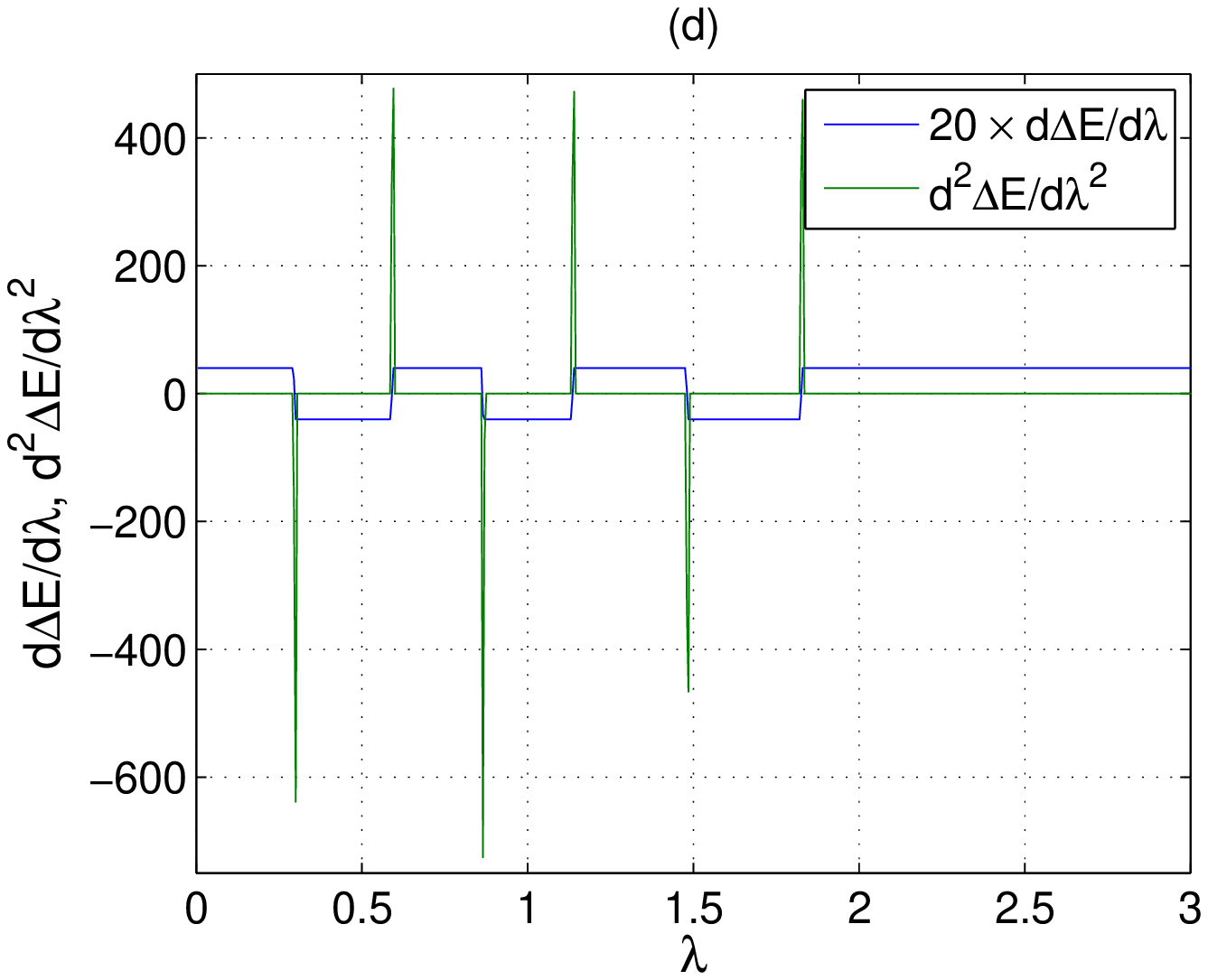}}
\caption{{\protect\footnotesize (a) The concurrence $C_{14}$ versus $\lambda$; (b) the first derivative of the concurrence $C_{14}$ with respect to $\lambda$ versus $\lambda$; (c) the energy gap between the ground state and first excited state versus $\lambda$; (d) the first derivative (in units of $J$) and second derivative (in units of $J^2$) of the energy gap with respect $\lambda$ versus $\lambda$ for the pure isotropic system ($\gamma=0$ and $\alpha=0$). Notice that the first derivative of energy gap is enlarged 20 times its actual scale for clearness.}}
\label{QQPT3}
\end{figure}
Critical quantum behavior in a many body system happens either when an actual crossing takes place between the excited state and the ground state or a limiting avoided level-crossing between them exists, i.e. an energy gab between the two states that vanishes in the infinite system size limit at the critical point \cite{Sachdev2001}. When a many body system crosses a critical point, significant changes in both its wave function and ground state energy takes place, which are manifested in the behavior of the entanglement function. The entanglement in one dimensional infinite spin systems, Ising and $XY$, was shown to demonstrate scaling behavior in the vicinity of critical points \cite{Osterloh2002}. The change in the entanglement across the critical point was quantified by considering the derivative of the concurrence with respect to the parameter $\lambda$. This derivative was explored versus $\lambda$ for different system sizes and although it didn't show divergence for finite system sizes, it showed clear anomalies which developed to a singularity at the thermodynamic limit. The ground state of the Heisenberg spin model is known to have a double degeneracy for an odd number of spins which is never achieved unless the thermodynamic limit is reached \cite{Sachdev2001}. Particularly, the Ising 1D spin chain in an external transverse magnetic field has doubly degenerate ground state in a ferromagnetic phase that is gapped from the excitation spectrum by $2 J (1-h/J)$, which is removed at the critical point and the system becomes in a paramagnetic phase. Now let us first consider our two-dimensional finite size Ising spin system. The concurrence $C_{14}$ and its first derivative are depicted versus $\lambda$ in figs.~\ref{QQPT1}(a) and (b) respectively. As one can see, the derivative of the concurrence shows strong tendency of being singular at $\lambda_c = 1.64$. The characteristics of the energy gap between the ground state and the first excited state as a function of $\lambda$ are explored in fig.~\ref{QQPT1}(c). The system shows strict double degeneracy, zero energy gap, only at $\lambda =0$  i.e. at zero magnetic field, but once the magnetic field is on the degeneracy is lifted and an extremely small energy gap develops, which increase very slowly for small magnetic field values but increases abruptly at certain $\lambda$ value. It is important to emphasis here that at $\lambda=0$, regardless of which one of the double ground states is selected for evaluating the entanglement, the same value is obtained. The critical point of a phase transition should be characterized by a singularity in the ground state energy, and an abrupt change in the energy gap of the system as a function of the system parameter as it crosses the critical point. To better understand the behavior of the energy gap across the prospective critical point and identify it, we plot the first and second derivatives of the energy gap as a function of $\lambda$ in fig.~\ref{QQPT1}(d). Interestingly, the first derivative $d\Delta E / d \lambda$ which represents the rate of change of the energy gap as a function of $\lambda$ starts with a zero value at $\lambda=0$ and then increase very slowly before it shows a great rate of change and finally reaches a saturation value. This behavior is best represented by the second derivative $d^2 \Delta E / d \lambda^2$, which shows strong tendency of being singular at $\lambda_c=1.8$, which indicates the highest rate of change the energy gap as a function of $\lambda$. The reason for the small discrepancy between the two values of the $\lambda_c$ extracted from the $dC / d \lambda$ plot and the one of $d^2 \Delta E / d \lambda^2$ is that the concurrence $C_{14}$ is only between two sites and does not represent the whole system in contrary to the energy gap. One can conclude that the rate of change of the energy gap as a function of the system parameter, $\lambda$ in our case, should be maximum across the critical point. Turning to the case of the partially anisotropic spin system, $\gamma=0.5$, presented in fig.~\ref{QQPT2}, one can notice from fig.~\ref{QQPT2}(a) that the concurrence shows few sharp changes, which is reflected in the energy gap plot as an equal number of minima as shown in fig.~\ref{QQPT2}(b). Nevertheless, again there is only one strict double degeneracy at $\lambda=0$ while the other three energy gap minima are non-zero and in the order of $10^{-5}$. It is interesting to notice that the anomalies in both $dC / d \lambda$ and $d^2 \Delta E / d \lambda^2$ are much stronger and sharper compared with the Ising case as shown in figs.~\ref{QQPT2}(c) and (d). Finally the isotropic system which is depicted in fig.~\ref{QQPT3}, shows even sharper energy gap changes as a result of the sharp changes in the concurrence and the anomalies in the derivatives $dC / d \lambda$ and $d^2 \Delta E / d \lambda^2$ are even much stronger than the previous two cases.
\begin{acknowledgments}
We would like to thank our collaborators Bedoor Alkurtass, Dr. Zhen Huang and Dr. Omar Aldossary  for their contributions to the studies of entanglement of formation and dynamics of one-dimensional magnetic systems.
We would like also to acknowledge the financial support of the Saudi NPST (project no. 11-MAT1492-02) and the deanship of scientific research, King Saud University . We are also grateful to the USA Army research office for partial support of this work at Purdue.
\end{acknowledgments}

\end{document}